\documentclass[prc,aps,float,nofootinbib,superscriptaddress,byrevtex,twocolumn]{revtex4-1}
\usepackage{amssymb}
\usepackage{amsmath}
\usepackage{amsfonts}
\usepackage{epsfig}
\usepackage{bbm}
\usepackage{comment}

\usepackage{multirow}
\usepackage{longtable}
\usepackage{array}
\usepackage{subfigure}
\usepackage{graphics}

\newcommand\bbone{\ensuremath{\mathbbm{1}}}

\usepackage[usenames,dvipsnames]{color}

\allowdisplaybreaks
\begin{document}
\title{Symmetry broken and restored coupled-cluster theory\\
II. Global gauge symmetry and particle number}
\author{T. Duguet}
\email{thomas.duguet@cea.fr} 
\affiliation{CEA-Saclay DSM/Irfu/SPhN, F-91191 Gif sur Yvette Cedex, France} 
\affiliation{KU Leuven, Instituut voor Kern- en Stralingsfysica, 3001 Leuven, Belgium}
\affiliation{National Superconducting Cyclotron Laboratory and Department of Physics and Astronomy, Michigan State University, East Lansing, MI 48824, USA}

\author{A. Signoracci}
\email{asignora@utk.edu}
\affiliation{Department of Physics and Astronomy, University of Tennessee, Knoxville, TN 37996, USA}
\affiliation{Physics Division, Oak Ridge National Laboratory, Oak Ridge, TN 37831, USA}

\date{\today}
%
%
\begin{abstract}
We have recently extended many-body perturbation theory and coupled-cluster theory performed on top of a Slater determinant breaking rotational symmetry to allow for the restoration of the angular momentum at any truncation order [T. Duguet, J. Phys. G: Nucl. Part. Phys. 42 (2015) 025107]. Following a similar route, we presently extend Bogoliubov many-body perturbation theory and Bogoliubov coupled cluster theory performed on top of a Bogoliubov reference state breaking global gauge symmetry to allow for the restoration of the particle number at any truncation order. Eventually, formalisms can be merged to handle $SU(2)$ and $U(1)$ symmetries at the same time. Several further extensions of the newly proposed many-body formalisms can be foreseen in the mid-term future. The long-term goal relates to the ab initio description of near-degenerate finite quantum systems with an open-shell character. 
\end{abstract}
\maketitle

%

\section{Introduction}
\label{introduction}

In Ref.~\cite{Duguet:2014jja}, hereafter referred to as Paper I, the motivations to tackle degenerate (or near-degenerate) finite quantum systems with an open-shell character via ab initio methods relying on the concept of symmetry breaking and restoration were explained at length. Dealing with singly-open shell atomic nuclei, i.e. nuclei displaying a good closed-shell character for either protons or neutrons, requires the breaking and the restoration of $U(1)$ global gauge symmetry associated with particle number conservation. Breaking the symmetry allows one to lift the degeneracy of the unperturbed reference state associated with the Cooper pair instability responsible for nuclear superfluidity. Extending the treatment to doubly open-shell nuclei demands to further break and restore $SU(2)$ symmetry associated with the conservation of angular momentum\footnote{Although it can be beneficial to indeed break and restore both symmetries, it is possible to limit oneself to breaking and restoring $SU(2)$ symmetry in this case. If breaking $U(1)$ symmetry, one must do it both for neutrons and protons.}. 

Standard single-reference Rayleigh-Schroedinger many-body perturbation theory (MBPT)~\cite{goldstone57a,hugenholtz57a,bloch58a,kutzelnigg09a,shavitt09a} and coupled cluster (CC) theory~\cite{coester58a,shavitt09a} expand the exact many-body ground-state energy around a reference state taking the form of a Slater determinant. Consequently, these methods do respect $U(1)$ global gauge symmetry all throughout. To allow for the breaking of $U(1)$ symmetry, one must expand the exact many-body state around a more general vacuum taking the form of a Bogoliubov product state~\cite{ring80a}. As for perturbation theory, this leads to formulating single-reference Bogoliubov many-body perturbation theory (SR-BMBPT) or its variant based on a simpler Bardeen-Cooper-Schrieffer reference state~\cite{mehta1,balian62a,henley}. In either form, such a particle-number-breaking many-body perturbation theory has been scarcely used in the physics literature. Going beyond perturbation theory, the single-reference Bogoliubov coupled-cluster (SR-BCC) theory was only recently formulated and applied~\cite{StolarczykMonkhorst,Signoracci:2014dia,Henderson:2014vka}. 

While BMBPT and BCC ab initio methods can efficiently access open-shell systems, it remains necessary to restore $U(1)$ symmetry when dealing with a finite quantum system such as the atomic nucleus. Consequently, it is the goal of the present paper to generalize BMBPT and BCC formalisms to allow for the exact restoration of good neutron (proton) number at any truncation order. This will lead to the design of {\it particle-number-restored Bogoliubov many-body perturbation theory} (PNR-BMBPT)\footnote{We will employ a general Rayleigh-Schroedinger scheme that can be eventually reduced to a Moller-Plesset scheme by using the Bogoliubov state solution of Hartree-Fock-Bogoliubov equations~\cite{ring80a} as a reference state.} and {\it particle-number-restored Bogoliubov coupled cluster} (PNR-BCC) theory. This is achieved by adapting the work done for the $SU(2)$ group in Paper I to the $U(1)$ group, which effectively requires the entire reformulation of the formalism on the basis of a Bogoliubov reference state and making use of Bogoliubov algebra. 

The paper is organized as follows. Section~\ref{definitions} provides the ingredients necessary to set up the formalism while Sec.~\ref{normkernel} elaborates on the general principles of the approach, independently of the actual many-body technique eventually employed to expand the exact solution of the Schroedinger equation. In Sec.~\ref{sectionMBPT}, a generalized BMBPT is developed and acts as the foundation for the generalized BCC approach introduced in Sec.~\ref{CCtheory}. It is shown how generalized energy and norm kernels at play in the formalism can be computed from {\it naturally terminating} BCC expansions. The way to recover SR-BMBPT and SR-BCC theory on the one hand and particle-number-projected Hartree-Fock-Bogoliubov theory on the other hand is illustrated. Eventually, the algorithm to be implemented by the owner of a BCC code to incorporate the particle-number restoration is highlighted. The body of the paper is restricted to discussing the overall scheme, limiting technical details to the minimum. Complete analytic results are provided in an extended set of appendices.

\section{Basic ingredients}
\label{definitions}

Let us introduce necessary ingredients to make the paper self-contained. Although pedestrian, this section displays definitions and identities that are crucial to the building of the formalism later on.

\subsection{Hamiltonian}

Let the Hamiltonian $H=T+V$ of the system be of the form\footnote{The formalism can be extended to a Hamiltonian containing three- and higher-body forces without running into any fundamental problem. Also, one subtracts the center of mass kinetic energy to the Hamiltonian in actual calculations of finite nuclei. As far as the present work is concerned, this simply leads to a redefinition of one- and two-body matrix elements $t_{pq}$ and $\bar{v}_{pqrs}$ in the Hamiltonian without changing any aspect of the many-body formalism that follows.}
\begin{equation}
H \equiv  \frac{1}{(1!)^2} \sum _{pq} t_{pq} c^{\dagger}_{p} c_{q}+\frac{1}{(2!)^2} \sum _{pqrs} \bar{v}_{pqrs}  c^{\dagger}_{p} c^{\dagger}_{q} c_{s} c_{r}  \, , \label{e:ham} 
\end{equation} 
where antisymmetric matrix elements of the two-body interaction are employed and where $\{c_{p};c^{\dagger}_{p}\}$ denote particle annihilation and creation operators associated with an arbitrary basis of the one-body Hilbert space ${\cal H}_1$.

\subsection{Bogoliubov algebra}

The unitary Bogoliubov transformation connects quasi-particle annihilation and creation operators $\{\beta_{k};\beta^{\dagger}_{k}\}$ to particle ones through~\cite{ring80a}
\begin{subequations}
\label{e:p2qp}
\begin{align}
\beta_{k} &= \sum_{p} U^{*}_{pk} \, c_{p} 
 + V^{*}_{pk} \,  c^{\dagger}_{p} \, , \\
\beta_{k}^{\dagger} &= \sum_{p} U_{pk} \, c^{\dagger}_{p} 
 + V_{pk} \,  c_{p} \, .
\end{align}
\end{subequations}
Both sets of operators obey anticommutation rules
\begin{subequations}
\label{anticom}
\begin{align}
\{c_{p},c_{q}\}=0 \,\,\,\,\,\,\,\, &; \,\,\, \{\beta_{k_1},\beta_{k_2}\}=0    \, , \\
\{c^{\dagger}_{p},c^{\dagger}_{q}\}=0 \,\,\,\,\,\,\,\, &; \,\,\, \{\beta^{\dagger}_{k_1},\beta^{\dagger}_{k_2}\}=0   \, , \\
\{c_{p},c^{\dagger}_{q}\} = \delta_{pq} \,\,\, &; \,\,\, \{\beta_{k_1},\beta^{\dagger}_{k_2}\} = \delta_{k_1k_2}   \, .
\end{align}
\end{subequations}

The Bogoliubov product state, which carries even number-parity as a quantum number, is defined as 
\begin{equation}
\label{e:bogvac}
| \Phi \rangle \equiv \mathcal{C} \displaystyle \prod_{k} \beta_{k} | 0 \rangle \, ,
\end{equation}
and is the vacuum of the quasiparticle operators, i.e. $\beta_k | \Phi \rangle =0$ for all $k$. In Eq.~\eqref{e:bogvac}, $\mathcal{C}$ is a complex normalization ensuring that $\langle \Phi | \Phi \rangle =1$. As quasiparticle operators mix particle creation and annihilation operators (see Eq.~\eqref{e:p2qp}), the Bogoliubov vacuum breaks  $U(1)$ symmetry associated with particle number conservation, i.e. $| \Phi \rangle$ is not an eigenstate of the particle-number operator $A$, except in the limit where it reduces to a Slater determinant.

The Bogoliubov transformation can be written in matrix form
\begin{equation}
\left(
\begin{array} {c}
\beta \\
\beta^{\dagger}
\end{array}
\right) = W^{\dagger} \left(
\begin{array} {c}
c \\
c^{\dagger}
\end{array}
\right) \, ,
\end{equation}
where
\begin{equation}
W \equiv \left(
\begin{array} {cc}
U & V^{\ast} \\
V &  U^{\ast}
\end{array}
\right) \, . \label{bogomatrix}
\end{equation}
One can further define the skew-symmetric matrix 
\begin{equation}
Z\equiv V^{\ast}[U^{\ast}]^{-1}
\end{equation}
in terms of which $| \Phi \rangle$ can be expressed by virtue of Thouless' theorem~\cite{thouless60}. The anticommutation rules obeyed by the quasi-particle operators relate to the unitarity of $W$ that leads to four relations 
\begin{subequations}
\label{unitarity}
\begin{eqnarray}
UU^{\dagger} + V^{\ast}V^{T} &=& 1 \, , \label{unitarityA} \\
VU^{\dagger} + U^{\ast}V^{T} &=& 0 \, , \label{unitarityB} \\
U^{\dagger}U + V^{\dagger}V  &=& 1 \, , \label{unitarityC} \\
V^{T}U + U^{T}V  &=& 0 \, , \label{unitarityD} 
\end{eqnarray}
originating from $W^{\dagger}W = 1$ and  four relations
\begin{eqnarray}
UV^{\dagger} + V^{\ast}U^{T} &=& 0 \, , \label{unitarityE} \\
VV^{\dagger} + U^{\ast}U^{T} &=& 1 \, , \label{unitarityF} \\
U^{\dagger}V^{\ast} + V^{\dagger}U^{\ast}  &=& 0 \, , \label{unitarityG} \\
V^{T}V^{\ast} + U^{T}U^{\ast}  &=& 1 \, , \label{unitarityH} 
\end{eqnarray}
\end{subequations}
originating from $WW^{\dagger} = 1$.

The Bogoliubov state $| \Phi \rangle$ is fully characterized by the generalized density matrix~\cite{ring80a}
\begin{subequations}
\label{generalizeddensitymatrix} 
\begin{eqnarray}
{\cal R} &\equiv& 
\left(
\begin{array} {cc}
\frac{\langle \Phi | c^{\dagger}c^{\phantom{\dagger}} | \Phi \rangle}{\langle \Phi | \Phi \rangle} & \frac{\langle \Phi | c^{\phantom{\dagger}}c^{\phantom{\dagger}} | \Phi \rangle}{\langle \Phi | \Phi \rangle} \\
\frac{\langle \Phi | c^{\dagger}c^{\dagger} | \Phi \rangle}{\langle \Phi | \Phi \rangle} &  \frac{\langle \Phi | c^{\phantom{\dagger}}c^{\dagger} | \Phi \rangle}{\langle \Phi | \Phi \rangle}
\end{array}
\right) \\
&\equiv& 
\left(
\begin{array} {cc}
+\rho & +\kappa \\
-\bar{\kappa}^{\ast} &  -\sigma^{\ast}
\end{array}
\right)  \\
&=& 
\left(
\begin{array} {cc}
V^{\ast}V^T & V^{\ast}U^T \\
U^{\ast}V^T &  U^{\ast}U^T
\end{array}
\right)
\, ,
\end{eqnarray}
\end{subequations}
where $\rho$ and $\kappa$ denote the normal one-body density matrix and the anomalous density matrix (or pairing tensor), respectively. Using anticommutation rules of particle creation and annihilation operators, one demonstrates that 
\begin{subequations}
\label{densitymatrices}
\begin{eqnarray}
\rho_{qp} &=& +\rho^{\ast}_{pq}  \, , \label{densitymatrices1} \\
\kappa_{qp} &=& - \kappa_{pq} \, , \label{densitymatrices2} \\
\sigma_{qp} &=& +\rho_{qp} - \delta_{qp} \, , \label{densitymatrices3} \\
\bar{\kappa}_{qp} &=& +\kappa_{qp} \, , \label{densitymatrices4} 
\end{eqnarray}
\end{subequations}
meaning that $\rho$ is hermitian (i.e. $\rho^{\dagger} = \rho$) while $\kappa$ is skew-symmetric (i.e. $\kappa^{T} = - \kappa$). Transforming the generalized density matrix to the quasi-particle basis via ${\bf R} \equiv W^{\dagger} {\cal R} W $ leads to
\begin{subequations}
\label{generalizeddensitymatrix2}
\begin{eqnarray}
{\bf R} &=& 
\left(
\begin{array} {cc}
\frac{\langle \Phi | \beta^{\dagger}\beta^{\phantom{\dagger}}\, | \Phi \rangle}{\langle \Phi | \Phi \rangle} & \frac{\langle \Phi | \beta^{\phantom{\dagger}}\beta^{\phantom{\dagger}} | \Phi \rangle}{\langle \Phi | \Phi \rangle} \\
\frac{\langle \Phi | \beta^{\dagger}\beta^{\dagger} | \Phi \rangle}{\langle \Phi | \Phi \rangle} &  \frac{\langle \Phi | \beta^{\phantom{\dagger}}\beta^{\dagger} | \Phi \rangle}{\langle \Phi | \Phi \rangle}
\end{array}
\right)  \\
&\equiv& 
\left(
\begin{array} {cc}
R^{+-} & R^{--} \\
R^{++} &  R^{-+}
\end{array}
\right)  \\
&=& 
\left(
\begin{array} {cc}
0 & 0 \\
0 &  1
\end{array}
\right)
\, , 
\end{eqnarray}
\end{subequations}
where the result, trivially obtained by considering the action of quasi-particle operators on the vacuum, can also be recovered starting from Eq.~\ref{generalizeddensitymatrix} and making use of Eqs.~\ref{e:p2qp} and~\ref{unitarity}.

\subsection{Normal ordering}

A Lagrange term is eventually required to constrain the particle number to the correct value on average, such that the grand potential $\Omega \equiv H - \lambda A$ is to be used in place of $H$, where the particle-number operator $A = \sum_{n=1}^{\text{A}} 1$ takes the second-quantized form
\begin{equation}
A = \sum_{p} c^{\dagger}_{p} c_{p} \, .
\end{equation}
The present formalism is best formulated in the quasiparticle basis introduced in Eq.~\eqref{e:p2qp} by normal ordering all operators at play with respect to $| \Phi \rangle$ via Wick's theorem~\cite{wick50a}. Taking $\Omega$ as an example, and as extensively discussed in Ref.~\cite{Signoracci:2014dia}, its normal-ordered form expressed in terms of fully antisymmetric matrix elements\footnote{Explicit expressions of $\Omega^{ij}_{k_1 \ldots k_i k_{i+1} \ldots k_{i+j}}$ in terms of matrix elements $t_{pq}$ and $\bar{v}_{pqrs}$ and of ($U,V$) matrix elements are provided in Ref.~\cite{Signoracci:2014dia}.} reads as
\begin{widetext}
\begin{subequations}
\label{e:h3qpas}
\begin{align}
\label{e:h3qpasa}
\Omega &\equiv \Omega^{[0]} + \Omega^{[2]} + \Omega^{[4]} \\
&\equiv \Omega^{00} + \big[\Omega^{20} + \Omega^{11} + \Omega^{02}\big]+\big[\Omega^{40} + \Omega^{31} + \Omega^{22} + \Omega^{13} + 
\Omega^{04}\big] \\
&= \Omega^{00} \\
& \: \: \: \: \: \: \: \: \: \: \: \: \: + \frac{1}{1!}\displaystyle\sum_{k_1 k_2} \Omega^{11}_{k_1 k_2}\beta^{\dagger}_{k_1} \beta_{k_2} \\
& \: \: \: \: \: \: \: \: \: \: \: \: \: + \frac{1}{2!}\displaystyle\sum_{k_1 k_2} \Big \{\Omega^{20}_{k_1 k_2} \beta^{\dagger}_{k_1}
 \beta^{\dagger}_{k_2} + \Omega^{02}_{k_1 k_2}   \beta_{k_2} \beta_{k_1} \Big \} \\
\label{e:h3qpase}
& \: \: \: \: \: \: \: \: \: \: \: \: \: + \frac{1}{(2!)^{2}} \displaystyle\sum_{k_1 k_2 k_3 k_4} \Omega^{22}_{k_1 k_2 k_3 k_4} 
   \beta^{\dagger}_{k_1} \beta^{\dagger}_{k_2} \beta_{k_4}\beta_{k_3} \\
   & \: \: \: \: \: \: \: \: \: \: \: \: \: + \frac{1}{3!}\displaystyle\sum_{k_1 k_2 k_3 k_4}\Big \{ \Omega^{31}_{k_1 k_2 k_3 k_4}
   \beta^{\dagger}_{k_1}\beta^{\dagger}_{k_2}\beta^{\dagger}_{k_3}\beta_{k_4} +
   \Omega^{13}_{k_1 k_2 k_3 k_4} \beta^{\dagger}_{k_1} \beta_{k_4} \beta_{k_3} \beta_{k_2}  \Big \} \\
  & \: \: \: \: \: \: \: \: \: \: \: \: \: +  \frac{1}{4!} \displaystyle\sum_{k_1 k_2 k_3 k_4}\Big \{ \Omega^{40}_{k_1 k_2 k_3 k_4}
   \beta^{\dagger}_{k_1}\beta^{\dagger}_{k_2}\beta^{\dagger}_{k_3}\beta^{\dagger}_{k_4}  + 
   \Omega^{04}_{k_1 k_2 k_3 k_4}  \beta_{k_4} \beta_{k_3} \beta_{k_2} \beta_{k_1}  \Big \}  \, ,
   \end{align}
   \end{subequations}
\end{widetext}
where
\begin{enumerate}
\item Each term $\Omega^{ij}$ is characterized by its number $i$ ($j$) of quasiparticle creation (annihilation) operators. Because $\Omega$ has been normal-ordered  with respect to $| \Phi \rangle$, all quasiparticle creation operators (if any) are located to the left of all quasiparticle annihilation operators (if any).  The class $\Omega^{[k]}$ groups all the terms $\Omega^{ij}$ with $i+j=k$. The first contribution 
\begin{equation}
\Omega^{[0]} = \Omega^{00} = \frac{\langle \Phi | \Omega | \Phi \rangle}{\langle \Phi | \Phi \rangle} 
\end{equation}
denotes the fully contracted part of $\Omega$ and is nothing but a (real) number. 
\item The subscripts of the matrix elements $\Omega^{ij}_{k_1 \ldots k_i k_{i+1} \ldots k_{i+j}}$ are ordered sequentially, independently of 
the creation or annihilation character of the operators the indices refer to. While quasiparticle creation operators themselves also follow 
sequential order, quasiparticle annihilation operators follow inverse sequential order. In Eq. \eqref{e:h3qpase}, for example, the two 
creation operators are ordered $\beta^{\dagger}_{k_1}\beta^{\dagger}_{k_2}$ while 
the two annihilation operators are ordered $\beta_{k_4}\beta_{k_3}$.
\item Matrix elements are fully antisymmetric, i.e.
\begin{eqnarray}
\Omega^{ij}_{k_1 \ldots k_{i} k_{i+1} \ldots k_{i+j}} &=& (-1)^{\sigma(P)}
\Omega^{ij}_{P(k_1 \ldots k_i | k_{i+1} \ldots k_{i+j})} 
\end{eqnarray}
where $\sigma(P)$ refers to the signature of the 
permutation $P$.  The notation $P(\ldots | \ldots)$ denotes a 
separation into the $i$ quasiparticle creation operators and the $j$ quasiparticle annihilation operators such that
permutations are only considered between members of the same group. 
\item As each  $\Omega^{[k]}$ component is hermitian, matrix elements exhibit the following behavior under hermitian conjugation
\begin{subequations}
\label{e:me3sym}
\begin{align}
\label{e:me3syma}
\Omega^{11}_{k_1 k_2} &= \Omega^{11*}_{k_2 k_1} \, ,\\
\label{e:me3symc}
\Omega^{20}_{k_1 k_2} &= \Omega^{02*}_{k_1 k_2} \, , \\
\label{e:me3symb}
\Omega^{22}_{k_1 k_2 k_3 k_4} &= \Omega^{22*}_{k_3 k_4 k_1 k_2} \, , \\
\label{e:me3symf}
\Omega^{31}_{k_1 k_2 k_3 k_4} &= \Omega^{13*}_{k_4 k_1 k_2 k_3} \, , \\
\label{e:me3symj}
\Omega^{40}_{k_1 k_2 k_3 k_4} &= \Omega^{04*}_{k_1 k_2 k_3 k_4} \, .
\end{align}
\end{subequations}
\end{enumerate}

Similarly to $\Omega$, the normal-ordered form of the particle-number operator is obtained as\footnote{Explicit expressions of $A^{00}$, $A^{11}_{k_1 k_2}$, $A^{20}_{k_1 k_2}$ and $A^{02}_{k_1 k_2}$ are provided in Ref.~\cite{Signoracci:2014dia}.}
\begin{subequations}
\label{normalorderedA}
\begin{eqnarray}
A &\equiv& A^{[0]} + A^{[2]}  \\
&\equiv& A^{00} + \big[A^{20} + A^{11} + A^{02}\big]\\
&\equiv& A^{00} \\
&& + \frac{1}{1!} \sum_{k_1 k_2} A^{11}_{k_1 k_2}\beta^{\dagger}_{k_1} \beta_{k_2} \\
&& + \frac{1}{2!}\sum_{k_1 k_2} \Big \{A^{20}_{k_1 k_2} \beta^{\dagger}_{k_1}
 \beta^{\dagger}_{k_2} + A^{02}_{k_1 k_2}   \beta_{k_2} \beta_{k_1} \Big \} 
\end{eqnarray}
\end{subequations}
and allows the extraction of the normal-ordered Hamiltonian itself whose various terms take the same form as those of $\Omega$ but with the modifications $H^{[0]} = \Omega^{[0]} + \lambda A^{[0]}$ and $H^{[2]} = \Omega^{[2]} + \lambda A^{[2]}$.

\subsection{Diagrammatic representation of an operator}
\label{diagramsforvertices}

\begin{figure}[t!]
\begin{center}
\includegraphics[width=\columnwidth]{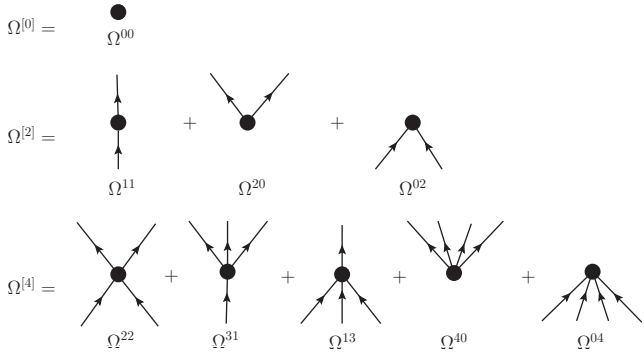}
\end{center}
\caption{
\label{variousvertices1}
Canonical diagrammatic representation of normal-ordered contributions to the grand potential $\Omega$ in the Schroedinger representation.}
\end{figure}

\begin{figure}[t!]
\begin{center}
\includegraphics[clip=,width=0.6\columnwidth]{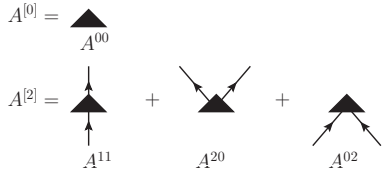}\\ 
\end{center}
\caption{
\label{variousvertices2}
Canonical diagrammatic representation of the normal-ordered contributions to the particle-number operator $A$ in the Schroedinger representation.}
\end{figure}

Normal-ordered operators in the Schroedinger representation can be represented diagrammatically. Taking the grand potential and the particle-number operators as typical examples, canonical diagrams representing their normal-ordered contributions $\Omega^{ij}$ and $A^{ij}$ are displayed in Figs.~\ref{variousvertices1} and~\ref{variousvertices2}, respectively. Focusing on $\Omega$ as an example, the various diagrams contributing to it must be understood in the following way.

\begin{figure}[t!]
\begin{center}
\includegraphics[clip=,width=0.5\textwidth,angle=0]{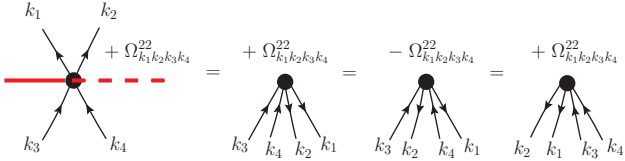}\\ 
\end{center}
\caption{
\label{variousvertices3}
Rules to apply when departing from the canonical diagrammatic representation of a normal-ordered operator. Oriented lines can be rotated through the dashed line but not through the full line.}
\end{figure}

\begin{enumerate}
\item One must associate the factor $\Omega^{ij}_{k_1 \ldots k_i k_{i+1} \ldots k_{i+j}}$ to the dot vertex, where $i$ denotes the number of lines traveling out of the vertex and representing quasiparticle creation operators while $j$ denotes the number of lines traveling into the vertex and representing quasiparticle annihilation operators. 
\item A factor $1/[i!j!]$ must multiply $\Omega^{ij}_{k_1 \ldots k_i k_{i+1} \ldots k_{i+j}}$ given that the corresponding diagram contain $j$ equivalent ingoing lines and $i$ equivalent outgoing lines.
\item In the canonical representation used in Figs.~\ref{variousvertices1} and~\ref{variousvertices2}, all oriented lines go up, i.e. lines representing quasiparticle creation (annihilation) operators appear above (below) the vertex. Accordingly, indices $k_1 \ldots k_i$ must be assigned consecutively from the leftmost to the rightmost line above the vertex, while $k_{i+1} \ldots k_{i+j}$ must be similarly assigned consecutively for lines below the vertex.
\item In the diagrammatic representation at play in the many-body formalism designed below, it is possible for a line to propagate downwards\footnote{As explained in Ref.~\cite{Signoracci:2014dia}, downwards quasi-particle lines do not occur in SR-BCC theory. This will be recovered in Section~\ref{CCtheory} as a particular case of the present diagrammatics.}. This can be obtained unambiguously starting from the canonical representation given in Figs.~\ref{variousvertices1} and~\ref{variousvertices2} at the price of adding a specific rule. As illustrated in Fig.~\ref{variousvertices3} for the diagram representing $\Omega^{22}$, lines must only be rotated through the right of the diagram, i.e. going through the dashed line, while it is forbidden to rotate them through the full line. Additionally, a minus sign must be added to the amplitude $\Omega^{ij}_{k_1 \ldots k_i k_{i+1} \ldots k_{i+j}}$ associated with the canonical diagram each time two lines cross as illustrated in Fig.~\ref{variousvertices3}.
\end{enumerate}

\subsection{$U(1)$ group}

We consider the abelian compact Lie group $U(1)\equiv \{S(\varphi), \varphi \in  [0,2\pi]\}$ associated with the global rotation of an A-body fermion system in gauge space. As $U(1)$ is considered to be a symmetry group of $H$, commutation relations
\begin{eqnarray}
\left[H,S(\varphi)\right]&=& \left[A,S(\varphi)\right] = \left[\Omega,S(\varphi)\right] =0 \, , \label{commutation}
\end{eqnarray}
hold for any $\varphi \in  [0,2\pi]$.

We utilize the unitary representation of $U(1)$ on Fock space ${\cal F}$ given by  
\begin{equation}
S(\varphi)  = e^{iA\varphi} \, .
\end{equation}
Matrix elements of the irreducible representations (IRREPs) of $U(1)$ are
\begin{equation}
\langle \Psi^{\text{A}}_{\mu} |  S(\varphi)  |\Psi^{\text{A}'}_{\mu'} \rangle \equiv e^{i\text{A}\varphi} \, \delta_{\text{A}\text{A}'} \, \delta_{\mu\mu'} \, , \label{irreps}
\end{equation}
where $| \Psi^{\text{A}}_{\mu} \rangle$ is an eigenstate of $A$
\begin{eqnarray}
A | \Psi^{\text{A}}_{\mu} \rangle &=&  \text{A} | \Psi^{\text{A}}_{\mu} \rangle  \, , \label{eigenequationA}
\end{eqnarray}
and, by virtue of Eq.~\ref{commutation}, of the Hamiltonian at the same time
\begin{eqnarray}
H | \Psi^{\text{A}}_{\mu} \rangle &=&  \text{E}^{\text{A}}_{\mu} \, | \Psi^{\text{A}}_{\mu} \rangle \,\,\, , \label{schroed}
\end{eqnarray}
where $\text{E}^{\text{A}}_{\mu}$, with $\mu=0, 1, 2\ldots$, orders increasing eigenenergies for a fixed $\text{A}$. From the group theory point of view, $\text{A} \in \mathbb{Z}$ on the right-hand side of Eq.~\ref{irreps}. Since $\text{A}$ actually represents the number of fermions in the system, its value is constrained from the physics point of view to $\text{A} \in \mathbb{N}$. Equations~\ref{eigenequationA} and~\ref{schroed} trivially lead to
\begin{eqnarray}
\Omega | \Psi^{\text{A}}_{\mu} \rangle &=&  \Omega^{\text{A}}_{\mu} \, | \Psi^{\text{A}}_{\mu} \rangle \,\,\, , \label{grandpotschroed}
\end{eqnarray}
where $\Omega^{\text{A}}_{\mu} \equiv \text{E}^{\text{A}}_{\mu} - \lambda \text{A}$. The volume of the group is
\begin{eqnarray*}
v_{U(1)} \equiv \int_{0}^{2\pi} \!d\varphi = 2\pi \,\,\, ,
\end{eqnarray*}
and the orthogonality of IRREPs reads as
\begin{equation}
\frac{1}{2}\int_{0}^{2\pi} \!d\varphi \,e^{-i\text{A}\varphi}\,e^{+i\text{A}'\varphi}  = \delta_{\text{A}\text{A}'} \, . \label{orthogonality}
\end{equation}%
A tensor operator $O$ of rank\footnote{A tensor operator of rank $\text{A}$ with respect to the $U(1)$ group is an operator that associates a state of the $(\text{N}\!+\!\text{A})$-body Hilbert space ${\cal H}_{\text{N}+\text{A}}$ to a state of the $\text{N}$-body Hilbert space ${\cal H}_{\text{N}}$, i.e. that changes the number of particles by $\text{A}$ units.} $\text{A}$ and a state $| \Psi^{\text{A}}_{\mu} \rangle$ transform under global gauge rotation according to
\begin{subequations}
\label{eq:ten:def}
\begin{eqnarray}
S(\varphi) \, O \, S(\varphi)^{-1}  &=& e^{i\text{A}\varphi} \, O \, \,\, ,\label{eq:ten:def1} \\
S(\varphi) \, | \Psi^{A}_{\mu} \rangle &=&  e^{i\text{A}\varphi} \, | \Psi^{\text{A}}_{\mu} \rangle \, \,\, . \label{eq:ten:def2}
\end{eqnarray}
\end{subequations}

A key feature for the following is that any integrable function $f(\varphi)$ defined on $[0,2\pi]$ can be expanded over the IRREPs of the $U(1)$ group. This constitutes nothing but the Fourier decomposition of the function
\begin{equation}
f(\varphi) \equiv \sum_{\text{A} \in \mathbb{Z}} \, f^{\text{A}} \, e^{i\text{A}\varphi} \, , \label{decomposition_general}
\end{equation}
which defines the set of expansion coefficients $\{f^{\text{A}}\}$. Last but not least, the IRREPs fulfill the first-order ordinary differential equation (ODE)
\begin{equation}
-i \frac{d}{d \varphi} e^{i\text{A}\varphi} = \text{A} \, e^{i\text{A}\varphi} \label{ODE} \, .
\end{equation}

\subsection{Time-dependent state}
\label{timedependentstate}

The many-body formalism proposed in the present work is conveniently formulated within an imaginary-time framework. We thus introduce the evolution operator in imaginary time as\footnote{The time is given in units of MeV$^{-1}$.}
\begin{equation}
{\cal U}(\tau) \equiv e^{-\tau \Omega} \, , \label{evoloperator}
\end{equation}
with $\tau$ real. A key quantity throughout the present study is the time-evolved many-body state defined as
\begin{eqnarray}
| \Psi (\tau) \rangle &\equiv& {\cal U}(\tau) | \Phi \rangle \nonumber \\
&=& \sum_{\text{A} \in \mathbb{N}}\sum_{\mu} e^{-\tau \Omega^{\text{A}}_{\mu}} \,  |  \Psi^{\text{A}}_{\mu} \rangle \, \langle \Psi^{\text{A}}_{\mu} |  \Phi \rangle  \, , \label{evolstate}
\end{eqnarray}
where we have inserted a completeness relationship on Fock space under the form
\begin{equation}
\bbone =  \sum_{\text{A} \in \mathbb{N}}\sum_{\mu} |  \Psi^{\text{A}}_{\mu} \rangle \, \langle \Psi^{\text{A}}_{\mu} | \, . \label{completeness}
\end{equation}
It is straightforward to demonstrate that $| \Psi (\tau) \rangle$ satisfies the time-dependent Schroedinger equation
\begin{equation}
\Omega \, | \Psi (\tau) \rangle = -\partial_{\tau} | \Psi (\tau) \rangle  \, . \label{schroedinger}
\end{equation}

\subsection{Large and infinite time limits}
\label{timelimits}

Below, we will be interested in first looking at the {\it large} $\tau$ limit of various quantities before eventually taking their {\it infinite} time limit. Although we utilize the same mathematical symbol ($\lim\limits_{\tau \to \infty}$) in both cases for simplicity, the reader must not be confused by the fact that there remains a residual $\tau$ dependence in the first case, which typically disappears by considering ratios before actually promoting the time to infinity. The large $\tau$ limit is essentially defined as $\tau \gg \Delta E^{-1}$, where $\Delta E$ is the energy difference between the ground state and the first excited state of $\Omega$. Depending on the system, the latter can be the first excited state in the IRREP of the ground state or the lowest state of another IRREP.

\subsection{Ground state}
\label{groundstateSeC}

Taking the large $\tau$ limit provides the ground state of $\Omega$ under the form\footnote{The chemical potential $\lambda$ is fixed such that $\Omega^{\text{A}_0}_0$ for the targeted particle number $\text{A}_0$ is the lowest value of all $\Omega^{\text{A}}_\mu$ over Fock space, i.e. it penalizes systems with larger number of particles such that $\Omega^{\text{A}_0}_0 < \Omega^{\text{A}}_\mu$ for all $\text{A}>\text{A}_0$ while maintaining at the same time that $\Omega^{\text{A}_0}_0 < \Omega^{\text{A}}_\mu$ for all $\text{A}<\text{A}_0$. This is practically achievable only if $E^{\text{A}}_0$ is strictly convex in the neighborhood of $\text{A}_0$, which is generally but not always true for atomic nuclei.}
\begin{subequations}
\label{groundstate}
\begin{eqnarray}
| \Psi^{\text{A}_0}_{0} \rangle &\equiv& \lim\limits_{\tau \to \infty} | \Psi (\tau) \rangle \\
&=& e^{-\tau \Omega^{\text{A}_0}_0}   |  \Psi^{\text{A}_0}_{0} \rangle \, \langle \Psi^{\text{A}_0}_{0} |  \Phi \rangle  \, .
\end{eqnarray}
\end{subequations}
As will become clear below, the many-body scheme developed in the present work relies on choosing the Bogoliubov product state $| \Phi \rangle$ as the ground state of an unperturbed grand potential $\Omega_0$ that breaks $U(1)$ symmetry. As such, $| \Phi \rangle$ mixes several IRREPS but is likely to contain a component belonging to the nucleus of interest given that it is typically chosen to have (close to) the number $\text{A}_0$ of particles in average. Eventually, one recovers from Eq.~\ref{schroedinger} that
\begin{eqnarray}
\Omega | \Psi^{\text{A}_0}_{0} \rangle &=&  \Omega^{\text{A}_0}_0 \, | \Psi^{\text{A}_0}_{0} \rangle \,\,\, , \label{schroedevolstate}
\end{eqnarray}
in the large $\tau$ limit.

\subsection{Off-diagonal kernels}
\label{transitionkernels}

We now introduce the off-diagonal, i.e. $\varphi$-dependent, time-dependent kernel of an operator\footnote{We are currently interested in operators that commute with $\Omega$ and that are scalars under transformations of the $U(1)$ group, i.e. that are of rank $\text{A}=0$. Dealing with operators of rank $\text{A}\neq 0$ and with amplitudes between different many-body eigenstates of $\Omega$ requires an extension of the presently developed formalism.} $O$ through
\begin{eqnarray}
O(\tau,\varphi) &\equiv& \langle \Psi (\tau)  | O | \Phi(\varphi) \rangle  \, ,
\end{eqnarray}
where $| \Phi(\varphi) \rangle \equiv S(\varphi) | \Phi \rangle$ denotes the {\it gauge-rotated} Bogoliubov state. Doing so for the identity, the Hamiltonian, the particle number and the grand potential operators, we introduce the set of off-diagonal kernels
\begin{subequations}
\label{defkernels}
\begin{eqnarray}
N(\tau,\varphi) &\equiv& \langle \Psi (\tau)  | \bbone | \Phi(\varphi) \rangle  \, , \label{defnormkernel} \\
H(\tau,\varphi) &\equiv& \langle \Psi (\tau) | H | \Phi(\varphi) \rangle  \, , \label{defenergykernel} \\
A(\tau,\varphi) &\equiv& \langle \Psi (\tau)  | A | \Phi(\varphi) \rangle  \, , \label{defAkernel} \\
\Omega(\tau,\varphi) &\equiv& \langle \Psi (\tau)  | \Omega | \Phi(\varphi) \rangle  \, , \label{defomegakernel} 
\end{eqnarray}
\end{subequations}
where the first one denotes the off-diagonal norm kernel and where the latter three are related through $\Omega(\tau,\varphi) = H(\tau,\varphi) - \lambda A(\tau,\varphi)$. Focusing on the grand potential operator as an example, its kernel can be split into various contributions associated with its normal-ordered components, i.e.
\begin{subequations}
\label{defkernels1and2body}
\begin{eqnarray}
\Omega(\tau,\varphi) &\equiv& \Omega^{00}(\tau,\varphi) \\
&&+ \Omega^{20}(\tau,\varphi) + \Omega^{11}(\tau,\varphi) + \Omega^{02}(\tau,\varphi)  \\
 &&+ \Omega^{40}(\tau,\varphi) + \Omega^{31}(\tau,\varphi) + \Omega^{22}(\tau,\varphi)  \\
 &&+  \Omega^{13}(\tau,\varphi) + \Omega^{04}(\tau,\varphi) \, ,
\end{eqnarray}
\end{subequations}
having trivially that $\Omega^{00}(\tau,\varphi) = \Omega^{00}\,N(\tau,\varphi)$. Similarly, the particle-number kernel can be split according to
\begin{subequations}
\label{defkernels1and2bodyB}
\begin{eqnarray}
A(\tau,\varphi) &\equiv& A^{00}(\tau,\varphi) \\
&&+ A^{20}(\tau,\varphi) + A^{11}(\tau,\varphi) + A^{02}(\tau,\varphi)  \, ,
\end{eqnarray}
\end{subequations}
with $A^{00}(\tau,\varphi) = A^{00}\,N(\tau,\varphi)$.

Finally, use will often be made of the {\it reduced} kernel of an operator $O$ defined through
\begin{equation}
{\cal O}(\tau,\varphi) \equiv \frac{O(\tau,\varphi)}{N(\tau,0)} \, , \label{reducedkernels}
\end{equation}
which leads, for $O=\bbone$, to working with {\it intermediate normalization} at $\varphi=0$, i.e. to having a norm kernel that satisfies ${\cal N}(\tau,0) \equiv 1$ for all $\tau$.

\section{Master equations}
\label{normkernel}

This section presents a set of master equations providing the basis of the newly proposed many-body formalism, i.e. they constitute {\it exact} equations of reference on top of which actual many-body expansion schemes will be designed in the remainder of the paper.

\subsection{Fourier expansion of the off-diagonal kernels}
\label{kernelexpansion}

Inserting twice Eq.~\ref{completeness} into Eqs.~\ref{defkernels} while making use of Eqs.~\ref{eigenequationA}, \ref{schroed} and~\ref{eq:ten:def2}, one obtains the Fourier decomposition of the kernels
\begin{subequations}
\label{expandedkernels}
\begin{eqnarray}
N(\tau,\varphi)  &=&  \sum_{\text{A} \in \mathbb{N}}\sum_{\mu} \phantom{E^{\text{A}}_\mu \,} e^{-\tau \Omega^{\text{A}}_\mu} |\langle \Phi | \Psi^{\text{A}}_{\mu} \rangle |^2 \, e^{i\text{A}\varphi}  \, , \label{expandedkernels1} \\
H(\tau,\varphi) &=&  \sum_{\text{A} \in \mathbb{N}}\sum_{\mu} \text{E}^{\text{A}}_\mu \, e^{-\tau \Omega^{\text{A}}_\mu} |\langle \Phi | \Psi^{\text{A}}_{\mu} \rangle |^2 \, e^{i\text{A}\varphi} \, , \label{expandedkernels2} \\
A(\tau,\varphi) &=&  \sum_{\text{A} \in \mathbb{N}}\sum_{\mu} \, \, \, \text{A} \, \, \, e^{-\tau \Omega^{\text{A}}_\mu} |\langle \Phi | \Psi^{\text{A}}_{\mu} \rangle |^2 \, e^{i\text{A}\varphi} \, , \label{expandedkernels3} \\
\Omega(\tau,\varphi) &=&  \sum_{\text{A} \in \mathbb{N}}\sum_{\mu} \Omega^{\text{A}}_\mu \, e^{-\tau \Omega^{\text{A}}_\mu} |\langle \Phi | \Psi^{\text{A}}_{\mu} \rangle |^2 \, e^{i\text{A}\varphi} \, , \label{expandedkernels4} 
\end{eqnarray}
\end{subequations}
where one trivially notices that contributions associated with $\text{A} < 0$ are zero.

\subsection{Ground-state energy}
\label{energy}

Defining the large $\tau$ limit of a kernel via
\begin{eqnarray}
O(\varphi) &\equiv & \lim\limits_{\tau \to \infty} O(\tau,\varphi)  \, , \label{limitoperator}
\end{eqnarray}
one obtains
\begin{subequations}
\label{limitkernels}
\begin{eqnarray}
N(\varphi) &=&  \phantom{E^{\text{A}_0}_0 \,} e^{-\tau \Omega^{\text{A}_0}_0} |\langle \Phi | \Psi^{\text{A}_0}_0 \rangle |^2 \, e^{i\text{A}_0\varphi}  \, , \label{limitnorm} \\
H(\varphi) &=& \text{E}^{\text{A}_0}_0 \, e^{-\tau \Omega^{\text{A}_0}_0} |\langle \Phi | \Psi^{\text{A}_0}_0 \rangle |^2 \, e^{i\text{A}_0\varphi}   \, , \label{limitenergy} \\
A(\varphi) &=& \, \, \, \text{A}_0 \, \,  e^{-\tau \Omega^{\text{A}_0}_0} |\langle \Phi | \Psi^{\text{A}_0}_0 \rangle |^2 \, e^{i\text{A}_0\varphi}   \, , \label{limitA} \\
\Omega(\varphi) &=& \Omega^{\text{A}_0}_0 \, e^{-\tau \Omega^{\text{A}_0}_0} |\langle \Phi | \Psi^{\text{A}_0}_0 \rangle |^2 \, e^{i\text{A}_0\varphi}   \, , \label{limitgrandpotential} 
\end{eqnarray}
\end{subequations}
where the residual time dependence typically disappears by eventually employing reduced kernels as defined in Eq.~\ref{reducedkernels}. Expressions~\ref{limitkernels} relate in the large-time limit off-diagonal operator kernels of interest to the off-diagonal norm kernel through eigenvalue-like equations
\begin{subequations}
\label{kernelequation}
\begin{eqnarray}
H(\varphi) &=&  \text{E}^{\text{A}_0}_0 \, N(\varphi) \,\, , \label{kernelequation1} \\
A(\varphi) &=&  \text{A}_0 \, N(\varphi) \,\, , \label{kernelequation2} \\
\Omega(\varphi) &=&  \Omega^{\text{A}_0}_0 \, N(\varphi) \,\, , \label{kernelequation3}
\end{eqnarray}
\end{subequations}
and similarly for reduced kernels. In Eq.~\ref{limitkernels}, the $\varphi$ dependence originally built into the time-dependent kernels reduces to that of the single IRREP $\text{A}_0$ of the target nucleus. Additionally, the expansion coefficient in the particle-number operator kernel equates the expected value $\text{A}_0$. These characteristic features, trivially valid for the exact kernels, testify that the selected eigenstate $| \Psi^{\text{A}_0}_{0} \rangle$ of $\Omega$ (and $H$) does carry good particle number $\text{A}_0$. Accordingly, Eq.~\ref{kernelequation} demonstrates that the straight ratio of the operator kernels to the norm kernel accesses, at any value of $\varphi$, the eigenvalues that are in one-to-one relationship with the physical IRREP. 

Let us now consider the case of actual interest where the kernels are approximated in a way that breaks $U(1)$ symmetry. In this situation, reduced kernels in the infinite time limit display the typical structure
\begin{subequations}
\label{limitkernelsapprox}
\begin{eqnarray}
{\cal N}_{\text{app}}(\varphi) &\equiv&   \sum_{\text{A} \in \mathbb{Z}} \textmd{N}^{\text{A}}_{\text{app}} \, e^{i\text{A}\varphi}  \, , \label{limitnormapprox} \\
{\cal H}_{\text{app}}(\varphi) &\equiv& \sum_{\text{A} \in \mathbb{Z}} \textmd{E}^{\text{A}}_{\text{app}}\,\textmd{N}^{\text{A}}_{\text{app}} \, e^{i\text{A}\varphi}   \, , \label{limitenergyapprox} \\
{\cal A}_{\text{app}}(\varphi) &\equiv&   \sum_{\text{A} \in \mathbb{Z}} \textmd{A}^{\text{A}}_{\text{app}}\,\textmd{N}^{\text{A}}_{\text{app}} \, e^{i\text{A}\varphi}  \, , \label{limitAapprox} 
\end{eqnarray}
\end{subequations}
where the condition ${\cal N}_{\text{app}}(0) = \sum_{\text{A} \in \mathbb{Z}} \textmd{N}^{\text{A}}_{\text{app}} = 1$ characterizes intermediate normalization at gauge angle $\varphi = 0$. In Eq.~\ref{limitkernelsapprox}, the remaining sum over $\text{A}$ signals the breaking of the symmetry induced by the approximation. The Fourier expansion~\ref{limitkernelsapprox} of the approximate kernels defined on $[0,2\pi]$ exists by virtue of Eq.~\ref{decomposition_general}. In the expansion, the sum over the IRREPS runs a priori through $\mathbb{Z}$. If the many-body approximation scheme is well behaved from the physics standpoint, coefficients corresponding to $\text{A} < 0$ must be zero, which acts as a check that the formalism is sensible~\cite{Bender:2008rn,Duguet:2013dga}.

Except for going back to an exact computation of the kernels, such that all the expansion coefficients but the physical one are zero in Eq.~\ref{limitkernelsapprox}, taking the straight ratio ${\cal H}_{\text{app}}(\varphi)/{\cal N}_{\text{app}}(\varphi)$ does not provide an approximate energy that is in one-to-one correspondence with the physical IRREP $\text{A}_0$. This materializes the contamination associated with the breaking of the symmetry. However, one can take advantage of the $\varphi$ dependence built into ${\cal N}_{\text{app}}(\varphi)$, ${\cal H}_{\text{app}}(\varphi)$ and ${\cal A}_{\text{app}}(\varphi)$ to extract the component associated with that physical IRREP.  Indeed, by virtue of the orthogonality of the IRREPs (Eq.~\ref{orthogonality}), the approximation to $\text{E}^{\text{A}_0}_0$ can be extracted as  
\begin{eqnarray}
\text{E}^{\text{A}_0}_0 &\approx& \frac{\int_{0}^{2\pi} \!d\varphi \,e^{-i\text{A}_0\varphi} \, {\cal H}_{\text{app}}(\varphi)}{\int_{0}^{2\pi} \!d\varphi \,e^{-i\text{A}_0\varphi} \,  \, {\cal N}_{\text{app}}(\varphi)} = \textmd{E}^{\text{A}_0}_{\text{app}} \, . \label{projected_energy}
\end{eqnarray}
Following the same line for the particle-number operator kernel provides a case of particular interest. Indeed, the integral over the domain of the group not only allows one to extract the component in one-to-one relationship with the physical IRREP but should be such that the expansion coefficient $\textmd{A}^{\text{A}_0}_{\text{app}}$ thus obtained is actually equal to the correct result $\text{A}_0$, i.e. it should be such that
\begin{eqnarray}
\textmd{A}^{\text{A}_0}_{\text{app}} &=& \frac{\int_{0}^{2\pi} \!d\varphi \,e^{-i\text{A}_0\varphi} \, {\cal A}_{\text{app}}(\varphi)}{\int_{0}^{2\pi} \!d\varphi \,e^{-i\text{A}_0\varphi} \,  \, {\cal N}_{\text{app}}(\varphi)} \,  \label{projected_A}
\end{eqnarray}
is indeed equal to $\text{A}_0$. This is a necessary condition to claim that the particle-number symmetry is indeed restored at any truncation order in the many-body expansion. We will see in Sec.~\ref{Secnormkernel} how this key demand constrains the many-body expansion scheme in a very specific way.

Whenever $| \Phi \rangle$ is taken to be a Slater determinant rather than a Bogoliubov vacuum, the targeted IRREP $\text{A}_0$ is selected {\it a priori} at the level of Eq.~\ref{expandedkernels}, i.e. the gauge-angle dependence of all the kernels reduces to the IRREP $e^{i\text{A}_0\varphi}$ at any finite time $\tau$. Correspondingly, the coefficients in the Fourier expansion of the approximate kernels in Eq.~\ref{limitkernelsapprox} directly provide $\textmd{E}^{\text{A}_0}_{\text{app}}$ and $\textmd{A}^{\text{A}_0}_{\text{app}}=\text{A}_0$ (after division by the coefficient in the norm kernel). The integration over the domain of the $U(1)$ group becomes obviously superfluous in this case as no symmetry was broken in the first place.

\subsection{Comparison with standard approaches}
\label{standardapproaches}

Applying particle-number breaking BMBPT~\cite{mehta1,balian62a,henley} or BCC theory~\cite{StolarczykMonkhorst,Signoracci:2014dia} amounts to expanding the {\it diagonal} kernels $N(0)$, $H(0)$, $A(0)$ and $\Omega(0)$  around the Bogoliubov state $| \Phi \rangle$. These methods can efficiently tackle systems characterized by a near degeneracy of the unperturbed ground state associated with a Cooper pair infra-red instability. This is done at the price of breaking global gauge invariance, even though the symmetry is restored by definition in the limit of exact calculations, i.e. when summing all diagrams. In practice, approximate kernels obtained via a truncation of the expansion mix components associated with {\it different} IRREPs of $U(1)$ and thus contain spurious contaminations from the symmetry standpoint. The difficulty resides here in the fact that the kernels at play do not carry any $\varphi$ dependence, i.e. Eq.~\ref{limitkernelsapprox} reduces in this case to
\begin{subequations}
\label{limitkernelsapprox0}
\begin{eqnarray}
{\cal N}_{\text{app}}(0) &=& \sum_{\text{A}  \in \mathbb{Z}} \textmd{N}^{\text{A}}_{\text{app}} \, , \label{limitnormapprox0} \\
{\cal H}_{\text{app}}(0) &=& \sum_{\text{A} \in \mathbb{Z}} \textmd{E}^{\text{A}}_{\text{app}}\,\textmd{N}^{\text{A}}_{\text{app}} \, , \label{limitenergyapprox0} \\
{\cal A}_{\text{app}}(0) &=& \sum_{\text{A} \in \mathbb{Z}} \textmd{A}^{\text{A}}_{\text{app}}\,\textmd{N}^{\text{A}}_{\text{app}} \, , \label{limitAapprox0}
\end{eqnarray}
\end{subequations}
such that the coefficients associated with the physical IRREP $\text{A}_0$ cannot be extracted via the integral over the domain of the $U(1)$ group. Accordingly, the key feature of the generalized approach presently proposed is to utilize {\it off-diagonal} kernels incorporating, from the outset, the effect of the gauge rotation $S(\varphi)$. The associated $\varphi$ dependence leaves a fingerprint of the artificial symmetry breaking built into approximate kernels that can be exploited to extract the physical components of interest through Eqs.~\ref{projected_energy} and~\ref{projected_A}, i.e. to remove the symmetry contaminants. 


\subsection{Accessing neighboring nuclei at once}
\label{yrast}

Now that the benefit of performing the integral over the domain of the $U(1)$ group has been highlighted for the lowest state of the target nucleus, let us step back to Eq.~\ref{expandedkernels} and slightly modify the procedure to access the lowest eigenenergy $\text{E}^{\text{A}}_0$ associated with {\it each} IRREP, i.e. to access within the same calculation the ground state of neighboring nuclei having a non-zero overlap with the reference state $| \Phi \rangle$. To do so, we invert the order in which the limit $\tau \to \infty$ and the integral over the domain of the group are performed. We first extract the component of the time-dependent kernels associated with the specific Hilbert space ${\cal H}_{\text{A}}$ of interest
\begin{subequations}
\label{integratedkernels}
\begin{eqnarray}
N^{\text{A}}(\tau) &\equiv &  \frac{1}{2\pi}  \int_{0}^{2\pi} \!d\varphi \, e^{-i\text{A}\varphi} \,  \, {\cal N}(\tau,\varphi) \nonumber \\
&=&  \sum_{\mu} e^{-\tau \Omega^{\text{A}}_{\mu}} \, |\langle \Phi | \Psi^{\text{A}}_{\mu} \rangle |^2 /N(\tau,0)  \, , \label{integratednorm} \\
H^{\text{A}}(\tau) &\equiv & \frac{1}{2\pi}  \int_{0}^{2\pi} \!d\varphi \, e^{-i\text{A}\varphi} \,   {\cal H}(\tau,\varphi)  \nonumber \\ 
&=& \sum_{\mu} e^{-\tau \Omega^{\text{A}}_{\mu}} E^{\text{A}}_{\mu}  \, |\langle \Phi | \Psi^{\text{A}}_{\mu} \rangle |^2 /N(\tau,0)   \, . \label{integratedenergy}
\end{eqnarray}
\end{subequations}
and take the limit $\tau \to \infty$ to access the lowest eigenenergy $\text{E}^{\text{A}}_{0}$ through
\begin{eqnarray}
\text{E}^{\text{A}}_{0} &=& \lim\limits_{\tau \to \infty} \frac{H^{\text{A}}(\tau)}{N^{\text{A}}(\tau)} . \label{yrast_projected_energy}
\end{eqnarray}
The above analysis is based on exact kernels respecting the symmetries and requires the extraction of the IRREP of interest prior to taking the large time limit. As explained above, the large time limit of {\it approximate} kernels based on a symmetry breaking reference state still mixes the IRREPS of $U(1)$. This can be used as an advantage to actually extract the ground state associated with various $\text{A}$ from the infinite time kernels, i.e. Eq.\ref{yrast_projected_energy} is eventually replaced by Eq.~\ref{projected_A} applied to $\text{A}\neq \text{A}_0$. In practice, this procedure is  limited to values of $\text{A}$ whose components in the infinite time kernels are larger than a given threshold.

Everything exposed so far is valid independently of the many-body method employed to expand and truncate the off-diagonal kernels. The remainder of the paper is devoted to the computation of $N(\tau,\varphi)$, $\Omega(\tau,\varphi)$ and $A(\tau,\varphi)$ via an extension of SR-BMBPT and SR-BCC theory.

\section{Perturbation theory}
\label{sectionMBPT}

Single-reference BCC theory starts from the similarity-transformed grand potential~\cite{Signoracci:2014dia} or could be formulated from an energy functional~\cite{shavitt09a}. The Baker-Campbell-Hausdorff identity applied to the similarity-transformed grand potential on the basis of standard Wick's theorem provides the naturally terminating expansion of the reduced diagonal grand-potential kernel. Unfortunately, this property cannot be obtained directly for the {\it off-diagonal} operator kernels presently at play. This is due to the fact that the off-diagonal Wick theorem~\cite{balian69a} we will rely on to expand off-diagonal matrix elements of strings of quasi-particle operators does not grant a normal ordering of the operators themselves. This feature prevents us from straightforwardly recovering the connected structure of the kernels associated with an underlying exponentiated connected cluster operator. Doing so will require a detour via the perturbative expansion of the off-diagonal kernels. With off-diagonal BMBPT at hand, it will be possible to design the off-diagonal BCC scheme in Sec.~\ref{CCtheory}.

\subsection{Unperturbed system}
\label{chap:slater}

The grand potential is split into an unperturbed part $\Omega_{0}$ and a residual part $\Omega_1$
\begin{equation}
\label{split1}
\Omega = \Omega_{0} + \Omega_{1} \, ,
\end{equation} 
such that 
\begin{subequations}
\begin{eqnarray}
\Omega_{0} &\equiv& \Omega^{00}+\bar{\Omega}^{11} \\
\Omega_{1} &\equiv& \Omega^{20} + \breve{\Omega}^{11} + \Omega^{02} \nonumber \\
  && + \Omega^{22} + \Omega^{31} + \Omega^{13} +  \Omega^{40} + \Omega^{04} \label{perturbation}
\end{eqnarray}
\end{subequations}
where $\breve{\Omega}^{11}\equiv\Omega^{11}- \bar{\Omega}^{11}$. The term $\bar{\Omega}^{11}$ has the same formal structure as $\Omega^{11}$ and  remains to be specified.

For a given number of interacting fermions, the key is to choose $\Omega_0$ with a low-enough symmetry for its ground state $| \Phi \rangle$ to be non-degenerate with respect to elementary excitations. For open-shell superfluid nuclei, this leads to choosing an operator $\Omega_0$ that breaks particle number conservation, i.e. while $\Omega$ commutes with transformations of $U(1)$, we are interested in the case where $\Omega_0$, and thus $\Omega_1$, do {\it not} commute with $S(\varphi)$, i.e.
\begin{subequations}
\label{commutators}
\begin{eqnarray}
\left[\Omega_0,S(\varphi)\right]&\neq& 0 \, , \label{commutators2} \\
\left[\Omega_1,S(\varphi)\right]&\neq& 0 \, . \label{commutators3} 
\end{eqnarray}
\end{subequations}
In this context, the vacuum $|  \Phi \rangle$ is a Bogoliubov state that is {\it deformed} in gauge space and that is thus not an eigenstate of $A$; i.e. it spans several IRREPs of $U(1)$.

The operator $\Omega_{0}$ can be written in diagonal form in terms of its one quasi-particle eigenstates
\begin{equation}
\Omega_{0} \equiv \Omega^{00} + \sum_{k} E_k \beta^{\dagger}_k \beta_k \label{hzero} \, ,
\end{equation}
with $E_k > 0$ for all $k$. Eventually, it remains to specify how the quasi-particle operators $\{\beta_k ; \beta^{\dagger}_k\}$ and energies $\{E_k\}$ are determined. This corresponds to fixing the Bogoliubov transformation $W$ (Eq.~\ref{bogomatrix}), and thus $|  \Phi \rangle$, along with $\bar{\Omega}^{11}$. The traditional choice consists in requiring that $|  \Phi \rangle$ minimizes $\Omega^{00}$, which amounts to solving so-called Hartree-Fock-Bogoliubov (HFB) equations~\cite{ring80a} to fix both $W$ and the set of $E_k$. This corresponds to working within a Moller-Plesset scheme. While this choice conveniently leads to canceling $\Omega^{20}$ and $\Omega^{02}$ in $\Omega$ while diagonalizing $\bar{\Omega}^{11}=\Omega^{11}$, we do not impose such a choice in the present work in order to design the formalism in its general Rayleigh-Schroedinger form. 

Introducing many-body states generated via an even number of quasi-particle excitations\footnote{The present many-body formalism only requires to consider Bogoliubov states with a given, i.e. odd or even, number parity. As such, Bogoliubov states involved necessarily differ from one another by an {\it even} number of quasi-particle excitations.} of the vacuum
\begin{eqnarray}
| \Phi^{k_1 k_2\ldots} \rangle &\equiv& {\cal B}_{k_1 k_2\ldots}  |  \Phi \rangle \, , 
\end{eqnarray}
where
\begin{eqnarray}
{\cal B}_{k_1 k_2\ldots} &\equiv& \beta^{\dagger}_{k_1} \, \beta^{\dagger}_{k_2} \,  \ldots  \, ,  \label{phexcitation}
\end{eqnarray}
the unperturbed system is fully characterized by its complete set of eigenstates
\begin{subequations}
\begin{eqnarray}
\Omega_{0}\, |  \Phi \rangle &=& \Omega^{00} \, |  \Phi \rangle \, , \\
\Omega_{0}\, |  \Phi^{k_1 k_2\ldots} \rangle &=& \left[\Omega^{00} \!+\! E_{k_1} \!+\! E_{k_2}\!+\!\ldots\right] |  \Phi^{k_1 k_2\ldots} \rangle  \label{phi} \, . \; \;
\end{eqnarray}
\end{subequations}
As mentioned above, the Bogoliubov vacuum $|  \Phi \rangle$ necessarily possesses a {\it closed-shell} character with respect to elementary (quasi-particle) excitations. This means that there exists a finite energy gap between the vacuum state and the lowest two quasi-particle excitations, i.e. $E_{k_1}+E_{k_1}\geq 2\Delta_{\text{F}} > 0$ for all $(k_1,k_2)$, where $\Delta_{\text{F}}$ is traditionally characterized as the pairing gap.

\subsection{Rotated reference state}
\label{chap:rotatedbasis}

Particle creation and annihilation operators are tensor operators of rank $+1$ and $-1$, respectively. As a result, they transform under gauge rotation according to
\begin{subequations}
\begin{eqnarray}
c_{\bar{p}} &\equiv& S(\varphi) \, c_{p} \, S^{-1}(\varphi) = e^{-i\varphi}  \, c_{p} \, , \\
c_{\bar{p}}^{\dagger} &\equiv& S(\varphi) \, c^{\dagger}_{p} \, S^{-1}(\varphi) = e^{+i\varphi} \, c_{p}^{\dagger} \, ,
\end{eqnarray}
\end{subequations}
where $S_{pq}(\varphi) \equiv \langle p | S(\varphi) | q \rangle = e^{i\varphi} \delta_{pq}$ is the unitary transformation matrix connecting the rotated particle basis to the unrotated one. This leads to rotated quasi-particle operators\footnote{Mixing particle creation and annihilation operators, quasi-particle operators do not correspond to tensor operator of specific rank.}
\begin{subequations}
\begin{eqnarray}
\beta_{\bar{k}} &\equiv& S(\varphi) \, \beta_{k} \, S^{-1}(\varphi) \nonumber \\
&=&  \sum_{p} U^{*}_{pk} \, e^{-i\varphi}  \, c_{p}  + V^{*}_{pk} \, e^{+i\varphi} \,  c^{\dagger}_{p} \, , \\
\beta_{\bar{k}}^{\dagger} &\equiv& S(\varphi) \, \beta_{k}^{\dagger} \, S^{-1}(\varphi) \nonumber \\
&=&  \sum_{p} U_{pk} \, e^{+i\varphi} \, c^{\dagger}_{p} + V_{pk} \, e^{-i\varphi} \,  c_{p} \, ,
\end{eqnarray}
\end{subequations}
which are used to specify the rotated partner of $|  \Phi \rangle$ under the form
\begin{subequations}
\begin{eqnarray}
|  \Phi (\varphi) \rangle &\equiv& S(\varphi) |  \Phi \rangle \\
&=& \mathcal{C} \displaystyle \prod_{\bar{k}} \beta_{\bar{k}} | 0 \rangle \, .
\end{eqnarray}
\end{subequations}
By virtue of Thouless' theorem~\cite{thouless60}, the rotated state $|  \Phi (\varphi) \rangle$ is itself a Bogoliubov state associated with the transformation
\begin{subequations}
\begin{eqnarray}
W^{\varphi} &\equiv&  \left(
\begin{array} {cc}
U^{\varphi}  & V^{\varphi \ast}  \\
V^{\varphi}  &  U^{\varphi \ast} 
\end{array}
\right) \\
&=&   
\left(
\begin{array} {cc}
e^{+i\varphi} U  & e^{+i\varphi} V^{\ast} \\
e^{-i\varphi} V &  e^{-i\varphi}U^{\ast} 
\end{array}
\right)\, , \label{rotbogomatrix}
\end{eqnarray}
\end{subequations}
that leads to defining the skew-symmetrix matrix 
\begin{subequations}
\begin{eqnarray}
Z^{\varphi} &\equiv& V^{\varphi \ast}[U^{\varphi \ast}]^{-1} \\
&=& e^{2i\varphi} Z\, .
\end{eqnarray}
\end{subequations}
State $| \Phi(\varphi) \rangle$ is the ground-state of the rotated Hamiltonian $\Omega_0(\varphi)\equiv S(\varphi) \Omega_0 S^{-1}(\varphi)$ with the $\varphi$-independent eigenvalue $\Omega^{00}$. This feature characterizes the fact that, while the unperturbed ground-state is non-degenerate with respect to quasi-particle excitations, there exists a degeneracy, i.e. a zero mode, in the manifold of its gauge rotated partners. In other words, breaking $U(1)$ symmetry commutes the degeneracy of the unperturbed state with respect to individual excitations into a degeneracy with respect to collective rotations in gauge space. Lifting the latter degeneracy is eventually necessary for a finite quantum systems and it is the objective of the present work to do so within the frame of BMBPT and BCC theory.

\subsection{Unperturbed off-diagonal norm kernel}
\label{transitionover}

As proven in Ref.~\cite{Robledo:2009yd}, the overlap between $|  \Phi \rangle$ and $|  \Phi (\varphi) \rangle$ is best expressed as a Pfaffian
\begin{equation}
\langle \Phi | \Phi(\varphi) \rangle = |\mathcal{C}|^{2} (-1)^{N(N+1)/2} \textrm{pf}\left(\begin{array}{cc}
Z^{\varphi} & -1\\
1 & -Z^{\ast}
\end{array}\right) \, , \label{kernel}
\end{equation}
with $N$ the (even) dimension of the (truncated) one-body Hilbert space ${\cal H}_1$ spanned by the basis $\{c_{p};c^{\dagger}_{p}\}$. In case both states $|  \Phi \rangle$ and $|  \Phi (\varphi) \rangle$ share a common discrete symmetry like simplex or time reversal\footnote{This will typically be the case when applying the present approach to even-even nuclei.}, the overlap can be reduced to a determinant~\cite{Robledo:2009yd} without any loss of its sign. 

In Sec.~\ref{newunperturbedkernel}, a new alternative to Eq.~\ref{kernel} will be proposed.

\subsection{Unperturbed off-diagonal density matrix}
\label{transitiondens}

The transformation that links $|  \Phi \rangle$ and $|  \Phi (\varphi) \rangle$ is itself a Bogoliubov transformation built as the product $W^{\varphi \dagger}W$ such that
\begin{subequations}
\label{e:p2qpAA}
\begin{align}
\beta_{k_1} &= \sum_{k_2} A^{*}_{k_2k_1}(\varphi) \, \beta_{\bar{k}_2} 
 + B^{*}_{k_2k_1}(\varphi) \,  \beta^{\dagger}_{\bar{k}_2} \, , \\
\beta_{k_1}^{\dagger} &= \sum_{k_2} A_{k_2k_1}(\varphi) \, \beta^{\dagger}_{\bar{k}_2} 
 + B_{k_2k_1}(\varphi) \,  \beta_{\bar{k}_2} \, ,
\end{align}
\end{subequations}
where
\begin{subequations}
\begin{eqnarray}
A(\varphi) &\equiv& U^{\varphi \dagger}U +  V^{\varphi \dagger}V  \, , \\
B(\varphi) &\equiv& V^{\varphi T}U +  U^{\varphi T}V \, .
\end{eqnarray}
\end{subequations}

Having defined this transitional Bogoliubov transformation, one introduces the {\it off-diagonal} generalized density matrix expressed in the one-body basis as~\cite{ring80a}
\begin{subequations}
\label{offdiaggeneralizeddensitymatrix}
\begin{eqnarray}
{\cal R}(\varphi) &\equiv& 
\left(
\begin{array} {cc}
\frac{\langle \Phi | c^{\dagger}c^{\phantom{\dagger}} | \Phi(\varphi) \rangle}{\langle \Phi | \Phi(\varphi) \rangle} & \frac{\langle \Phi | c^{\phantom{\dagger}}c^{\phantom{\dagger}} | \Phi(\varphi) \rangle}{\langle \Phi | \Phi(\varphi) \rangle} \\
\frac{\langle \Phi | c^{\dagger}c^{\dagger} | \Phi(\varphi) \rangle}{\langle \Phi | \Phi(\varphi) \rangle} &  \frac{\langle \Phi | c^{\phantom{\dagger}}c^{\dagger} | \Phi(\varphi) \rangle}{\langle \Phi | \Phi(\varphi) \rangle}
\end{array}
\right)  \\
&\equiv& 
\left(
\begin{array} {cc}
+\rho(\varphi) & +\kappa(\varphi) \\
-\bar{\kappa}^{\ast}(\varphi) &  -\sigma^{\ast}(\varphi)
\end{array}
\right)  \\
&=& 
\left(
\begin{array} {cc}
V^{\varphi \ast}[A^{T}(\varphi)]^{-1}V^T & V^{\varphi \ast}[A^{T}(\varphi)]^{-1}U^T \\
U^{\varphi \ast}[A^{T}(\varphi)]^{-1}V^T &  U^{\varphi \ast}[A^{T}(\varphi)]^{-1}U^T
\end{array}
\right)
\, , 
\end{eqnarray}
\end{subequations}
which, after transformation to the quasi-particle basis of $| \Phi \rangle$, becomes
\begin{subequations}
\label{offdiaggeneralizeddensitymatrix2}
\begin{eqnarray}
{\bf R}(\varphi) &=& 
\left(
\begin{array} {cc}
\frac{\langle \Phi | \beta^{\dagger}\beta^{\phantom{\dagger}} | \Phi(\varphi) \rangle}{\langle \Phi | \Phi(\varphi) \rangle} & \frac{\langle \Phi | \beta^{\phantom{\dagger}}\beta^{\phantom{\dagger}} | \Phi(\varphi) \rangle}{\langle \Phi | \Phi(\varphi) \rangle} \\
\frac{\langle \Phi | \beta^{\dagger}\beta^{\dagger} | \Phi(\varphi) \rangle}{\langle \Phi | \Phi(\varphi) \rangle} &  \frac{\langle \Phi | \beta^{\phantom{\dagger}}\beta^{\dagger} | \Phi(\varphi) \rangle}{\langle \Phi | \Phi(\varphi) \rangle}
\end{array}
\right)  \\
&\equiv& 
\left(
\begin{array} {cc}
R^{+-}(\varphi) & R^{--}(\varphi) \\
R^{++}(\varphi) &  R^{-+}(\varphi)
\end{array}
\right)  \\
&=& 
\left(
\begin{array} {cc}
0 & B^{\dagger}(\varphi)[A^{T}(\varphi)]^{-1} \\
0 &  1
\end{array}
\right)
\, , 
\end{eqnarray}
\end{subequations}
with
\begin{eqnarray}
R^{--}(\varphi)&=& B^{\dagger}(\varphi)[A^{T}(\varphi)]^{-1} \label{upperright} \\
&=& V^{\dagger} (1-e^{2i\varphi})(1-e^{2i\varphi}Z^{\ast}Z)^{-1}[U^{T}]^{-1}\, . \nonumber
\end{eqnarray}
We note that ${\bf R}(0) = {\bf R}$ and that
\begin{subequations}
\label{propR}
\begin{eqnarray}
R^{--}(\varphi)&=& - R^{-- T}(\varphi) \, , \\
R^{--}(0) &=& 0 \, .
\end{eqnarray}
\end{subequations}

\subsection{Unperturbed off-diagonal propagator}
\label{prop}

Quasi-particle creation and annihilation operators read in the interaction representation
\begin{subequations}
\label{aalphatau}
\begin{eqnarray}
\beta_{k} \left(\tau\right)  &\equiv& e^{+\tau \Omega_{0}} \, \beta_{k} \, e^{-\tau \Omega_{0}}=e^{-\tau E_k} \, \beta_{k} \label{aalphatau2} \, , \\
\beta_{k}^{\dagger}\left(\tau\right)  &\equiv& e^{+\tau \Omega_{0}} \, \beta_{k}^{\dagger} \, e^{-\tau
\Omega_{0}}=e^{+\tau E_{k}} \, \beta_{k}^{\dagger} \, . \label{aalphatau1}
\end{eqnarray}
\end{subequations}

\begin{figure}[t!]
\begin{center}
\includegraphics[clip=,width=0.3\textwidth,angle=0]{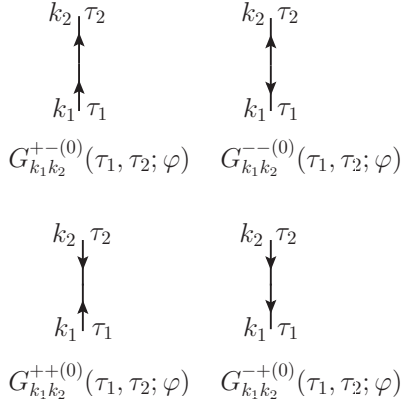}\\
\end{center}
\caption{
\label{prop1}
Diagrammatic representation of the four unperturbed elementary off-diagonal one-body propagators $G^{gg' (0)}(\varphi)$.}
\end{figure}

The generalized unperturbed off-diagonal one-body propagator is introduced as a $2\times 2$ matrix in Bogoliubov space 
\begin{eqnarray}
\bold{G}^{0}(\varphi) 
&\equiv& 
\left(
\begin{array} {cc}
G^{+- (0)}(\varphi) & G^{-- (0)}(\varphi) \\
G^{++ (0)}(\varphi) &  G^{-+ (0)}(\varphi)
\end{array}
\right)   \label{offdiaggeneralizedprog}
\end{eqnarray}
whose four components are defined through their matrix elements in the quasi-particle basis $\{\beta_{k};\beta^{\dagger}_{k}\}$ according to
\begin{subequations}
\label{propagatorsA}
\begin{eqnarray}
G^{+- (0)}_{k_1k_2}(\tau_1, \tau_2 ; \varphi) &\equiv& \frac{\langle \Phi |  \textmd{T}[\beta^{\dagger}_{k_1}(\tau_1) \beta_{k_2}(\tau_2)] | \Phi(\varphi) \rangle}{\langle \Phi | \Phi(\varphi) \rangle}  \, , \label{propagatorsA1} \; \; \;\\
G^{-- (0)}_{k_1k_2}(\tau_1, \tau_2 ; \varphi) &\equiv& \frac{\langle \Phi |  \textmd{T}[\beta_{k_1}(\tau_1) \beta_{k_2}(\tau_2)] | \Phi(\varphi) \rangle}{\langle \Phi | \Phi(\varphi) \rangle}  \, , \label{propagatorsA2} \; \; \;\\
G^{++ (0)}_{k_1k_2}(\tau_1, \tau_2 ; \varphi) &\equiv& \frac{\langle \Phi |  \textmd{T}[\beta^{\dagger}_{k_1}(\tau_1) \beta^{\dagger}_{k_2}(\tau_2)] | \Phi(\varphi) \rangle}{\langle \Phi | \Phi(\varphi) \rangle}  \, , \label{propagatorsA3} \; \; \;\\
G^{-+ (0)}_{k_1k_2}(\tau_1, \tau_2 ; \varphi) &\equiv& \frac{\langle \Phi |  \textmd{T}[\beta_{k_1}(\tau_1) \beta^{\dagger}_{k_2}(\tau_2)] | \Phi(\varphi) \rangle}{\langle \Phi | \Phi(\varphi) \rangle}  \, , \label{propagatorsA4} \; \; \;
\end{eqnarray}
\end{subequations}
where $\textmd{T}$ denotes the time ordering operator. The diagrammatic representation of the four elementary propagators $G^{gg' (0)}(\varphi)$, with $g\equiv \pm$ and $g'\equiv \pm$, is provided in Fig.~\ref{prop1}. The above definition of propagators implies the relations 
\begin{subequations}
\label{relat}
\begin{eqnarray}
G^{+- (0)}_{k_1k_2}(\tau_1, \tau_2 ; \varphi) &=& -G^{+- (0)}_{k_2k_1}(\tau_2, \tau_1 ; \varphi)  \, , \label{relat1} \; \; \;\\
G^{-- (0)}_{k_1k_2}(\tau_1, \tau_2 ; \varphi) &=& -G^{-- (0)}_{k_2k_1}(\tau_2, \tau_1 ; \varphi) \, , \label{relat2} \; \; \;\\
G^{++ (0)}_{k_1k_2}(\tau_1, \tau_2 ; \varphi) &=& -G^{++ (0)}_{k_2k_1}(\tau_2, \tau_1 ; \varphi). \; \; \;
\end{eqnarray}
\end{subequations}
Combining Eqs.~\ref{offdiaggeneralizeddensitymatrix2}, \ref{propR} and \ref{aalphatau}, together with anticommutation rules of quasi-particle creation and annihilation operators, one obtains that
\begin{subequations}
\label{propagatorsB}
\begin{eqnarray}
G^{+- (0)}_{k_1k_2}(\tau_1, \tau_2 ; \varphi) &=& - e^{-(\tau_2-\tau_1)E_{k_1}} \theta(\tau_2-\tau_1) \delta_{k_1k_2} \, , \label{propagatorsB1}\; \; \; \; \; \; \\
G^{-- (0)}_{k_1k_2}(\tau_1, \tau_2 ; \varphi) &=& + e^{-\tau_1 E_{k_1}} e^{-\tau_2 E_{k_2}} R^{--}_{k_1k_2}(\varphi) \, , \label{propagatorsB2} \\
G^{++ (0)}_{k_1k_2}(\tau_1, \tau_2 ; \varphi) &=& 0  \, , \label{propagatorsB3} \\
G^{-+ (0)}_{k_1k_2}(\tau_1, \tau_2 ; \varphi) &=& + e^{-(\tau_1-\tau_2)E_{k_1}} \theta(\tau_1-\tau_2) \delta_{k_1k_2}   \label{propagatorsB4} \, , \; \; \; \; \; \;
\end{eqnarray}
\end{subequations}
where only $G^{-- (0)}(\varphi)$ actually depends on the gauge angle $\varphi$ and is such that $G^{-- (0)}(0)=0$.

The equal-time unperturbed propagator deserves special attention. Equal-time propagators will solely arise from contracting two quasi-particle operators belonging to the same normal-ordered operator displaying creation operators to the left of annihilation ones. In both $G^{+- (0)}_{k_1k_2}(\tau, \tau ; \varphi)$ and $G^{-+ (0)}_{k_1k_2}(\tau, \tau ; \varphi)$, this necessarily leads to selecting a contraction associated with $R^{+-}(\varphi)$ that is identically zero. As a result, a non-zero equal-time propagator is always of the anomalous type\footnote{As a general wording, {\it normal} contractions involve one (quasi-)particle creation operator and one (quasi-)particle annihilation operator whereas {\it anomalous} contractions involve two (quasi-)particle operators of the same type.}, i.e.
\begin{subequations}
\label{propagatorsC}
\begin{eqnarray}
G^{+- (0)}_{k_1k_2}(\tau, \tau ; \varphi) &\equiv& 0 \, , \label{propagatorsC1} \\
G^{-- (0)}_{k_1k_2}(\tau, \tau ; \varphi) &\equiv& + e^{-\tau (E_{k_1}+E_{k_2})} R^{--}_{k_1k_2}(\varphi) \, , \label{propagatorsC2} \\
G^{++ (0)}_{k_1k_2}(\tau, \tau ; \varphi) &\equiv& 0  \, , \label{propagatorsC3} \\
G^{-+ (0)}_{k_1k_2}(\tau, \tau ; \varphi) &\equiv& 0  \label{propagatorsC4} \, ,
\end{eqnarray}
\end{subequations}
such that no equal-time contraction, and thus no contraction of an interaction vertex onto itself, can occur in the diagonal case, i.e. for $\varphi=0$.

\subsection{Expansion of the evolution operator}

As recalled in App.~\ref{perturbativeannexe}, the evolution operator can be expanded in powers of $\Omega_{1}$ under the form
\begin{eqnarray}
{\cal U}(\tau) &=& e^{-\tau \Omega_{0}} \, \textmd{T}e^{-\int_{0}^{\tau}dt \Omega_{1}\left(t\right) \, ,
} \label{evol1}%
\end{eqnarray}
where\footnote{A time-dependent {\it operator} $O(\tau)$  should not be confused with the gauge-angle dependent {\it kernel} $O(\varphi)$ (or its time-dependent partner $O(\tau,\varphi)$). Later on (see Sec.~\ref{transformedoperator}), the notion of transformed, gauge-angle dependent, {\it operator} $\tilde{O}(\varphi)$ will be introduced and distinguished by the "tilde". It should be confused neither with the kernel $O(\varphi)$ nor with the operator $O(\tau)$.}
\begin{equation}
\Omega_{1}\left( \tau\right)  \equiv e^{\tau \Omega_{0}}\Omega_{1}e^{-\tau \Omega_{0}} \, ,
\end{equation}
defines the perturbation in the interaction representation. 

\subsection{Off-diagonal norm kernel}
\label{normkernel2}

\subsubsection{Off-diagonal BMBPT expansion}

Expressing $\Omega_1$ in the eigenbasis of $\Omega_0$ and expanding the exponential in Eq.~\ref{evol1} in power series, one obtains the perturbative expansion of the off-diagonal norm kernel
\begin{widetext}
\begin{eqnarray}
N(\tau,\varphi) &=& \langle \Phi | e^{-\tau \Omega_{0}} \, \textmd{T}e^{-\int_{0}^{\tau}dt \Omega_{1}\left(t\right)} | \Phi(\varphi) \rangle \label{expansionnormkernel}  \\
&=& e^{-\tau\Omega^{00}} \langle \Phi |\Big\{  1-\int_{0}^{\tau}d\tau_1 \Omega_{1}\left(  \tau_1\right)  +\frac{1}{2!}\int_{0}^{\tau}d\tau_{1}d\tau
_{2}\textmd{T}\left[  \Omega_{1}\left(  \tau_{1}\right)  \Omega_{1}\left(  \tau_{2}\right)  \right]
+... \Big\}|  \Phi(\varphi) \rangle \nonumber\\
&=& e^{-\tau\Omega^{00}}\Big\{  \sum_{p=0}^{\infty}\frac{(-1)^{p}}{p!} \sum_{\substack{i_1+j_1=2,4 \\ \vdots \\i_p+j_p=2,4}} \int_{0}^{\tau}\!\!d\tau_{1}\ldots d\tau_{p} \hspace{-0.3cm}\sum_{\substack{k_1 \ldots k_{i_1} \\k_{i_1+1} \ldots k_{i_1+j_1} \\ \vdots \\ l_1 \ldots l_{i_p} \\l_{i_p+1} \ldots l_{i_p+j_p}}} \hspace{-0.3cm} \frac{\Omega^{i_1j_1}_{k_1 \ldots k_{i_1} k_{i_1+1} \ldots k_{i_1+j_1}}}{(i_1)!(j_1)!} \ldots  \frac{\Omega^{i_pj_p}_{l_1 \ldots l_{i_p} l_{i_p+1} \ldots l_{i_p+j_p}}}{(i_p)!(j_p)!}  \nonumber\\
&& \hspace{0.2cm} \times \langle \Phi |  \textmd{T}\left[  \beta^{\dagger}_{k_1}\left(  \tau_{1}\right)\ldots \beta^{\dagger}_{k_{i_1}}\left(  \tau_{1}\right)\beta_{k_{i_1+j_1}}\left(  \tau_{1}\right) \ldots \beta_{k_{i_1+1}}\left(  \tau_{1}\right) \ldots \beta^{\dagger}_{l_1}\left(  \tau_{p}\right)\ldots \beta^{\dagger}_{l_{i_p}}\left(  \tau_{p}\right)\beta_{l_{i_p+j_p}}\left(  \tau_{p}\right) \ldots \beta_{l_{i_p+1}}\left(  \tau_{p}\right) 
 \right]  |  \Phi(\varphi) \rangle  \Big\} . \nonumber 
\end{eqnarray}
\end{widetext}
where $\breve{\Omega}^{11}$ must be understood in place of $\Omega^{11}$ whenever necessary.

The off-diagonal matrix elements of products of time-dependent field operators appearing in Eq.~\ref{expansionnormkernel} can be
expressed as the sum of all possible systems of products of elementary contractions $G^{gg' (0)}_{k_1k_2}(\tau_1, \tau_2 ; \varphi)$ (Eqs.~\ref{propagatorsA}-\ref{propagatorsC}), eventually multiplied by the unperturbed norm kernel $\langle \Phi | \Phi(\varphi) \rangle$ (Eq.~\ref{kernel}). This derives from a generalized Wick theorem~\cite{balian69a} applicable to matrix elements between {\it different} (non-orthogonal) left and right vacua, i.e. presently $\langle \Phi |$ and $| \Phi(\varphi) \rangle$, which constitutes a powerful way to deal exactly with the presence of the rotation operator $S(\varphi)$ in off-diagonal kernels. Eventually, this makes possible to represent $N(\tau,\varphi)$ diagrammatically following techniques~\cite{blaizot86} usually applied to the diagonal norm kernel $N(\tau,0)$~\cite{bloch58a}.

\subsubsection{Off-diagonal BMBPT diagrammatic rules}
\label{diagrulenormMBPT}

Equation~\ref{expansionnormkernel} for $N(\tau,\varphi)$ can be translated into an infinite set of vacuum-to-vacuum diagrams. The rules to build and compute those diagrams are now detailed. 
\begin{enumerate}
\item A vacuum-to-vacuum, i.e. closed, Feynman diagram of order $p$ consists of $p$ vertices $\Omega^{ij}(\tau_k)$  connected by fermionic quasi-particle lines, i.e. elementary propagators $G^{gg' (0)}(\varphi)$, forming a set of closed loops. 
\item Each vertex is labeled by a time variable while each line is labeled by two quasi-particle indices and two time labels at its ends, the latter being associated with the two vertices the line is attached to. Each vertex contributes a factor $\Omega^{ij}_{k_1 \ldots k_i k_{i+1} \ldots k_{i+j}}$ with the sign convention detailed in Sec.~\ref{diagramsforvertices}. Each line contributes a factor $G^{gg' (0)}_{k_1k_2}(\tau_k, \tau_{k'} ; \varphi)$, where $g=\pm$  and $g'=\pm$ characterize the type of elementary propagator the line corresponds to\footnote{A normal line can be interpreted as $G^{-+(0)}(\varphi)$ or $G^{+-(0)}(\varphi)$ depending on the ascendant or descendant reading of the diagram. Similarly, the ordering of quasi-particle and time labels of a propagator depends on the ascendant or descendant reading of the diagram. While both ways are allowed, one must consistently interpret {\it all} the lines involved in a given diagram in the {\it same} way, i.e. sticking to an ascendant or descendant way of reading the diagram all throughout.}.
\item The contributions to $N(\tau,\varphi)$ of order $p$ are generated by drawing all possible vacuum-to-vacuum diagrams involving $p$ operators $\Omega_1(\tau_k)$. This is done by contracting the quasi-particle lines attached to the vertices in all possible ways, allowing both for normal and anomalous propagators. Eventually, the set of diagrams must be limited to {\it topologically distinct} diagrams, i.e. diagrams that cannot be obtained from one another via a mere displacement, i.e. translation, of the vertices.
\item All quasi-particle labels must be summed over while all time variables must be integrated over from $0$ to $\tau$.
\item A sign factor $(-1)^{p+n_c}$, where $p$ denotes the order of the diagram and $n_c$ denotes the number of crossing lines in the diagram, must be considered. The overall sign results from multiplying the latter factor with the sign associated with each factor $\Omega^{ij}_{k_1 \ldots k_i k_{i+1} \ldots k_{i+j}}$ as discussed above.
\item Each diagram comes with a numerical prefactor obtained from the following combination
\begin{itemize}
\item A factor $1/(n_e)!$ must be considered for {\it each} group of $n_e$ equivalent lines. Equivalent lines must all begin and end at the same vertices (or vertex, for anomalous propagators starting and ending at the same vertex), and must correspond to the same type of contractions, i.e. they must all correspond to propagators characterized by the same superscripts $g$ and $g'$ in addition to having identical time labels.
\item Given the previous rule, an extra factor $1/2$ must be considered for {\it each} anomalous propagator that starts and ends at the same vertex. The proof of this unusual\footnote{This rule is "unusual" only because many-body methods based on diagrammatic techniques invoking anomalous contractions are scarce in the physics literature.} diagrammatic rule, already used in Ref.~\cite{soma11a}, is given in App.~\ref{ruleanomalouscontraction}.
\item A symmetry factor $1/n_s$ must be considered in connection with exchanging the time labels of the vertices in all possible ways. The factor $n_s$ corresponds to the number of ways exchanging the time labels provides a diagram that is topologically equivalent to the original one.
\end{itemize} 
\end{enumerate} 

As each operator $\Omega_1(\tau)$ actually contains eight normal-ordered operators $\Omega^{ij}(\tau)$, with $i+j=2,4$, and given that four types of propagators must be considered, one may be worried about the proliferation of diagrams. Whereas the number of diagrams to be considered is indeed significantly larger than in standard, i.e. diagonal ($\varphi=0$), BMBPT, several "selection rules" can be identified by virtue of Eqs.~\ref{propagatorsB} and~\ref{propagatorsC} that limit drastically the number of non-zero diagrams. Let us detail those additional rules.
\begin{enumerate}
\item As $G^{++ (0)}(\varphi)$ is identically zero, non-zero anomalous off-diagonal contractions necessarily involve two quasi-particle {\it annihilation} operators, i.e. the diagram is identically zero anytime a contraction between two creation operators is considered. Whenever a string of operators contain more creation operators than annihilation operators, the result is thus necessarily zero, i.e. for an arbitrary matrix element $\langle \Phi | \Omega^{i_1j_1}(\tau_1) \Omega^{i_2j_2}(\tau_2) \ldots \Omega^{i_pj_p}(\tau_p) |  \Phi(\varphi) \rangle$ to give non-zero contributions (diagrams), it is mandatory that $n_a = \sum_{k=1}^{p}(j_k-i_k) \ge 0$. Corresponding diagrams must contain exactly $n_a$ anomalous contractions to provide a non-zero result. Diagrams at play in diagonal ($\varphi=0$) BMBPT reduce to those characterized by $n_a=0$ as no anomalous contraction occurs in this case (i.e. $G^{-- (0)}(0)=0$).
\item Normal lines linking two given vertices $\Omega^{i_kj_k}(\tau_k)$  and $\Omega^{i_{k'}j_{k'}}(\tau_{k'})$ must propagate in the same direction. Equations~\ref{propagatorsB1} and~\ref{propagatorsB4} indeed indicate that two normal lines propagating in opposite directions induce a factor $\theta(\tau_k-\tau_{k'})\theta(\tau_{k'}-\tau_k)$ that makes the diagram to be zero.
\item As Eq.~\ref{propagatorsC} demonstrates, propagators starting and ending at the same vertex are necessarily of anomalous, i.e. $G^{-- (0)}(\varphi)$, type.
\end{enumerate}

\subsubsection{Exponentiation of connected diagrams}

Diagrams representing the off-diagonal norm kernel are vacuum-to-vacuum diagrams, i.e. diagrams with no incoming or outgoing external lines. In general, a diagram consists of disconnected parts which are joined neither by vertices nor by propagators. Consider a diagram $\Gamma(\tau,\varphi)$ contributing to Eq.~\ref{expansionnormkernel} and consisting of $n_{1}$ identical connected parts $\Gamma_{1}(\tau,\varphi)$, of $n_{2}$ identical connected parts $\Gamma_{2}(\tau,\varphi)$, and so on. Using for simplicity the same symbol to designate both the diagram and its contribution, the whole diagram gives
\begin{equation}
\Gamma(\tau,\varphi) = \frac{\left[\Gamma_{1}(\tau,\varphi)\right]^{n_{1}}}{n_{1}!}\frac{\left[  \Gamma_{2}(\tau,\varphi)\right]^{n_{2}}}{n_{2}!}...
\end{equation}
The factor $n_{i}!$ is the symmetry factor due to the exchange of time labels among the $n_{i}$ identical diagrams $\Gamma_{i}(\tau,\varphi)$. It follows that the sum of all vacuum-to-vacuum diagrams is equal to the exponential of the sum of {\it connected} vacuum-to-vacuum diagrams
\begin{eqnarray}
\sum_{\Gamma} \Gamma(\tau,\varphi) &=& \sum_{n_{1}n_{2}...}\frac{\left[  \Gamma_{1}(\tau,\varphi)\right]^{n_{1}}}{n_{1}!}\frac{\left[\Gamma_{2}(\tau,\varphi)\right]^{n_{2}}}{n_{2}!}... \nonumber \\
&=& e^{\Gamma_{1}(\tau,\varphi)+\Gamma_{2}(\tau,\varphi)+...}  \label{condiag} \, .
\end{eqnarray}
Consequently, the norm can be written as
\begin{equation}
N(\tau,\varphi) =   e^{-\tau \Omega^{00} + n(\tau,\varphi)}  \, \langle \Phi |  \Phi(\varphi) \rangle \label{wexp3} \, ,
\end{equation}
where $n(\tau,\varphi)\equiv\sum^{\infty}_{n=1}n^{(n)}(\tau,\varphi)$, with $n^{(n)}(\tau,\varphi)$ the sum of all $\varphi$-dependent connected vacuum-to-vacuum diagrams of order $n$. By virtue of Eqs.~\ref{condiag}-\ref{wexp3}, only connected diagrams have to be eventually considered in practice.

\subsubsection{Computing diagrams}
\label{computingdiagramsNc}

\begin{figure}[t!]
\begin{center}
\includegraphics[width=1.03\columnwidth]{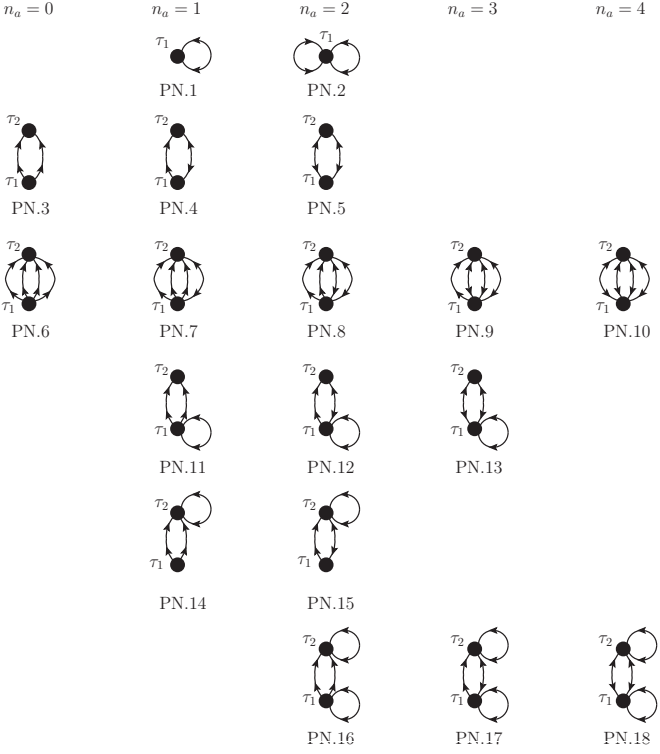}
\end{center}
\caption{
\label{diagramsNc}
First- and second-order connected Feynman diagrams contributing to $n(\tau,\varphi)$. Vertices referring to $\Omega^{11}$, i.e. the lower vertex in diagrams PN.4 and PN.15, must actually be understood as referring to $\breve{\Omega}^{11}$.}
\end{figure}

The eighteen non-zero first- and second-order connected vacuum-to-vacuum diagrams contributing to $n(\tau,\varphi)$ are displayed in Fig.~\ref{diagramsNc}, where they are classified according to the value of $n_a$, i.e. according to the number of anomalous lines they contain. 

Choosing the reference state $|\Phi\rangle$ to be the solution of HFB equations amounts to setting $\breve{\Omega}^{11}=\Omega^{20}=\Omega^{02}=0$ such that diagrams PN.1, PN.3-PN.5 and PN.11-PN.15 are zero in the Moller-Plesset scheme, i.e. the set reduces from eighteen non-zero diagrams to nine non-zero diagrams at second order. Finally, $n(\tau,0)$ at play in diagonal BMBPT reduces to PN.3 and PN.6 ($n_a=0$) diagrams at second order (only PN.6 in the Moller-Plesset scheme).

\begin{figure}[t!]
\begin{center}
\includegraphics[width=0.35\columnwidth]{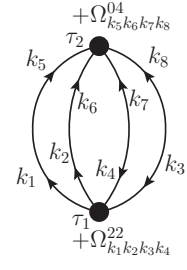}
\end{center}
\caption{
\label{diagramsNcex}
Example of a fully-labeled off-diagonal BMBPT diagram contributing to $n(\tau,\varphi)$, i.e. the second-order diagram labeled PN.8 in Fig.~\ref{diagramsNc}.}
\end{figure}

While the full analytic expression of each of these diagrams is provided in App.~\ref{diagramsN}, we presently detail the calculation of one of them for illustration. The second-order diagram labeled as PN.8 in Fig.~\ref{diagramsNc} is displayed in detail in Fig.~\ref{diagramsNcex}.  It contains one $\Omega^{22}$ vertex and one $\Omega^{04}$ vertex. The diagram contains two anomalous lines ($n_a=2$), two vertices and no crossing lines ($(-1)^{p+n_c}=+1$), two equivalent lines of normal type propagating in the same direction along with two equivalent anomalous lines ($n_e=4$), and a symmetry factor $n_s=1$ as exchanging the time labels of the two vertices gives topologically distinct diagrams. Last but not least, the sign convention for the vertices requires to associate the factors $+\Omega^{22}_{k_1k_2k_3k_4}$ and $+\Omega^{04}_{k_5k_6k_7k_8}$ to the vertices as they appear on the diagram drawn in Fig.~\ref{diagramsNcex}. Eventually, diagram PN.8 reads as
 \begin{widetext}
 \begin{align}
 \text{PN}.8=& +\left(\frac{1}{2}\right)^2 \displaystyle\sum_{\substack{k_1 k_2 k_3 k_4 \\ k_5 k_6k_7k_8}} \hspace{-0.3cm} \Omega^{22}_{k_1k_2k_3k_4} \Omega^{04}_{k_5k_6k_7k_8}  \int\limits_0^{\tau} \mathrm{d} \tau_1 \mathrm{d} \tau_2  G^{+-(0)}_{k_1k_5}(\tau_1, \tau_2; \varphi) G^{+-(0)}_{k_2k_6}(\tau_1, \tau_2; \varphi) G^{--(0)}_{k_3k_8}(\tau_1, \tau_2; \varphi) G^{--(0)}_{k_4k_7}(\tau_1, \tau_2; \varphi)  \nonumber \\
=& +\frac{1}{4} \displaystyle\sum_{\substack{k_1 k_2 k_3 k_4 \\ k_5 k_6}} \hspace{-0.3cm}\Omega^{22}_{k_1k_2k_3k_4} \Omega^{04}_{k_1k_2k_5k_6} R^{--}_{k_3k_6}(\varphi) R^{--}_{k_4k_5}(\varphi)  \int\limits_0^{\tau} \mathrm{d} \tau_1 \mathrm{d} \tau_2 \theta(\tau_2-\tau_1) e^{-\tau_2(E_{k_1}+E_{k_2}+E_{k_5}+E_{k_6}) -\tau_1(E_{k_3}+E_{k_4}-E_{k_1}-E_{k_2})} \nonumber \\
=& +\frac{1}{4} \displaystyle\sum_{\substack{k_1 k_2 k_3 k_4 \\ k_5 k_6}} \frac{\Omega^{22}_{k_1k_2k_3k_4} \Omega^{04}_{k_1k_2k_5k_6}}{E_{k_1}+E_{k_2}-E_{k_3}-E_{k_4}}\left[ \frac{1- e^{-\tau(E_{k_3}+E_{k_4}+E_{k_5}+E_{k_6})}}{E_{k_3}+E_{k_4}+E_{k_5}+E_{k_6}} \right.\nonumber \\
  & \hspace{7cm} \left.- \frac{1- e^{-\tau(E_{k_1}+E_{k_2}+E_{k_5}+E_{k_6})}}{E_{k_1}+E_{k_2}+E_{k_5}+E_{k_6}}\right] R^{--}_{k_3k_6}(\varphi) R^{--}_{k_4k_5}(\varphi)  \, ,
\label{PNex1}
\end{align}
where use was made of the identities provided in App.~\ref{usefulID}. We note that the first line of Eq.~\ref{PNex1} was obtained by reading the diagram from bottom to top, i.e. in a ascendant fashion, in Fig.~\ref{diagramsNcex}. In the infinite $\tau$ limit, the result reduces to
 \begin{align}
 \text{PN}.8=&  +\frac{1}{4} \displaystyle\sum_{\substack{k_1 k_2 k_3 k_4 \\ k_5 k_6}} \frac{\Omega^{22}_{k_1k_2k_3k_4} \Omega^{04}_{k_1k_2k_5k_6}}{(E_{k_3}+E_{k_4}+E_{k_5}+E_{k_6})(E_{k_1}+E_{k_2}+E_{k_5}+E_{k_6})}R^{--}_{k_3k_6}(\varphi) R^{--}_{k_4k_5}(\varphi) \, .
\label{PNex2}
\end{align}

\end{widetext}
This diagram is zero in diagonal BMBPT as $R^{--}(0)=0$.

\subsubsection{Dependence on $\tau$ and $\varphi$}
\label{structure1}

The set of connected diagrams contributing to $n\left(\tau,\varphi\right)$ can be split according to
\begin{equation}
n\left(\tau,\varphi\right) \equiv n\left(\tau ; n_a =0\right)  + n\left(\tau,\varphi ; n_a > 0 \right) \, , \label{split1a}
\end{equation}
where $n\left(\tau ; n_a =0\right) \equiv n\left(\tau,0\right)$ is the sum of connected vacuum-to-vacuum diagrams containing no anomalous propagator and arising in standard, i.e. diagonal, BMBPT. The term $n\left(\tau,\varphi ; n_a > 0 \right)$ gathers all diagrams containing at least one anomalous propagator and carries the full gauge angle dependence of $n\left(\tau,\varphi\right)$. As a consequence of Eq.~\ref{split1a}, Eq.~\ref{wexp3} becomes
\begin{eqnarray}
N(\tau,\varphi) &\equiv& N(\tau,0) \, e^{n\left(\tau,\varphi ; n_a > 0 \right)}  \, \langle \Phi |  \Phi(\varphi) \rangle  \, , \label{splittedN} 
\end{eqnarray}
with 
\begin{eqnarray}
N(\tau,0) &=& e^{-\tau\Omega^{00}+n\left(\tau, 0\right)}   \label{betalimtruc1} \, , 
\end{eqnarray}
and where $n\left(\tau,0 ; n_a > 0 \right)=0$.  

In view of Eq.~\ref{limitnorm}, one is interested in the large $\tau$ limit 
\begin{subequations}
\label{betalim}
\begin{eqnarray}
\underset{\tau\rightarrow\infty}{\lim} n\left(\tau, 0\right) &\equiv& -\tau \Delta \Omega^{A_0}_0 + \ln |\langle \Phi |  \Psi^{A_0}_{0} \rangle|^2 \label{betalim1} \, , \; \; \; \; \\
 \underset{\tau\rightarrow\infty}{\lim} n\left(\tau,\varphi ; n_a > 0 \right) &\equiv& n\left(\varphi ; n_a > 0 \right) \label{betalim2} \, .
\end{eqnarray}
\end{subequations}
Equation~\ref{betalim1} relates to the known result applicable to the logarithm of the diagonal, i.e. $\varphi=0$, norm kernel whose part proportional to $\tau$ provides the correction to the unperturbed ground-state eigenvalue of $\Omega$
\begin{eqnarray}
\Delta \Omega^{A_0}_0 &\equiv& \Omega^{A_0}_0-\Omega^{00} \nonumber \\
&=& \langle \Phi | \Omega_{1} \sum_{k=1}^{\infty} \left(\frac{1}{\Omega^{00}-\Omega_0} \Omega_{1} \right)^{k-1} | \Phi \rangle_{c} \, , \label{frombloch}%
\end{eqnarray}
given under the form of Goldstone's formula~\cite{goldstone57a}, which is here computed relative to the superfluid (i.e. Bogoliubov) reference state $| \Phi \rangle$ breaking global gauge symmetry. This expansion of $\Delta \Omega^{A_0}_0$ based on diagonal BMBPT does not constitute the solution to the problem of present interest but is anyway recovered as a byproduct. Relation~\ref{betalim1} recalls that, in the large $\tau$ limit, the $\varphi$-independent part $n\left(\tau, 0\right)$ gathers a term independent of $\tau$ and a term linear in $\tau$. Contrarily, Eq.~\ref{betalim2} states that the $\varphi$-dependent counterpart $n\left(\tau,\varphi ; n_a > 0 \right)$ is independent of $\tau$ in that limit, i.e. it converges to a finite value when $\tau$ goes to infinity. These characteristic behaviors at large imaginary time can be proven for any arbitrary order by trivially adapting the proof given in App.~B.7 of Paper I.

In Eq.~\ref{betalim1}, the contribution that does not depend on $\tau$ provides the overlap between the unrotated unperturbed state and the correlated ground-state. This overlap is not equal to $1$, which underlines that the expansion of $N(\tau,\varphi)$ does not rely on intermediate normalization at $\varphi=0$. Equation~\ref{betalim2} only contains a term independent of $\tau$ because the presence of the operator $S(\varphi)$ in the off-diagonal norm kernel does not modify $\Delta \Omega^{A_0}_0$ but simply provides the overlap with the ground state selected in the large $\tau$ limit with a dependence on $\varphi$.

\subsubsection{Particle-number conserving case}
\label{SD1}

If the reference state $| \Phi \rangle$ is chosen to be a Slater determinant, i.e. to be an eigenstate of $A$ with eigenvalue $\text{A}_0$, one trivially finds from Eq.~\ref{offdiaggeneralizeddensitymatrix2} that $R^{--}(\varphi)=0$  for all $\varphi$. This leads to the fact that $n\left(\tau,\varphi ; n_a > 0 \right)=0$ for all $\tau$ and $\varphi$. At the same time, the unperturbed off-diagonal norm kernel becomes $\langle \Phi |  \Phi(\varphi) \rangle = e^{i\text{A}_0\varphi}$ such that the off-diagonal norm kernel reduces to
\begin{eqnarray}
N(\tau,\varphi) &\equiv& N(\tau,0) \, e^{i\text{A}_0\varphi}  \, . \label{splittedN} 
\end{eqnarray}
When particle-number symmetry is not broken by the reference state, the introduction of the rotation operator $S(\varphi)$ in the definition of the norm kernel simply leads to an overall phase and the particle-number-conserving MBPT of $N(\tau,0)$ is trivially recovered. As a matter of fact, Eq.~\ref{splittedN} complies with Eq.~\ref{expandedkernels1} as $| \Phi \rangle$ is orthogonal to all the eigenstates of $\Omega$ characterized by $\text{A}\neq \text{A}_0$ in this case.  

\subsection{Off-diagonal grand-potential kernel}
\label{energykernel}

\subsubsection{Off-diagonal BMBPT expansion}

Proceeding similarly to $N(\tau,\varphi)$, and taking the grand potential as a particular example, one obtains the perturbative expansion of an operator kernel according to
\begin{widetext}
\begin{eqnarray}
\Omega(\tau,\varphi) &=& \langle \Phi | e^{-\tau \Omega_{0}} \, \textmd{T}e^{-\int_{0}^{\tau}dt \Omega_{1}\left(t\right)} \Omega | \Phi(\varphi) \rangle \label{expansionenergykernel}  \\
&=& e^{-\tau\Omega^{00}} \langle \Phi |\Big\{ \Omega(0)-\int_{0}^{\tau}d\tau_1 \textmd{T}\left[  \Omega_{1}\left(  \tau_1\right)\Omega(0)\right]  +\frac{1}{2!}\int_{0}^{\tau}d\tau_{1}d\tau
_{2}\textmd{T}\left[  \Omega_{1}\left(  \tau_{1}\right)  \Omega_{1}\left(  \tau_{2}\right)  \Omega(0)\right]
+... \Big\}|  \Phi(\varphi) \rangle , \nonumber 
\end{eqnarray}
\end{widetext}
where each term in the matrix element can be fully expanded in the way that was done for the norm kernel in Eq.~\ref{expansionnormkernel}. The one key difference with the norm kernel relates to the presence of the time-independent operator $\Omega$ to which a fixed time $t=0$ is attributed in order to insert it inside the time ordering at no cost. 

As for $N(\tau,\varphi)$, $\Omega(\tau,\varphi)$ can be expressed diagrammatically according to
\begin{eqnarray}
\Omega(\tau,\varphi) &\equiv&  e^{-\tau\Omega^{00}} \!\!\!  \sum_{i+j=0,2,4} \;  \sum^{\infty}_{n=0} \Omega^{ij \, (n)}(\tau,\varphi) \langle \Phi |  \Phi(\varphi) \rangle \,  , \nonumber
\end{eqnarray}
where $\Omega^{ij \, (n)}(\tau,\varphi)$  denotes the sum of all vacuum-to-vacuum diagrams of order $n$ including the operator $\Omega^{ij}$ at fixed time $t=0$. The convention is that the zero-order diagram $\Omega^{ij \, (0)}(\tau,\varphi)$ solely contains the fixed-time operator $\Omega^{ij}(0)$, i.e. the latter must not be considered when counting the order of the diagram to apply the diagrammatic rules listed in Sec.~\ref{diagrulenormMBPT}.

\subsubsection{Exponentiation of disconnected diagrams}

Any diagram $\Omega^{ij \, (n)}(\tau,\varphi)$ consists of a part that is {\it linked} to the operator $\Omega^{ij}(0)$, i.e. that results from contractions involving the creation and annihilation operators of $\Omega^{ij}(0)$, and parts that are disconnected. In the infinite series of diagrams obtained via the off-diagonal BMBPT expansion of $\Omega^{ij}(\tau,\varphi)$, each vacuum-to-vacuum diagram linked to $\Omega^{ij}(0)$ effectively multiplies the complete set of vacuum-to-vacuum diagrams making up $N(\tau,\varphi)$. Gathering those infinite sets of diagrams accordingly leads to the remarkable factorization
\begin{equation}
\Omega^{ij}(\tau,\varphi) \equiv \omega^{ij}(\tau,\varphi) \, N(\tau,\varphi) \label{linked1} \, ,
\end{equation}
where
\begin{subequations}
\label{linked2}
\begin{eqnarray}
\omega^{ij}(\tau,\varphi) &\equiv&  \sum^{\infty}_{n=0} \omega^{ij \, (n)}(\tau,\varphi) \label{linked2a}  
\end{eqnarray}
\end{subequations}
sums all connected vacuum-to-vacuum diagrams of order $n$ {\it linked} to $\Omega^{ij}(0)$. 

The fact that the (reduced) kernel $O(\tau,\varphi)$ (${\cal O}(\tau,\varphi)$) of any normal-ordered operator $O$ factorizes into its linked/connected part $o(\tau,\varphi)$ times the (reduced) norm kernel $N(\tau,\varphi)$ (${\cal N}(\tau,\varphi)$) similarly to Eq.~\ref{linked1} is a fundamental result that will be exploited extensively in the remainder of the paper.

\subsubsection{Computing diagrams}
\label{computingdiagramsEL}

\begin{figure}[t!]
\begin{center}
\includegraphics[width=\columnwidth]{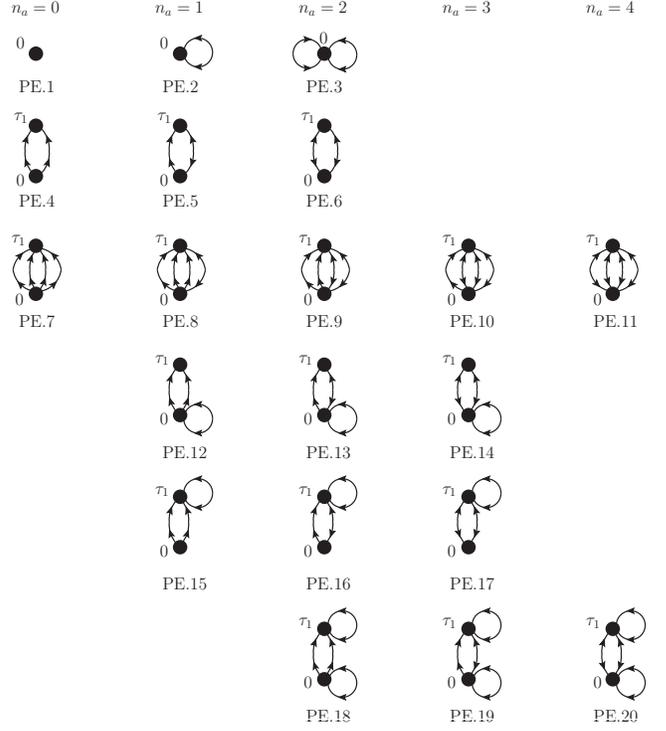}\\
\end{center}
\caption{
\label{diagramsTL}
Zero- and first-order connected Feynman off-diagonal BMBPT diagrams contributing to $\omega(\tau,\varphi)$. All vertices referring to $\Omega^{11}$, i.e. the lower vertex in diagrams PE.5 and PE.16, must be understood as indeed referring to the full $\Omega^{11}=\bar{\Omega}^{11} +\breve{\Omega}^{11}$.}
\end{figure}

The twenty non-zero connected/linked zero- and first-order diagrams contributing to $\omega(\tau,\varphi)$ are displayed in Fig.~\ref{diagramsTL}, where they are classified according to the value of $n_a$, i.e. according to the number of anomalous lines they contain. Given that all first-order diagrams involve $\Omega(\tau_1)$ and $\Omega(0)$ with the constraint that $\tau_1 > 0$, normal lines not only propagate in the same direction but are also limited to propagate upward.  

Choosing the reference state $|\Phi\rangle$ to be the solution of HFB equations amounts to setting $\breve{\Omega}^{11}=\Omega^{20}=\Omega^{02}=0$ such that diagrams PE.2, PE.4-PE.6, PE.12-PE.15 and PE.17 are zero in the Moller-Plesset scheme, i.e. the set reduces from twenty non-zero diagrams to eleven non-zero diagrams at first order. Finally, $\omega(\tau,0)$ at play in diagonal BMBPT reduces to PE.1, PE.4 and PE.7 ($n_a=0$) diagrams at first order (only PE.1 and PE.7 in the Moller-Plesset scheme).

\begin{figure}[t!]
\begin{center}
\includegraphics[width=0.35\columnwidth]{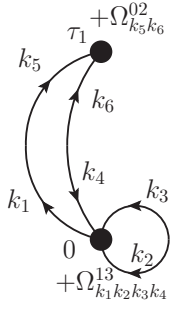}\\
\end{center}
\caption{
\label{diagramsTLex}
Example of a fully-labeled Feynman off-diagonal BMBPT diagram contributing to $\omega(\tau,\varphi)$, i.e. the diagram labeled PE.13 in Fig.~\ref{diagramsTL}.}
\end{figure}

While the full analytic expression of the twenty diagrams is provided in App.~\ref{diagramsE}, we presently detail the calculation of one of them for illustration. The first-order connected/linked diagram labeled as PE.13 in Fig.~\ref{diagramsTL} and displayed in details in Fig.~\ref{diagramsTLex} contains one $\Omega^{02}$ vertex at running time $\tau_1$ coming from the perturbative expansion of the evolution operator and one vertex $\Omega^{13}$ at fixed time 0, i.e. this diagram contributes to $\omega^{13 \, (1)}(\tau,\varphi)$. The diagram contains two anomalous lines ($n_a=2$), one vertex and no crossing line ($(-1)^{p+n_c}=-1$), one anomalous line beginning and ending at the $\Omega^{13}$ vertex, and a symmetry factor $n_s=1$ as only one vertex carries a running time and thus cannot be exchanged with any other. Last but not least, the sign convention requires to associate the factors $+\Omega^{02}_{k_5k_6}$ and $+\Omega^{13}_{k_1k_2k_3k_4}$ to the vertices as they appear in Fig~\ref{diagramsTLex}. Eventually, diagram PE.13 reads as
 \begin{widetext}
 \begin{align}
 \text{PE}.13=& -\frac{1}{2} \displaystyle\sum_{\substack{k_1 k_2 k_3 k_4 \\ k_5 k_6}} \hspace{-0.3cm} \Omega^{13}_{k_1k_2k_3k_4} \Omega^{02}_{k_5k_6}  \int\limits_0^{\tau} \mathrm{d} \tau_1  G^{-+(0)}_{k_5k_1}(\tau_1, 0; \varphi) G^{--(0)}_{k_6k_4}(\tau_1, 0; \varphi) G^{--(0)}_{k_3k_2}(0, 0; \varphi)  \nonumber \\
=& -\frac{1}{2} \displaystyle\sum_{k_1 k_2 k_3 k_4k_5} \hspace{-0.3cm} \Omega^{13}_{k_1k_2k_3k_4} \Omega^{02}_{k_1k_5} R^{--}_{k_5k_4}(\varphi) R^{--}_{k_3k_2}(\varphi) \int\limits_0^{\tau} \mathrm{d} \tau_1 e^{-\tau_1(E_{k_1}+E_{k_5})} \nonumber \\
=& -\frac{1}{2} \displaystyle\sum_{k_1 k_2 k_3 k_4k_5} \frac{\Omega^{13}_{k_1k_2k_3k_4} \Omega^{02}_{k_1k_5} }{E_{k_1}+E_{k_5}}\left[ 1- e^{-\tau(E_{k_1}+E_{k_5})}\right] R^{--}_{k_5k_4}(\varphi)R^{--}_{k_3k_2}(\varphi) \, ,
\label{PEex1B}
\end{align}
where use was made of the identities provided in App.~\ref{usefulID}. We note that, at variance with the example worked out in Sec.~\ref{computingdiagramsNc}, the first line of Eq.~\ref{PEex1B} was obtained by reading the diagram from top to bottom, i.e. in a descendant fashion, in Fig.~\ref{diagramsTLex}. In the infinite $\tau$ limit, this reduces to
 \begin{align}
 \text{PE}.13=&  -\frac{1}{2} \displaystyle\sum_{k_1 k_2 k_3 k_4k_5} \frac{\Omega^{13}_{k_1k_2k_3k_4} \Omega^{02}_{k_1k_5} }{E_{k_1}+E_{k_5}} R^{--}_{k_3k_2}(\varphi) R^{--}_{k_5k_4}(\varphi)  \, .
\label{PEex2B}
\end{align}
\end{widetext}
This diagram is zero in diagonal BMBPT as $R^{--}(0)=0$.

\subsubsection{Large $\tau$ limit and $\varphi$ dependence}
\label{phidepatlargetau}

According to Eq.~\ref{limitkernels}, $N(\tau,\varphi)$ and $\Omega(\tau,\varphi)$ carry the same dependence on $\varphi$ in the large $\tau$ limit, which leads to the remarkable result that the complete sum $\omega(\varphi)$ of {\it all} vacuum-to-vacuum diagrams linked to the fixed-time operator $\Omega(0)$ is actually independent of $\varphi$ in this limit. This corresponds to the fact that the expansion does fulfill the symmetry in the exact limit independently of whether the expansion is performed around a particle-number conserving Slater determinant or a particle-number breaking Bogoliubov state. In the latter case, however, each individual contribution $\omega^{(n)}(\varphi)$ or any partial sum of diagrams carries a dependence on $\varphi$ as a fingerprint of the particle-number breaking. To conclude, while the dependence of $n\left(\varphi ; n_a > 0 \right)$ on $\varphi$ is genuine, the dependence of $\omega(\varphi)$ is not and must be dealt with to restore the symmetry.

\subsubsection{Particle-number conserving case}
\label{SD2}

If the reference state $| \Phi \rangle$ is chosen to be a Slater determinant, $\omega(\tau,\varphi)$ is independent of $\varphi$ for all $\tau$ at any truncation order. This relates to the fact that $R^{--}(\varphi)=0$ for all $\varphi$ in this case. It is thus a situation where $N(\tau,\varphi)$ and $\Omega(\tau,\varphi)$ carry the same dependence on $\varphi$ for all $\tau$ independently of the truncation employed. As discussed in Sec.~\ref{SD1}, this dependence is trivial and reduces to the overall phase $e^{i\text{A}_0\varphi}$ in compliance with Eqs.~ \ref{expandedkernels1} and~\ref{expandedkernels4}. The expansion of $\omega(\tau, \varphi)$ is, at any $\varphi$, nothing but the standard particle-number conserving MBPT in this case.

\section{Coupled cluster theory}
\label{CCtheory}

Having the off-diagonal BMBPT expansion of $\Omega(\tau,\varphi)$ and $N(\tau,\varphi)$ at hand, we are now in position to design their off-diagonal BCC expansion.

\subsection{Off-diagonal grand-potential kernel}
\label{energykernelMBPTtoCC}

We first demonstrate that the perturbative expansion of the linked/connected kernel $\omega(\tau,\varphi)$ can be recast in terms of an exponentiated cluster operator whose expansion naturally terminates. 

\subsubsection{From off-diagonal BMBPT to off-diagonal BCC}

\begin{figure}[t!]
\begin{center}
\includegraphics[clip=,width=0.16\textwidth,angle=0]{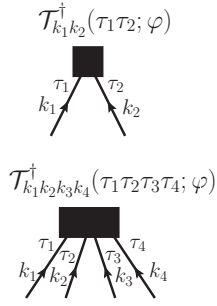}
\end{center}
\caption{
\label{TD1B2BCA}
Feynman diagrams representing  one- (first line) and two-body (second line) cluster amplitudes.}
\end{figure}

We introduce the $\tau$- and $\varphi$-dependent $n$-body Bogoliubov cluster operator through
\begin{widetext}
\begin{equation}
\mathcal{T}^{\dagger}_n (\tau, \varphi) = \frac{1}{(2n)!} \displaystyle\sum_{k_1 \ldots k_{2n}} \int\limits_0^{\tau} \mathrm{d} \tau_1 \ldots
\mathrm{d} \tau_{2n} \mathcal{T}^{\dagger}_{k_1 \ldots k_{2n}}(\tau_1 \ldots \tau_{2n}; \varphi) T \Big[ \beta_{k_{2n}}(\tau_{2n})
\ldots \beta_{k_1}(\tau_1) \Big] , \label{clusteroperators}
\end{equation}
\end{widetext}
where the Feynman amplitude $\mathcal{T}^{\dagger}_{k_1 \ldots k_{2n}}(\tau_1 \ldots \tau_{2n}; \varphi)$ is antisymmetric under the exchange of $(k_i,\tau_i)$ and $(k_j,\tau_j)$ for any $(i,j)\in\{1,\ldots 2n\}^2$. One- and two-body cluster amplitudes are represented diagrammatically in Fig.~\ref{TD1B2BCA}. For historical reasons, the operators introduced in Eq.~\ref{clusteroperators} reduce to the {\it Hermitian conjugate} of the traditional cluster operators appearing in diagonal BCC theory~\cite{Signoracci:2014dia}.

\begin{figure}[t!]
\begin{center}
\includegraphics[clip=,width=0.53\textwidth,angle=0]{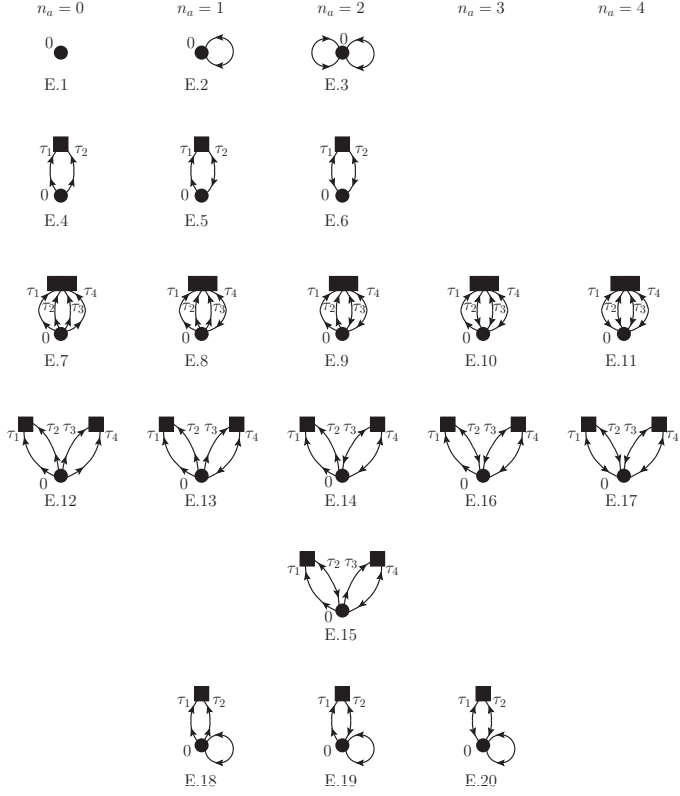}\\ 
\end{center}
\caption{
\label{T1contribtokinetic}
Off-diagonal Feynman diagrams representing the twenty BCC contributions to $\omega(\tau,\varphi)$. Only the time labels (i.e.  not the quasi-particle labels) are indicated next to the lines/vertices. Refer to Fig.~\ref{pnrbccex} to see a fully-labeled diagram.}
\end{figure}

As discussed in Sec.~\ref{energykernel}, $\omega(\tau,\varphi)$ represents the infinite set of connected off-diagonal BMBPT diagrams linked at time zero to the operator $\Omega$. By virtue of their linked character, diagrams entering $\omega(\tau,\varphi)$ necessarily possess the topology of one of the twenty diagrams represented in Fig.~\ref{T1contribtokinetic} and ordered according to the value of $n_a$. The restriction to these twenty topologies are dictated by the diagrammatic rules detailed in Sec.~\ref{diagrulenormMBPT} and by the fact that normal lines attached to an operator at fixed time $0$ necessarily propagate upward as already mentioned. 

The first three diagrams displayed in Fig.~\ref{T1contribtokinetic} isolate the contributions with no lines propagating in time, i.e the zero-order contributions associated with the matrix element of $\Omega$ between the reference state $| \Phi \rangle$ and its rotated partner $| \Phi(\varphi) \rangle$. Non-zero contributions of this type are limited to contributions originating from $\Omega^{00}$, $\Omega^{02}$ and $\Omega^{04}$. 

All diagrams entering $\omega(\tau,\varphi)$ beyond zero order are captured by the remaining seventeen topologies. This leads to defining the one-body cluster amplitude ${\cal T}^{\dagger}_{k_1k_2}(\tau_1 \tau_2 ; \varphi)$ as the {\it complete} sum of connected off-diagonal BMBPT diagrams with one line entering at an arbitrary time $\tau_1$ and another line entering at an arbitrary time $\tau_2$. In Fig.~\ref{T1contribtokinetic}, these two lines contract with lines arising from the various components $\Omega^{ij}$ of $\Omega$ at time zero. Covering the remaining topologies requires the introduction of the two-body cluster amplitude ${\cal T}^{\dagger}_{k_1k_2k_3k_4}(\tau_1 \tau_2  \tau_3 \tau_4 ; \varphi)$ defined as the {\it complete} sum of connected off-diagonal BMBPT diagrams with four lines entering at arbitrary times $\tau_1$, $\tau_2$, $\tau_3$ and $ \tau_4$. In Fig.~\ref{T1contribtokinetic}, these four lines contract with lines arising from the various components $\Omega^{ij}$ of $\Omega$ at time zero. This definition trivially extends to higher-body cluster operators. First-order expressions of ${\cal T}^{\dagger}_{k_1k_2}(\tau_1 \tau_2 ; \varphi)$ and ${\cal T}^{\dagger}_{k_1k_2k_3k_4}(\tau_1 \tau_2 \tau_3 \tau_4 ; \varphi)$ are provided in Sec.~\ref{firstorderToperators}.

Thus, the introduction of cluster operators allows one to group the complete set of linked/connected vacuum-to-vacuum diagrams making up $\omega(\tau,\varphi)$ under the form
\begin{widetext}
\begin{equation}
\label{CCenergyequation}
\omega(\tau, \varphi) = \langle \Phi | \big[1+\mathcal{T}^{\dagger}_1(\tau, \varphi) + \tfrac{1}{2}\mathcal{T}^{\dagger \; 2}_1(\tau, \varphi) + \mathcal{T}^{\dagger}_2(\tau,\varphi) \big] \Omega | \Phi (\varphi) \rangle _{\text{c}}
\langle \Phi | \Phi(\varphi) \rangle ^{-1} ,
\end{equation}
which translates into the twenty different terms displayed in Fig.~\ref{T1contribtokinetic} when expanding $\Omega$ in terms of its normal-ordered components $\Omega^{ij}$ and only retaining the non-zero contributions
\begin{subequations}
\label{CCenergyequation2}
\begin{eqnarray}
\omega^{00}(\tau, \varphi) &=& \Omega^{00} \, , \label{CCenergyequation2A} \\
\omega^{20}(\tau, \varphi) &=& \langle \Phi | \mathcal{T}^{\dagger}_1(\tau, \varphi)\Omega^{20} | \Phi (\varphi) \rangle _{\text{c}}
\langle \Phi | \Phi(\varphi) \rangle ^{-1} \, , \label{CCenergyequation2B} \\
\omega^{11}(\tau, \varphi) &=& \langle \Phi | \mathcal{T}^{\dagger}_1(\tau, \varphi)  \Omega^{11} | \Phi (\varphi) \rangle _{\text{c}}
\langle \Phi | \Phi(\varphi) \rangle ^{-1} \, , \label{CCenergyequation2C} \\
\omega^{02}(\tau, \varphi) &=& \langle \Phi | \big[1+\mathcal{T}^{\dagger}_1(\tau, \varphi)  \big] \Omega^{02} | \Phi (\varphi) \rangle _{\text{c}}
\langle \Phi | \Phi(\varphi) \rangle ^{-1} \, , \label{CCenergyequation2D} \\
\omega^{40}(\tau, \varphi) &=& \langle \Phi | \big[\tfrac{1}{2}\mathcal{T}^{\dagger \; 2}_1(\tau, \varphi) + \mathcal{T}^{\dagger}_2(\tau,\varphi) \big] \Omega^{40} | \Phi (\varphi) \rangle _{\text{c}}
\langle \Phi | \Phi(\varphi) \rangle ^{-1} \, , \\
\omega^{31}(\tau, \varphi) &=& \langle \Phi | \big[\tfrac{1}{2}\mathcal{T}^{\dagger \; 2}_1(\tau, \varphi) + \mathcal{T}^{\dagger}_2(\tau,\varphi) \big] \Omega^{31} | \Phi (\varphi) \rangle _{\text{c}}
\langle \Phi | \Phi(\varphi) \rangle ^{-1} \, , \\
\omega^{22}(\tau, \varphi) &=& \langle \Phi | \big[\mathcal{T}^{\dagger}_1(\tau, \varphi) + \tfrac{1}{2}\mathcal{T}^{\dagger \; 2}_1(\tau, \varphi) + \mathcal{T}^{\dagger}_2(\tau,\varphi) \big] \Omega^{22} | \Phi (\varphi) \rangle _{\text{c}}
\langle \Phi | \Phi(\varphi) \rangle ^{-1} \, , \\
\omega^{13}(\tau, \varphi) &=& \langle \Phi | \big[\mathcal{T}^{\dagger}_1(\tau, \varphi) + \tfrac{1}{2}\mathcal{T}^{\dagger \; 2}_1(\tau, \varphi) + \mathcal{T}^{\dagger}_2(\tau,\varphi) \big] \Omega^{13} | \Phi (\varphi) \rangle _{\text{c}}
\langle \Phi | \Phi(\varphi) \rangle ^{-1} \, , \\
\omega^{04}(\tau, \varphi) &=& \langle \Phi | \big[1+\mathcal{T}^{\dagger}_1(\tau, \varphi) + \tfrac{1}{2}\mathcal{T}^{\dagger \; 2}_1(\tau, \varphi) + \mathcal{T}^{\dagger}_2(\tau,\varphi) \big] \Omega^{04} | \Phi (\varphi) \rangle _{\text{c}}
\langle \Phi | \Phi(\varphi) \rangle ^{-1} \, , 
\end{eqnarray}
\end{subequations}
\end{widetext}
In Eqs.~\ref{CCenergyequation} and~\ref{CCenergyequation2}, the subscript $c$ means that (i) cluster operators must all be linked to $\Omega$ through strings of contractions and that (ii)  no contraction can occur among cluster operators or within a given cluster operator. Contracting quasiparticle annihilation operators originating from different cluster operators (e.g. from ${\cal T}^{\dagger}_1$ and ${\cal T}^{\dagger}_2$) or within the same cluster operator generate diagrams that are already contained in a connected cluster of lower rank and would thus lead to double counting. As off-diagonal contractions within a given cluster operator are {\it not} zero a priori, the rule that those contractions are to be excluded when computing contributions to Eq.~\ref{CCenergyequation} must indeed be stated explicitly. The grand potential being of two-body character, the sum of terms in Eq.~\ref{CCenergyequation} does exhaust exactly the complete set of diagrams generated through perturbation theory. 

The $1/2$ factor made explicit in the third term of Eq.~\ref{CCenergyequation} can be justified order by order by considering a contribution ${\cal T}^{\dagger(n)}_{1}(\tau,\varphi)$ extracted from an arbitrary diagram of order $n$ having the topology of, e.g., diagram E5 in Fig.~\ref{T1contribtokinetic}. The corresponding contribution to $\omega(\tau,\varphi)$ of order $2n$ associated with the third term of Eq.~\ref{CCenergyequation2} acquires a factor $1/2$ because exchanging at once all time labels entering the two identical ${\cal T}^{\dagger (n)}_{1}(\tau,\varphi)$ pieces provides an equivalent contribution. This, combined with the other diagrammatic rules, will eventually result into the rule detailed in Sec.~\ref{1stCCdiagrammaticrules} dealing with so-called equivalent ${\cal T}^{\dagger}_{m}(\tau,\varphi)$ vertices.

Eventually, one can rewrite Eq.~\ref{CCenergyequation} under the characteristic form
\begin{subequations}
\label{clusterexpansion}
\begin{eqnarray}
\omega(\tau,\varphi) &=& \frac{\langle \Phi | e^{{\cal T}^{\dagger}(\tau,\varphi)} \Omega  |\Phi (\varphi) \rangle_{c}}{\langle \Phi |  \Phi (\varphi) \rangle} \, , \label{clusterexpansiona} \\
{\cal T}^{\dagger}(\tau,\varphi) &\equiv& \sum_{n \in \mathbb{N}} {\cal T}^{\dagger}_{n}(\tau,\varphi) \, , \label{clusterexpansionb}
\end{eqnarray}
\end{subequations}
given that no cluster operator beyond ${\cal T}^{\dagger}_{1}(\tau,\varphi)$ and ${\cal T}^{\dagger}_{2}(\tau,\varphi)$ can actually contribute to $\omega(\tau,\varphi)$ in view of its linked/connected character. The fact that Eq.~\ref{clusterexpansiona} does indeed reduce to Eq.~\ref{CCenergyequation} generalizes to the off-diagonal grand potential kernel the natural termination of the BCC expansion displayed by the diagonal one~\cite{Signoracci:2014dia}. As mentioned earlier, the termination and the specific connected structure of the resulting terms are traditionally obtained for the diagonal kernel from the similarity transformed grand potential $\bar{\Omega}\equiv e^{-T}\Omega e^{T}$ on the basis of the Baker-Campbell-Hausdorff identity and of standard Wick's theorem~\cite{shavitt09a}. In the present case, the long detour via perturbation theory applied to off-diagonal kernels was necessary to obtain the same connected structure as in the diagonal case, including the necessity to omit contractions within a cluster operator or among two different cluster operators. 

\subsubsection{Computation of diagrams}
\label{1stCCdiagrammaticrules}

We now compute the algebraic BCC contributions to $\omega(\tau,\varphi)$. The diagrammatic rules to obtain them are essentially the same as those detailed in Sec.~\ref{diagrulenormMBPT} for BMBPT Feynman diagrams. The only modifications are that
\begin{enumerate}
\item One must attribute a factor $\mathcal{T}^{\dagger}_{k_1 \ldots k_{2n}}(\tau_1 \ldots \tau_{2n}; \varphi)$ to any vertex representing an $n$-body cluster operator (i.e. its hermitian conjugate). Indices $k_1 \ldots k_{2n}$ and $\tau_1 \ldots \tau_{2n}$ must be assigned consecutively from the leftmost to the rightmost line below the vertex.
\item All diagrams are {\it connected}, i.e. each contributing $\mathcal{T}^{\dagger}_{n}(\tau , \varphi)$ operator is contracted at least once with $\Omega$. No line may connect two cluster operators while lines belonging to a given cluster operator cannot be contracted together.
\item Following the above rule, construct all possible independent closed diagrams from the building blocks. Doing so typically limits which parts $\Omega^{ij}$ of $\Omega$ contribute to a given term.
\item The symmetry factor $1/n_s$ must be replaced by a factor $(\ell_m!)^{-1}$ for each set of $\ell_m$ equivalent  ${\cal T}^{\dagger}_{m}(\tau,\varphi)$ vertices.  Two  ${\cal T}^{\dagger}_{m}(\tau,\varphi)$ vertices are equivalent if they have the same number of quasi-particle lines $n_l \, (n_l \leq 2m)$ connected to the interaction vertex via propagators of the {\it same} type.
\item The sign of the diagram is obtained by combining the factor $(-1)^{n_c}$ with the sign rule associated with each factor $\Omega^{ij}_{k_1 \ldots k_i k_{i+1} \ldots k_{i+j}}$.
\end{enumerate}

The above rules result into the twenty non-zero BCC diagrams contributing to $\omega(\tau,\varphi)$ and displayed in Fig.~\ref{T1contribtokinetic} where they are classified according to the value of $n_a$, i.e. according to the number of anomalous lines they contain. 

Choosing the reference state $|\Phi\rangle$ to be the solution of HFB equations amounts to setting $\Omega^{20}=\Omega^{02}=0$ such that diagrams E.2, E.4 and E.6 are zero in the Moller-Plesset scheme, i.e. the set reduces from twenty non-zero diagrams to seventeen non-zero diagrams in this case. Finally, $\omega(\tau,0)$ at play in diagonal BCC reduces to E.1, E.4, E.7 and E.12 ($n_a=0$) diagrams (only E.1, E.7 and E.12 in the Moller-Plesset scheme).

\begin{figure}[t!]
\begin{center}
\includegraphics[clip=,width=0.35\columnwidth,angle=0]{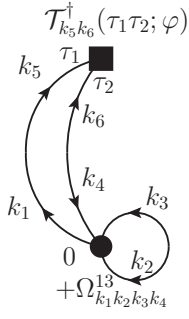}\\ 
\end{center}
\caption{
\label{pnrbccex}
Example of fully-labeled BCC diagram contributing to $\omega(\tau,\varphi)$, i.e. diagram labeled E.19 in Fig.~\ref{T1contribtokinetic}.}
\end{figure}

While the full analytic expression of each of the twenty diagrams is provided in App.~\ref{CCenergycontrib}, we presently detail the calculation of one of them for illustration. We calculate diagram E.19 that is displayed in complete detail in Fig.~\ref{pnrbccex}.  According to the diagrammatic rules, and reading the diagram in a descendant fashion, its contribution reads as
\begin{widetext}
\begin{align}
 \nonumber
\text{E}.19 &= + \frac{1}{2} \displaystyle\sum_{\substack{k_1 k_2 k_3 k_4 \\ k_5 k_6}} \Omega^{13}_{k_1k_2k_3 k_4}  \int\limits_0^{\tau} \mathrm{d} \tau_1 \mathrm{d} \tau_2
 \mathcal{T}^{\dagger}_{k_5 k_6}(\tau_1 \tau_2; \varphi) G^{-+(0)}_{k_5 k_1}(\tau_1,0; \varphi) G^{--(0)}_{k_6 k_4}(\tau_2,0;\varphi) G^{--(0)}_{k_3 k_2}(0,0;\varphi)\\
&= + \frac{1}{2} \displaystyle\sum_{k_1 k_2 k_3 k_4 k_5} \Omega^{13}_{k_1 k_2 k_3 k_4} \int\limits_0^{\tau} \mathrm{d} \tau_1 \mathrm{d} \tau_2 
\mathcal{T}^{\dagger}_{k_1 k_5} (\tau_1 \tau_2; \varphi) e^{-E_{k_1} \tau_1}e^{-E_{k_5} \tau_2} R^{--}_{k_5k_4}(\varphi)R^{--}_{k_3k_2}(\varphi)\, .
\label{diag1}
\end{align}
\end{widetext}
Defining a time-integrated one-body cluster amplitude
\begin{equation}
\mathcal{T}^{\dagger}_{k_1 k_2}(\tau,\varphi) \equiv \int\limits_0^{\tau} \mathrm{d} \tau_1 \mathrm{d} \tau_2 
\mathcal{T}^{\dagger}_{k_1 k_2} (\tau_1 \tau_2; \varphi) e^{-E_{k_1} \tau_1}e^{-E_{k_2} \tau_2} \, , \label{T1matrixelements}
\end{equation}
such that
\begin{eqnarray}
{\cal T}^{\dagger}_{1}(\tau,\varphi) &=& \frac{1}{2!} \sum_{k_1 k_2} \mathcal{T}^{\dagger}_{k_1 k_2}(\tau,\varphi) \, \beta_{k_2} \, \beta_{k_1}  \, , \label{T1} 
\end{eqnarray}
one can finalize Eq.~\ref{diag1} under the form
\begin{equation}
\text{E}.19 = +\frac{1}{2} \displaystyle\sum_{k_1 k_2 k_3 k_4 k_5} \mathcal{T}^{\dagger}_{k_1 k_5}(\tau,\varphi) \Omega^{13}_{k_1 k_2 k_3 k_4}R^{--}_{k_5k_4}(\varphi)R^{--}_{k_3k_2}(\varphi) \, .
\end{equation}
Further introducing
\begin{widetext}
\begin{subequations}
\begin{eqnarray}
\mathcal{T}^{\dagger}_{k_1 k_2 k_3 k_4}(\tau,\varphi) &\equiv& \int\limits_0^{\tau} \mathrm{d} \tau_1 \mathrm{d} \tau_2 
 \mathrm{d} \tau_3 \mathrm{d} \tau_4 \mathcal{T}^{\dagger}_{k_1 k_2 k_3 k_4} (\tau_1 \tau_2 \tau_3 \tau_4; \varphi) e^{-E_{k_1} \tau_1}e^{-E_{k_2} \tau_2}  e^{-E_{k_3} \tau_3}e^{-E_{k_4} \tau_4} \, \label{T2matrixelements} \\
{\cal T}^{\dagger}_{2}(\tau,\varphi) &=& \frac{1}{4!} \sum_{k_1 k_2 k_3 k_4} \mathcal{T}^{\dagger}_{k_1 k_2 k_3 k_4}(\tau,\varphi) \, \beta_{k_4} \, \beta_{k_3} \, \beta_{k_2} \, \beta_{k_1}  \, , \label{T2} 
\end{eqnarray}
\end{subequations}
\end{widetext}
all twenty contributions to $\omega(\tau,\varphi)$ can be derived.

\subsection{Determining off-diagonal BCC amplitudes}
\label{dynamicalequations}

In order to effectively compute the various contributions to $\omega(\tau,\varphi)$, one must have the matrix elements of the $\varphi$-dependent cluster operators at hand.

\subsubsection{First-order in off-diagonal BMBPT}
\label{firstorderToperators}

\begin{figure}[t!]
\begin{center}
\includegraphics[clip=,width=0.44\textwidth,angle=0]{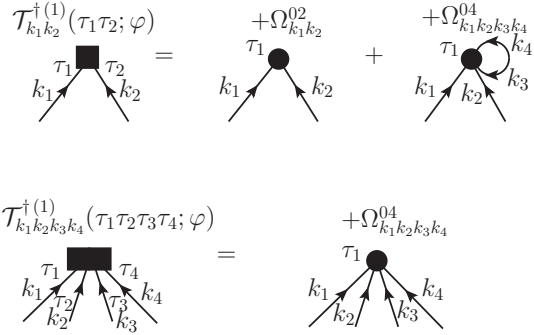}
\end{center}
\caption{
\label{1storderTamplitudes}
Feynman one-body (first line) and two-body (second line) cluster amplitudes at first order in off-diagonal BMBPT.}
\end{figure}

The first option consists of determining the cluster amplitudes via off-diagonal BMBPT. Feynman diagrams contributing to one- and two-body cluster amplitudes at first order in off-diagonal BMBPT are displayed in Fig.~\ref{1storderTamplitudes} and give
\begin{widetext}
\begin{subequations}
\label{1stordercluster}
\begin{eqnarray}
\mathcal{T}^{\dagger (1)}_{k_1k_2}(\tau_1 \tau_{2}; \varphi) &=& - \Omega^{02}_{k_1k_2} \delta(\tau_1 - \tau_2) - \frac{1}{2}
 \displaystyle\sum_{k_3k_4} \Omega^{04}_{k_1k_2k_3k_4} G^{--(0)}_{k_4k_3} (\tau_1,\tau_1; \varphi) \delta(\tau_1 - \tau_2) \, , \label{1storderclustera} \\
\mathcal{T}^{\dagger (1)}_{k_1k_2k_3k_4}(\tau_1 \tau_{2} \tau_3 \tau_4; \varphi) &=& 
 - \Omega^{04}_{k_1k_2k_3k_4} \delta(\tau_1 - \tau_2)\delta(\tau_2 - \tau_3)\delta(\tau_3 - \tau_4) \, . \label{1storderclusterb}
\end{eqnarray}
\end{subequations}

Inserting these expressions into Eqs.~\ref{T1matrixelements} and~\ref{T2matrixelements} provides associated Goldstone amplitudes
\begin{subequations}
\label{1storderclusterB}
\begin{eqnarray}
\mathcal{T}^{\dagger (1)}_{k_1k_2}(\tau,\varphi) &=& - \frac{\Omega^{02}_{k_1k_2}}{E_{k_1}+E_{k_2}} \big[ 1- e^{- \tau(E_{k_1}+E_{k_2})} \big] \nonumber \\
&&-\frac{1}{2} 
 \displaystyle\sum_{k_3k_4} \frac{\Omega^{04}_{k_1k_2k_3k_4}}{E_{k_1}+E_{k_2}+E_{k_3}+E_{k_4}}
 \big[ 1- e^{- \tau(E_{k_1}+E_{k_2}+E_{k_3}+E_{k_4})} \big] R^{--}_{k_4k_3} (\varphi) \, , \label{1storderclusterBa}  \\
\mathcal{T}^{\dagger (1)}_{k_1k_2k_3k_4}(\tau,\varphi) &=& - \frac{\Omega^{04}_{k_1k_2k_3k_4}}{E_{k_1}+E_{k_2}+E_{k_3}+E_{k_4}}
 \big[ 1- e^{- \tau(E_{k_1}+E_{k_2}+E_{k_3}+E_{k_4})} \big] \, \, , \label{1storderclusterBb}
\end{eqnarray}
\end{subequations}
\end{widetext}
such that ${\cal T}^{\dagger \, (1)}_{2}(\tau,\varphi)$ does not depend on $\varphi$. One can check that ${\cal T}^{\dagger \, (1)}_{1}(0,\varphi)={\cal T}^{\dagger \, (1)}_{2}(0,\varphi)=0$ as it should be. 

\begin{figure}[t!]
\begin{center}
\includegraphics[clip=,width=0.16\textwidth,angle=0]{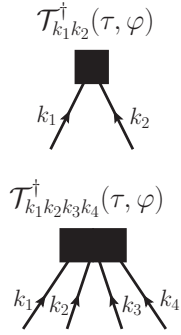}
\end{center}
\caption{
\label{diagbetadependentclusters}
Goldstone diagrams representing  one- (first line) and two-body (second line) cluster amplitudes.}
\end{figure}

\subsubsection{Off-diagonal BCC amplitude equations}
\label{amplitudeequations}

To work within a non-perturbative BCC framework, one must derive equations of motion for the $\varphi$-dependent cluster amplitudes. To do so, we introduce $n$-tuply excited off-diagonal norm, grand-potential and particle-number kernels through
\begin{subequations}
\label{amplitudekernels}
\begin{eqnarray}
N_{k_1k_2\ldots}(\tau,\varphi) &\equiv& \langle \Psi (\tau) | \bbone | \Phi^{k_1k_2\ldots}(\varphi) \rangle   \,\, , \label{amplitudekernels1} \\
\Omega_{k_1k_2\ldots}(\tau,\varphi) &\equiv& \langle \Psi (\tau) | \Omega | \Phi^{k_1k_2\ldots}(\varphi) \rangle   \,\, , \label{amplitudekernels2} \\
A_{k_1k_2\ldots}(\tau,\varphi) &\equiv& \langle \Psi (\tau) | A | \Phi^{k_1k_2\ldots}(\varphi) \rangle   \,\, , \label{amplitudekernels3}
\end{eqnarray}
\end{subequations}
where 
\begin{equation}
| \Phi^{k_1k_2\ldots}(\varphi) \rangle \equiv {\cal B}_{k_1k_2\ldots} | \Phi(\varphi) \rangle \, ,
\end{equation}
with the operator ${\cal B}_{k_1k_2\ldots}$ defined in Eq.~\ref{phexcitation}. From Eq.~\ref{schroedinger}, one obtains that
\begin{equation}
\Omega_{k_1k_2\ldots}(\tau,\varphi) = -\partial_{\tau} N_{k_1k_2\ldots}(\tau,\varphi)  \, . \label{dynamicalkernels}
\end{equation}
In App.~\ref{amplitudeequations}, we demonstrate in detail how Eq.~\ref{dynamicalkernels} eventually provides the equations of motion satisfied by the $n$-body  ($\tau$- and $\varphi$-dependent) cluster amplitudes under the form
\begin{equation}
\omega_{k_1k_2\ldots}(\tau,\varphi) = -\partial_{\tau} {\cal T}^{\dagger}_{k_1k_2\ldots}(\tau,\varphi)  \,  \, , \label{reduceddynamicalkernels}
\end{equation}
where the $n$-tuply excited {\it connected} grand potential kernel is defined through
\begin{equation}
\omega_{k_1k_2\ldots}(\tau,\varphi) \equiv \frac{\langle \Phi | e^{{\cal T}^{\dagger}(\tau,\varphi)} \Omega  | \Phi^{k_1k_2\ldots}(\varphi) \rangle_{c}}{\langle \Phi |  \Phi(\varphi) \rangle} \, , \label{CCamplitudekernels}
\end{equation}
and whose connected character denotes that (i) cluster operators are all connected to $\Omega$ and that (ii) no contraction is to be considered among cluster operators or within any given cluster operator such that the power series of the exponential naturally terminates.

As a result of this termination of the exponential, singly- and doubly-excited off-diagonal connected grand potential kernels read as
\begin{widetext}
\begin{subequations}
\label{termination}
\begin{eqnarray}
\omega_{k_1k_2}(\tau,\varphi) &=& \langle \Phi | \big[1+\mathcal{T}^{\dagger}_1(\tau, \varphi) + \frac{1}{2}\mathcal{T}^{\dagger  2}_1(\tau, \varphi) + 
\frac{1}{3!}\mathcal{T}^{\dagger  3}_1(\tau, \varphi)  \nonumber \\
&& \,\,\,\,\,\,\,\,\,\, + \mathcal{T}^{\dagger}_2(\tau, \varphi) + 
\mathcal{T}^{\dagger}_1(\tau, \varphi)\mathcal{T}^{\dagger}_2(\tau, \varphi) \big] \Omega | \Phi^{k_1 k_2} (\varphi) \rangle_{\text{c}}
\langle \Phi | \Phi(\varphi) \rangle^{-1} \, , \label{termination1} \\ 
\omega_{k_1k_2k_3k_4}(\tau,\varphi) &=& \langle \Phi | \big[1+\mathcal{T}^{\dagger}_1(\tau, \varphi) + \frac{1}{2}\mathcal{T}^{\dagger  2}_1(\tau, \varphi) + 
\frac{1}{3!}\mathcal{T}^{\dagger  3}_1(\tau, \varphi) + \mathcal{T}^{\dagger}_2(\tau, \varphi) + 
\mathcal{T}^{\dagger}_1(\tau, \varphi)\mathcal{T}^{\dagger}_2(\tau, \varphi)   \nonumber \\
&& \,\,\,\,\,\,\,\,\,\, + \frac{1}{2} \mathcal{T}^{\dagger  2}_2(\tau, \varphi)+ \frac{1}{4!}\mathcal{T}^{\dagger  4}_1(\tau, \varphi) +
\frac{1}{2} \mathcal{T}^{\dagger  2}_1(\tau, \varphi) \mathcal{T}^{\dagger}_2 (\tau, \varphi) \big]
 \Omega | \Phi^{k_1 k_2 k_3 k_4} (\varphi) \rangle_{\text{c}}
\langle \Phi | \Phi(\varphi) \rangle^{-1} \, , \label{termination2}
\end{eqnarray}
\end{subequations}
\end{widetext}
respectively. Note that we refer to singly- and doubly-excited kernels to connect to the standard CC formalism as two-quasiparticle (four-quasiparticle) excitations reduce in the Slater determinant limit to one (two) particle-hole excitations.

\subsubsection{Computation of diagrams}

To compute the algebraic contributions to $\omega_{k_1\ldots k_{2n}}(\tau,\varphi)$, a few diagrammatic rules beyond those stated in Sec.~\ref{1stCCdiagrammaticrules} to determine $\omega(\tau,\varphi)$ must be added
\begin{enumerate}
\item Diagrams making up $\omega_{k_1\ldots k_{2n}}(\tau,\varphi)$ are {\it linked} with $2n$ external quasi-particle lines, exiting from below. External lines must be labeled with quasi-particle indices $k_1\ldots k_{2n}$ coinciding with the left-right ordering of the indices observed in the ket defining the $n$-tuply excited kernel. Internal lines must be labeled with different quasi-particle indices.
\item Only internal quasi-particle line indices must be summed over. 
\item All distinct permutations $P$ of labels of inequivalent external lines must be summed over, 
including a parity factor $(-1)^{\sigma(P)}$ from the signature of the permutation.  External lines are equivalent 
if and only if they connect to the same vertex.
\end{enumerate}

\begin{figure}[t!]
\begin{center}
\includegraphics[clip=,width=0.3\textwidth,angle=0]{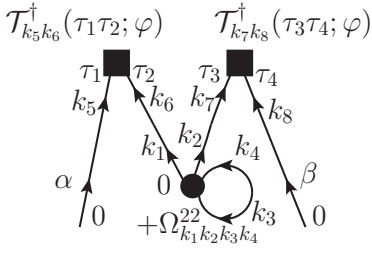}\\ 
\end{center}
\caption{
\label{pnrbccsingex}
Example of a fully-labeled diagram contributing to $\omega_{k_1 k_2}(\tau,\varphi)$, i.e. the diagram which results in expression S.25 in Appendix G.}
\end{figure}

One example diagram contributing to the singly-excited off-diagonal connected grand potential kernel is displayed in Fig.~\ref{pnrbccsingex} with explicit labeling.  Following the diagrammatic rules, its contribution reads as
\begin{widetext}
\begin{align}
 \nonumber
\text{S}.25 &= + \frac{1}{4} P(\alpha / \beta) \displaystyle\sum_{\substack{k_1 k_2 k_3 k_4 \\ k_5 k_6 k_7 k_8}} \Omega^{22}_{k_1k_2k_3 k_4}  \int\limits_0^{\tau} \mathrm{d} \tau_1 \mathrm{d} \tau_2 \mathrm{d} \tau_3 \mathrm{d} \tau_4
 \mathcal{T}^{\dagger}_{k_5 k_6}(\tau_1 \tau_2; \varphi) \mathcal{T}^{\dagger}_{k_7 k_8}(\tau_3 \tau_4; \varphi)G^{-+(0)}_{k_5 \alpha}(\tau_1,0;\varphi) \nonumber \\
 & \hspace{7cm} \times  G^{-+(0)}_{k_6 k_1}(\tau_2,0;\varphi) G^{--(0)}_{k_4 k_3}(0,0;\varphi)G^{-+(0)}_{k_7 k_2}(\tau_3,0;\varphi) G^{-+(0)}_{k_8 \beta}(\tau_4,0;\varphi) \nonumber \\
&= + \frac{1}{4} P(\alpha / \beta) \displaystyle\sum_{k_1 k_2 k_3 k_4} \Omega^{22}_{k_1 k_2 k_3 k_4} \int\limits_0^{\tau} \mathrm{d} \tau_1 \mathrm{d} \tau_2  \mathrm{d} \tau_3 \mathrm{d} \tau_4 \mathcal{T}^{\dagger}_{\alpha k_1} (\tau_1 \tau_2; \varphi) \mathcal{T}^{\dagger}_{k_2 \beta} (\tau_3 \tau_4; \varphi) e^{-E_{\alpha} \tau_1}e^{-E_{k_1} \tau_2}\nonumber \\
 & \hspace{10cm} \times e^{-E_{k_2} \tau_3}  e^{-E_{\beta} \tau_4}R^{--}_{k_4k_3}(\varphi) \nonumber \\
&= +\frac{1}{4}P(\alpha / \beta) \displaystyle\sum_{k_1 k_2 k_3 k_4} \mathcal{T}^{\dagger}_{\alpha k_1}(\tau,\varphi) 
\mathcal{T}^{\dagger}_{k_2 \beta}(\tau,\varphi) \Omega^{22}_{k_1 k_2 k_3 k_4}R^{--}_{k_4k_3}(\varphi) \, ,
\label{diagsing1}
\end{align}
\end{widetext} 
where $P(\alpha / \beta) \equiv 1-P_{\alpha\beta}$, with $P_{\alpha\beta}$ the operator exchanging quasi-particle indices $\alpha$ and $\beta$ labelling the two external lines.

Although the operator form of $\omega_{k_1k_2}(\tau,\varphi)$ and $\omega_{k_1k_2k_3k_4}(\tau,\varphi)$ as displayed in Eq.~\ref{termination} is formally identical to diagonal ($\varphi=0$) BCC expressions~\cite{Signoracci:2014dia}, their expanded algebraic expressions are much lengthier. This translates into the fact that $\omega_{k_1k_2}(\tau,\varphi)$ ($\omega_{k_1k_2k_3k_4}(\tau,\varphi)$) is made out of fifty-seven (seventy-seven) diagrams at the BCCSD level that reduce to only ten (fourteen) diagrams in the diagonal ($\varphi =0$) limit. While the explicit algebraic expressions of the fifty-seven contributions to $\omega_{k_1k_2}(\tau,\varphi)$ are provided in App.~\ref{CCsinglecontrib}, the contributions to $\omega_{k_1k_2k_3k_4}(\tau,\varphi)$ are too numerous and lengthy to be reported here. This is anyway unnecessary given that a compact form of these expressions containing just as many terms as in the diagonal ($\varphi=0$) limit will be identified in Sec.~\ref{compact} below.  For this reason, we do not produce the full diagrammatic description of the off-diagonal amplitude equations at the BCCSD level as well.

In the end, one is only interested in the infinite imaginary-time limit. In this limit, the scheme becomes stationary such that the static amplitude equations are obtained  by setting the right-hand side of Eq.~\ref{reduceddynamicalkernels} to zero, i.e.
\begin{equation}
\omega_{k_1 k_2\ldots}(\varphi) = 0  \, , \label{staticamplitudeequations}
\end{equation}
which naturally extend diagonal BCC amplitude equations. Coupled Eqs.~\ref{staticamplitudeequations} must be solved iteratively for each $\varphi \in [0,2\pi]$, typically employing first-order perturbation theory expressions as an initial guess (see Eq.~\ref{1storderclusterB} for single and double amplitudes). Once $\varphi$-dependent cluster amplitudes have been obtained, the off-diagonal connected grand-potential kernel $\omega(\tau,\varphi)$ can be computed on the basis of  the expressions provided in App.~\ref{CCenergycontrib}. 

\subsection{Compact formulation}
\label{compact}

As alluded to above, the algebraic expressions of  $\omega(\tau,\varphi)$, $\omega_{k_1k_2}(\tau,\varphi)$ and $\omega_{k_1k_2k_3k_4}(\tau,\varphi)$ (see Apps.~ \ref{CCenergycontrib} and~\ref{CCsinglecontrib} for the first two) are lengthy and translate into a large number of diagrams. This is not optimal, both for bookkeeping and from the numerical implementation viewpoint. It happens that the generalized BCC expansion of those off-diagonal kernels can eventually be reformulated in terms of a transformed, $\varphi$-dependent, grand potential operator such that their algebraic expressions are not only made much more compact but formally identical to their diagonal counterpart.

\subsubsection{Transformed grand-potential operator}
\label{transformedoperator}

We introduce a {\it non-unitary} Bogoliubov transformation that transforms quasi-particle operators defining the vacuum $| \Phi \rangle$ into a new set of quasi-particle operators according to
\begin{eqnarray}
\left(
\begin{array} {c}
\tilde{\beta} \\
\tilde{\beta}^{\dagger}
\end{array}
\right)(\varphi) &\equiv&  M(\varphi) \left(
\begin{array} {c}
\beta \\
\beta^{\dagger}
\end{array}
\right) M^{-1}(\varphi) \nonumber \\
&=& {\cal M}^{\dagger}(\varphi) \left(
\begin{array} {c}
\beta \\
\beta^{\dagger}
\end{array}
\right) \, ,
\end{eqnarray}
where 
\begin{equation}
{\cal M}(\varphi) \equiv \left(
\begin{array} {cc}
1 & 0 \\
R^{-- \ast}(\varphi) & 1
\end{array}
\right) \, , 
\end{equation}
such that
\begin{subequations}
\label{newbogo}
\begin{align}
\tilde{\beta}_{k_1}(\varphi) &= \beta_{k_1} + \sum_{k_2} R^{--}_{k_2k_1}(\varphi) \beta^{\dagger}_{k_2} \, , \\
\tilde{\beta}^{\dagger}_{k_1}(\varphi) &= \beta^{\dagger}_{k_1} \, .
\end{align}
\end{subequations}

Next, we introduce the non-hermitian transformed grand potential operator $\tilde{\Omega}(\varphi) \equiv M(\varphi) \Omega M^{-1}(\varphi)$. Starting from the normal-ordered form of $\Omega$, performing the non-unitary Bogoliubov transformation, normal-ordering the resulting $\tilde{\Omega}(\varphi)$ with respect to $| \Phi \rangle$ and gathering appropriately the terms thus generated allows one to write $\tilde{\Omega}(\varphi)$ under the typical form
\begin{widetext}
\begin{subequations}
\label{e:h3qpastransformed}
\begin{align}
\tilde{\Omega}(\varphi) &\equiv \tilde{\Omega}^{[0]}(\varphi) + \tilde{\Omega}^{[2]}(\varphi) + \tilde{\Omega}^{[4]}(\varphi)  \\
&\equiv \tilde{\Omega}^{00}(\varphi) +  \big[\tilde{\Omega}^{20}(\varphi) + \tilde{\Omega}^{11}(\varphi) + \tilde{\Omega}^{02}(\varphi)\big] +  \big[\tilde{\Omega}^{40}(\varphi) + \tilde{\Omega}^{31}(\varphi) + \tilde{\Omega}^{22}(\varphi) + \tilde{\Omega}^{13}(\varphi) + \tilde{\Omega}^{04}(\varphi)\big] \\
&= \tilde{\Omega}^{00}(\varphi) \\*
& \: \: \: \: \: \: \: \: \: \: \: \: \: + \frac{1}{1!}\displaystyle\sum_{k_1 k_2} \tilde{\Omega}^{11}_{k_1 k_2}(\varphi)\beta^{\dagger}_{k_1} \beta_{k_2} \\*
& \: \: \: \: \: \: \: \: \: \: \: \: \: + \frac{1}{2!}\displaystyle\sum_{k_1 k_2} \Big \{\tilde{\Omega}^{20}_{k_1 k_2}(\varphi) \beta^{\dagger}_{k_1}
 \beta^{\dagger}_{k_2} + \tilde{\Omega}^{02}_{k_1 k_2}(\varphi)   \beta_{k_2} \beta_{k_1} \Big \} \\*
& \: \: \: \: \: \: \: \: \: \: \: \: \: + \frac{1}{(2!)^{2}} \displaystyle\sum_{k_1 k_2 k_3 k_4} \tilde{\Omega}^{22}_{k_1 k_2 k_3 k_4}(\varphi) 
   \beta^{\dagger}_{k_1} \beta^{\dagger}_{k_2} \beta_{k_4}\beta_{k_3} \\*
   & \: \: \: \: \: \: \: \: \: \: \: \: \: + \frac{1}{3!}\displaystyle\sum_{k_1 k_2 k_3 k_4}\Big \{ \tilde{\Omega}^{31}_{k_1 k_2 k_3 k_4}(\varphi)
   \beta^{\dagger}_{k_1}\beta^{\dagger}_{k_2}\beta^{\dagger}_{k_3}\beta_{k_4} +
   \tilde{\Omega}^{13}_{k_1 k_2 k_3 k_4}(\varphi) \beta^{\dagger}_{k_1} \beta_{k_4} \beta_{k_3} \beta_{k_2}  \Big \} \\*
  & \: \: \: \: \: \: \: \: \: \: \: \: \: +  \frac{1}{4!} \displaystyle\sum_{k_1 k_2 k_3 k_4}\Big \{ \tilde{\Omega}^{40}_{k_1 k_2 k_3 k_4}(\varphi)
   \beta^{\dagger}_{k_1}\beta^{\dagger}_{k_2}\beta^{\dagger}_{k_3}\beta^{\dagger}_{k_4}  + 
   \tilde{\Omega}^{04}_{k_1 k_2 k_3 k_4}(\varphi)  \beta_{k_4} \beta_{k_3} \beta_{k_2} \beta_{k_1}  \Big \}  \, .
\end{align}
\end{subequations}
\end{widetext}
The expressions of the transformed matrix elements in terms of the original ones are provided in App.~\ref{transformME}. As a testimony of the non-hermitian character of $\tilde{\Omega}(\varphi)$, itself the consequence of the non-unitary character of $M(\varphi)$, matrix elements $\tilde{\Omega}^{ij}_{k_1 \ldots k_{i} k_{i+1}}(\varphi)$ do {\it not} display relationships characterized by Eq.~\ref{e:me3sym}. One also notices that transformation~\ref{newbogo} reduces to the identity for $\varphi=0$, i.e. $\tilde{\Omega}(0)=\Omega$.

\subsubsection{Compact algebraic expressions}
\label{compactexpressions}

It is tedious but straightforward to demonstrate that the algebraic expressions of $\omega(\tau,\varphi)$, $\omega_{k_1k_2}(\tau,\varphi)$ and $\omega_{k_1k_2k_3k_4}(\tau,\varphi)$ obtained through the application of the off-diagonal Wick theorem are formally identical to diagonal ($\varphi=0$) BCC formulae~\cite{shavitt09a}, as long as one uses the {\it transformed} grand potential $\tilde{\Omega}(\varphi)$ in place of the original one $\Omega$ on the basis of standard Wick's theorem, i.e.
\begin{subequations}
\label{compactification}
\begin{eqnarray}
\omega_{k_1k_2\ldots}(\tau,\varphi) &=& \frac{\langle \Phi | e^{{\cal T}^{\dagger}(\tau,\varphi)} \Omega  | \Phi^{k_1k_2\ldots}(\varphi) \rangle_{c}}{\langle \Phi |  \Phi(\varphi) \rangle} \label{CCamplitudekernelsbis0} \\
&=& \langle \Phi | e^{{\cal T}^{\dagger}(\tau,\varphi)} \tilde{\Omega}(\varphi) | \Phi_{k_1k_2\ldots} \rangle_{c} \, . \label{CCamplitudekernelsbis}
\end{eqnarray}
\end{subequations}

\begin{figure}[t!]
\begin{center}
\includegraphics[clip=,width=0.14\textwidth,angle=0]{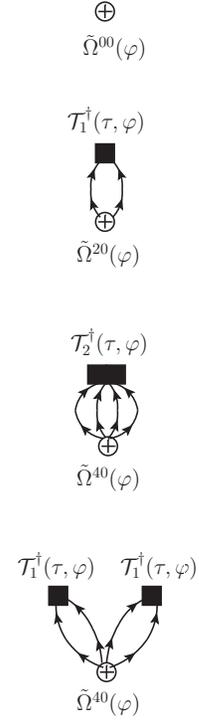}\\ 
\end{center}
\caption{
\label{diagramst}
Goldstone $n_a=0$ BCC diagrams contributing to $\omega(\tau,\varphi)$ and involving the transformed grand potential $\tilde{\Omega}(\varphi)$.}
\end{figure}

To illustrate the severe shortening of the algebraic expressions  accomplished by employing Eq.~\ref{CCamplitudekernelsbis} instead of Eq.~\ref{CCamplitudekernelsbis0}, let us focus on the energy kernel $\omega(\tau,\varphi)$ and refer to Ref.~\cite{Signoracci:2014dia} for single- and double-amplitude equations. While the lengthy expression associated with Eq.~\ref{CCamplitudekernelsbis0}  (App.~\ref{CCenergycontrib}) has been derived from twenty different diagrams, the compact form associated with Eq.~\ref{CCamplitudekernelsbis} reads as
\begin{subequations}
\label{rewritingomega}
\begin{eqnarray}
\omega(\tau,\varphi) &=& \tilde{\Omega}^{00}(\varphi) \label{rewritingomega1} \hspace{6.5cm} \\
&+& \frac{1}{2} \displaystyle\sum_{k_1 k_2} \mathcal{T}^{\dagger}_{k_1 k_2}(\tau,\varphi) \, \tilde{\Omega}^{20}_{k_1 k_2}(\varphi) \label{rewritingomega2} \\
&+& \frac{1}{8} \displaystyle\sum_{k_1 k_2 k_3 k_4} \mathcal{T}^{\dagger}_{k_1 k_2}(\tau,\varphi)\mathcal{T}^{\dagger}_{k_3 k_4}(\tau,\varphi) \, \tilde{\Omega}^{40}_{k_1 k_2 k_3 k_4}(\varphi) \label{rewritingomega3} \\
& +& \frac{1}{4!} \displaystyle\sum_{k_1 k_2 k_3 k_4} \mathcal{T}^{\dagger}_{k_1 k_2 k_3 k_4}(\tau,\varphi) \, \tilde{\Omega}^{40}_{k_1 k_2 k_3 k_4}(\varphi) \, ,  \label{rewritingomega4}
\end{eqnarray}
\end{subequations}
and relates to the four Goldstone, i.e. time-independent, diagrams displayed in Fig.~\ref{diagramst} and involving vertices $\tilde{\Omega}^{ij}(\varphi)$ of the transformed grand potential. Those four Goldstone diagrams, along with the associated expression~\ref{rewritingomega}, are indeed formally identical to the four $n_a=0$ diagrams at play in diagonal BCC theory~\cite{Signoracci:2014dia}. As a matter of fact, one can now entirely rephrase the generalized BCC formalism applicable to off-diagonal kernels in terms of a time-independent diagrammatic technique that parallels exactly BCC theory, i.e. diagrammatic rules, diagrams and algebraic expressions are identical except that transformed interaction vertices $\tilde{\Omega}^{ij}_{k_1 \ldots k_{i} k_{i+1} \ldots k_{i+j}}(\varphi)$ must be used in place of $\Omega^{ij}_{k_1 \ldots k_{i} k_{i+1} \ldots k_{i+j}}$, such that cluster amplitudes acquire an explicit $\varphi$ dependence. 

The result obtained in Eq.~\ref{compactification} is remarkable and constitutes a drastic simplification both from a formal and a practical standpoint. Regarding the latter, it means that a previously built single-reference BCC code can be employed almost straightforwardly. The additional cost, which is not negligible, consists of building and storing matrix elements of $\tilde{\Omega}(\varphi)$ for each $\varphi \in [0,2\pi]$ according to App.~\ref{transformME}. With those transformed matrix elements in input, the BCC code can be used essentially as it is.

\subsection{Norm kernel}
\label{Secnormkernel}

In Sec.~\ref{normkernel2}, we obtained the perturbative expansion of $N(\tau,\varphi)$ that eventually led to the connected expansion of $\ln N(\tau,\varphi)$. In the present section, we wish to identify a method to compute $N(\tau,\varphi)$ non-perturbatively in a way that is consistent with the BCC expansion of the connected grand potential kernel $\omega(\tau,\varphi)$. More specifically, we wish to design a naturally terminating expansion of $\ln N(\tau,\varphi)$. It happens that a naturally terminating expansion does not trivially emerge from the perturbative expansion of $\ln N(\tau,\varphi)$ as the corresponding diagrams are not linked to an operator at a fixed time. We now explain how this apparent difficulty can be overcome by following an alternative route from the outset.

\subsubsection{Key property}

In the case of {\it exact} kernels, Eqs.~\ref{expandedkernels1} and~\ref{expandedkernels3} trivially lead, for any $\tau$, to
\begin{eqnarray}
\frac{\int_{0}^{2\pi} \!d\varphi \,e^{-i\text{A}\varphi} \, {\cal A}(\tau,\varphi)}{\int_{0}^{2\pi} \!d\varphi \,e^{-i\text{A}\varphi} \,  \, {\cal N}(\tau,\varphi)} &=& \text{A} \, , \label{restoreLieoperators1} 
\end{eqnarray}
which testifies that the implicit many-body state at play is indeed an eigenstate of the particle-number operator with eigenvalue A. Equation~\ref{restoreLieoperators1} stresses the fact that we {\it know} a priori the value that must be obtained through the integral over the domain of the $U(1)$ group for the coefficient associated with the physical IRREP in the Fourier expansion of the particle number operator kernel (once it is divided by the corresponding expansion coefficient in the norm kernel). This differs from $H$ or $\Omega$ for which we can only require to extract the expansion coefficient that is in one-to-one correspondence with the physical IRREP of interest without knowing the {\it value} this coefficient should take (once it is divided by the corresponding expansion coefficient in the norm kernel).

Consequently, the key question is: what happens to Eq.~\ref{restoreLieoperators1} when ${\cal A}(\tau,\varphi)$ and ${\cal N}(\tau,\varphi)$ are approximated? Or rephrasing the question more appropriately: what constraint(s) does restoring the symmetry, i.e. fulfilling  Eqs.~\ref{restoreLieoperators1}, impose on the truncation scheme used to approximate the kernels? Addressing this question below delivers the proper approach to the reduced norm kernel. 

\subsubsection{Differential equation}

\begin{figure}[t!]
\begin{center}
\includegraphics[clip=,width=0.08\textwidth,angle=0]{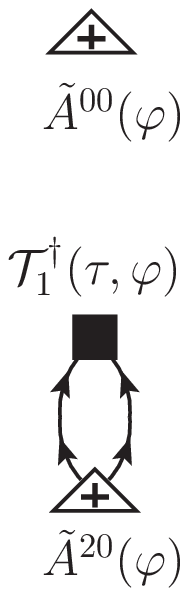}
\end{center}
\caption{
\label{diagramsa}
Goldstone $n_a=0$ BCC diagrams contributing to $a(\tau,\varphi)$ and involving the transformed particle number operator $\tilde{A}(\varphi)$.}
\end{figure}

We derive a first-order ordinary differential equation (ODE) fulfilled, at each imaginary time $\tau$, by ${\cal N}(\tau,\varphi)$. To do so, we employ Eq.~\ref{ODE} to relate ${\cal N}(\tau,\varphi)$ (Eqs.~\ref{expandedkernels1}) and ${\cal A}(\tau,\varphi)$  (Eqs.~\ref{expandedkernels3}). Exploiting that the reduced kernel of the operator $A$ can be factorized according to ${\cal A}(\tau,\varphi)=a(\tau,\varphi) {\cal N}(\tau,\varphi)$, where $a(\tau,\varphi)$ denotes the corresponding linked/connected kernel, we arrive at the first-order ODE
\begin{equation}
\frac{d}{d \varphi} \, {\cal N}(\tau,\varphi)  - i \, a(\tau,\varphi) \, {\cal N}(\tau,\varphi) = 0 \, ,\label{NormkernelODE1} 
\end{equation}
with the initial condition ${\cal N}(\tau,0)=1$ associated with intermediate normalization at $\varphi = 0$. Equation~\ref{NormkernelODE1} possesses a closed-form solution
\begin{equation}
{\cal N}(\tau,\varphi)  = e^{i \int_{0}^{\varphi} \!d\phi \, a(\tau,\phi)} \, , \label{solNormkernelODE} 
\end{equation}
which demonstrates that the logarithm of the off-diagonal  norm kernel can be related to the linked/connected kernel of $A$ via an integral over the gauge angle. The linked/connected kernel of $A$ possesses a naturally terminating BCC expansion
\begin{equation}
a(\tau,\varphi) = a^{00}(\tau,\varphi) + a^{20}(\tau,\varphi) + a^{11}(\tau,\varphi) + a^{02}(\tau,\varphi) \, ,\nonumber
\end{equation}
which is obtained by substituting $\Omega^{ij}$ with $A^{ij}$ in Eqs.~\ref{CCenergyequation2A}-\ref{CCenergyequation2D}. Employing the transformed particle-number operator, the algebraic expression takes the compact form
\begin{equation}
a(\tau,\varphi) =  \tilde{A}^{00}(\varphi)  + \frac{1}{2} \displaystyle\sum_{k_1 k_2} \mathcal{T}^{\dagger}_{k_1 k_2}(\tau,\varphi) \, \tilde{A}^{20}_{k_1 k_2}(\varphi)\, ,\label{linkedkernelA2} 
\end{equation}
where the formulae for $\tilde{A}^{00}(\varphi)$ and $\tilde{A}^{20}(\varphi)$ are provided in App~\ref{transformME}. The diagrams corresponding to Eq.~\ref{linkedkernelA2} are displayed in Fig.~\ref{diagramsa} for illustration.

In addition to authorizing the computation of ${\cal N}(\tau,\varphi)$ from a kernel displaying a naturally terminating BCC expansion, the scheme proposed above ensures that the particle number is indeed restored at any truncation order in the proposed many-body method. Indeed, $a(\tau,\varphi)$ and ${\cal N}(\tau,\varphi)$ being related through Eq.~\ref{NormkernelODE1}, one has that
\begin{eqnarray}
\int_{0}^{2\pi} \!d\varphi \,e^{-i\text{A}\varphi} \, {\cal A}(\tau,\varphi)  &=& -i\int_{0}^{2\pi} \!d\varphi \,e^{-i\text{A}\varphi} \,  \frac{d}{d \varphi} \, {\cal N}(\tau,\varphi) \nonumber \\
&=& +i \int_{0}^{2\pi} \!d\varphi \,\frac{d}{d \varphi} \, e^{-i\text{A}\varphi} \,  {\cal N}(\tau,\varphi) \nonumber \\
&=& \text{A} \int_{0}^{2\pi} \!d\varphi \, e^{-i\text{A}\varphi} \,  {\cal N}(\tau,\varphi) \, , \nonumber
\end{eqnarray}
where an integration by part was performed to go from the first to the second line. This demonstrates that, if the reduced norm kernel satisfies Eq.~\ref{NormkernelODE1}, then Eq.~\ref{restoreLieoperators1} is fulfilled {\it independently of the approximation made on} $a(\tau,\varphi)$, i.e. independently of the order at which the BCC expansion of linked/connected operator kernels is truncated. Eventually, the fact that ${\cal N}(\tau,\varphi)$ is determined from the structure of the $U(1)$ group, i.e. from the kernel of $A$, is  very natural in the present context. Once extracted through Eq.~\ref{solNormkernelODE} at a given BCC order, the off-diagonal reduced norm kernel can be consistently used in the computation of the energy as is discussed in Sec.~\ref{symrestE} below.

\subsubsection{Particle-number conserving case}
\label{normnobreaking}

In case the expansion is performed around a Slater determinant, one must recover the trivial behavior displayed by the norm kernel when global gauge symmetry is conserved. It is indeed easy to demonstrate that
\begin{subequations}
\begin{eqnarray}
\tilde{A}^{00}(\varphi) &=& A^{00}=\text{A}_0 \, , \\
\tilde{A}^{20}_{k_1 k_2}(\varphi) &=& A^{20}_{k_1 k_2}=0 \, ,
\end{eqnarray}
\end{subequations}
in this case, such that $a(\tau,\varphi) =\text{A}_0$. As a result, Eq.~\ref{solNormkernelODE} does provide the expected result for the reduced norm kernel
\begin{equation}
{\cal N}(\tau,\varphi)  = e^{i \text{A}_0 \varphi} \, , \label{normSD} 
\end{equation}
independently of the truncation employed in the many-body expansion.

\subsubsection{Lowest order}
\label{newunperturbedkernel}

Reducing the calculation to lowest order, i.e. at order $n=0$ in off-diagonal BMBPT or or equivalently taking ${\cal T}^{\dagger}_{n}(\tau,\varphi)=0$ for all $n$ in the off-diagonal BCC scheme, one obtains 
\begin{eqnarray}
a^{(0)}(\tau,\varphi) &=& \frac{\langle \Phi | A | \Phi(\varphi) \rangle}{\langle \Phi | \Phi(\varphi) \rangle} = -i \frac{d}{d \varphi} \ln \langle \Phi | \Phi(\varphi) \rangle \, , 
\end{eqnarray}
and thus recovers from Eq.~\ref{solNormkernelODE} that
\begin{eqnarray}
{\cal N}^{(0)}(\tau,\varphi) &=& e^{i \int_{0}^{\varphi} \!d\phi \, a^{(0)}(\tau,\phi)}  \nonumber \\
&=& e^{[\ln \langle \Phi | \Phi(\phi) \rangle]^{\varphi}_{0}} \nonumber \\
&=& \langle \Phi | \Phi(\varphi) \rangle \, . \label{solNormkernelODE0order} 
\end{eqnarray}
Given that at lowest order one has
\begin{subequations}
\begin{eqnarray}
a^{(0)}(\tau,\varphi) &=&  \tilde{A}^{00}(\varphi) \\
&=& A^{00} + \frac{1}{2} \displaystyle\sum_{k_1 k_2} A^{02}_{k_1 k_2} R^{--}_{k_2 k_1}(\varphi) \, , 
\end{eqnarray}
\end{subequations}
Eq.~\ref{solNormkernelODE0order} provides a way to compute the overlap between a HFB state and its gauge rotated partner according to
 \begin{eqnarray}
\langle \Phi | \Phi(\varphi) \rangle &=& e^{i \langle \Phi | A |  \Phi \rangle \varphi} \, e^{\frac{i}{2} \sum_{k_1 k_2} A^{02}_{k_1 k_2}  \int_{0}^{\varphi} \!d\phi \,R^{--}_{k_2 k_1}(\phi)}  \, , \nonumber
\end{eqnarray}
as $A^{00}=\langle \Phi | A |  \Phi \rangle$. The above expression constitutes an interesting alternative to the Pfaffian formula~\cite{Robledo:2009yd} provided in Eq.~\ref{kernel} or to even older approaches to the unperturbed norm kernel.

\section{Energy}
\label{symrestE}

\subsection{Particle-number-restored energy}

The particle-number-restored energy is computed according to Eq.~\ref{yrast_projected_energy2nd}, now written as
\begin{eqnarray}
\text{E}^{\text{A}}_{0} &=& \frac{\int_{0}^{2\pi} \!d\varphi \,e^{-i\text{A}\varphi} \, \, h(\varphi) \,\, {\cal N}(\varphi)}{\int_{0}^{2\pi} \!d\varphi \,e^{-i\text{A}\varphi} \, \, {\cal N}(\varphi)} , \label{projected_energy_MBPT}
\end{eqnarray}
where $h(\varphi) \equiv \omega(\varphi) + \lambda a(\varphi)$. Expressing the energy in terms of the {\it reduced} norm kernel in Eq.~\ref{projected_energy_MBPT} is essential. Indeed, the fact that ${\cal N}(\tau,\varphi)$ goes to a finite number in the large $\tau$ limit, contrarily to $N(\tau,\varphi)$ that goes exponentially to zero, is mandatory to make the ratio in Eq.~\ref{projected_energy_MBPT} well defined and numerically controllable. The connected/linked kernels $\omega(\varphi)$ and $a(\varphi)$ are to be truncated consistently, i.e. at a given order $n$ in off-diagonal BMBPT or at a given order in off-diagonal BCC amplitudes (including singles, doubles, triples\ldots). The approximate kernel $a(\varphi)$ is also employed to access ${\cal N}(\varphi)$ consistently via Eq.~\ref{solNormkernelODE} .

If one were to sum all diagrams in the computation of $h(\varphi)$ and ${\cal N}(\varphi)$ or were to expand them around a Slater determinant, the symmetry restoration would become dispensable by definition. This relates to the fact that $h(\varphi)$ becomes independent of $\varphi$  while ${\cal N}(\varphi)$ becomes trivially proportional to the targeted IRREP $e^{i \text{A}_0 \varphi}$ in these two cases. As a result, Eq.~\ref{projected_energy_MBPT} reduces to
\begin{eqnarray}
\text{E}^{\text{A}}_0 &=&  h(0) \, , \label{standard_MBPT}
\end{eqnarray}
for $\text{A}=\text{A}_0$ and zero otherwise.

The benefit of the method arises when the many-body expansion is performed around a Bogoliubov state and is eventually truncated. Indeed, $h(\varphi)$ acquires a dependence on $\varphi$ that signals the breaking of the symmetry generated by that truncation. The method authorizes the summation of standard sets of diagrams (i.e. dealing with so-called {\it dynamical} correlations) while leaving the non-perturbative symmetry-restoration process (i.e. dealing with so-called {\it static} correlations) to be achieved at each truncation order through the integration over the domain of the group. As a matter of fact, one can rewrite
\begin{eqnarray}
\text{E}^{\text{A}}_0 &=&  h(0) + \frac{\int_{0}^{2\pi} \!d\varphi \,e^{-i\text{A}\varphi} \,\, \left[h(\varphi)-h(0) \right] \,\, {\cal N}(\varphi)}{\int_{0}^{2\pi} \!d\varphi \,e^{-i\text{A}\varphi}  \,\,  {\cal N}(\varphi)} \, , \label{rewritingresult}
\end{eqnarray}
such that the effect of the particle-number restoration itself can be viewed, at any truncation order, as a {\it correction} to the particle-number-breaking BMBPT or BCC results provided by $h(0)$.


\subsection{Particle-number projected HFB theory}
\label{PHFsection}

Reducing the calculation to lowest order, i.e. at order $n=0$ in off-diagonal BMBPT or or equivalently taking ${\cal T}^{\dagger}_{n}(\tau,\varphi)=0$ for all $n$ in the off-diagonal BCC scheme, one recovers  particle-number projected Hartree-Fock-Bogoliubov (PNP-HFB) theory~\cite{ring80a,blaizot86} (assuming that the reference state $| \Phi \rangle$ is obtained from a HFB calculation). The associated kernels read as 
\begin{subequations}
\label{PHFkernels}
\begin{eqnarray}
h^{(0)}(\tau,\varphi) &=& \frac{\langle \Phi | H | \Phi(\varphi) \rangle}{\langle \Phi  | \Phi(\varphi) \rangle} \, ,  \label{PHFkernelsH} \\
{\cal N}^{(0)}(\tau,\varphi) &=& \langle \Phi  | \Phi(\varphi) \rangle \label{PHFkernelsN} \, ,
\end{eqnarray}
\end{subequations}
for all $\tau$, such that the symmetry-restored energy becomes
\begin{subequations}
\label{projected_HF}
\begin{eqnarray*}
\text{E}^{\text{A} (0)}_{0} &=& \frac{\int_{0}^{2\pi} \!d\varphi \,e^{-i\text{A}\varphi} \, \, \langle \Phi | H | \Phi(\varphi) \rangle}{\int_{0}^{2\pi} \!d\varphi \,e^{-i\text{A}\varphi}  \,\, \langle \Phi  | \Phi(\varphi) \rangle} \label{projected_HF1} \\
&=& \frac{\langle \Phi | H | \Theta^{\text{A}} \rangle}{\langle \Phi | \Theta^{\text{A}}  \rangle} \label{projected_HF2} \\
&=& \frac{\langle \Theta^{\text{A}}  | H | \Theta^{\text{A}}  \rangle}{\langle \Theta^{\text{A}}  | \Theta^{\text{A}}  \rangle} \, ,
\end{eqnarray*}
\end{subequations}
where the (un-normalized) PNP-HFB wave-function $| \Theta^{\text{A}}  \rangle \equiv P^{\text{A}} | \Phi \rangle$ is defined from the projection operator
\begin{equation}
P^{\text{A}} \equiv  \frac{1}{2\pi}\int_{0}^{2\pi} \!d\varphi \,e^{-i\text{A}\varphi}  \, S(\varphi) \, ,
\end{equation}
satisfying 
\begin{subequations}
\label{projector}
\begin{eqnarray}
P^{\text{A}}\,P^{\text{A}} &=& P^{\text{A}} \, , \label{projector1} \\
P^{\text{A} \dagger} &=& P^{\text{A}} \, , \label{projector2} \\ 
\left[H,P^{\text{A}}\right] &=& \left[A,P^{\text{A}}\right] = \left[\Omega,P^{\text{A}}\right] = 0 \, .
\end{eqnarray}
\end{subequations}

\section{Discussion}

Let us provide one last set of comments
\begin{itemize}
\item In the end, off-diagonal BMBPT and BCC schemes only need to be applied at $\tau=+\infty$, i.e. the imaginary-time formulation becomes superfluous and one is left with the static version of the many-body formalisms.
\item PNR-BMBPT and PNR-BCC formalisms are of multi-reference character but reduce in practice to a set of $N_{\text{sym}}$ single-reference-like off-diagonal BMBPT and BCC calculations, where $N_{\text{sym}}\sim 10$. The factor of $10$ is an estimation based on the discretization of the integral over the gauge angle in Eq.~\ref{projected_energy_MBPT} typically needed to achieve convergence in the ground-state computation of even-even nuclei at the PNP-HFB level. Beyond PNP-HFB, the appropriate value of $N_{\text{sym}}$ will have to be validated through numerical tests.
\item It could be of interest to design an approximation of the presently proposed many-body formalisms based on a Lipkin expansion method~\cite{lipkin60a}. This would permit the extension of this well-performing approximation to particle-number restoration before variation beyond PNP-HFB~\cite{Wang:2014nza}.
\item The chemical potential $\lambda$ entering the definition of $\Omega$ must be specified in the calculation. In principle, it must be such that the exact ground state of the $\text{A}_0$-body system of interest is the eigenstate of $\Omega$ with the lowest eigenergy. In practice, $\lambda$ is to be fixed at the single reference level, e.g. when solving the BCC equations ($\varphi =0$) at the chosen level  of many-body truncation, e.g. BCCSD. As a further simplification, one can envision to fix  $\lambda$ to the value obtained by solving the simpler HFB equations at $\varphi =0$. 
It remains to be seen how much the choice made to fix $\lambda$ impacts PNR-BMBPT and PNR-BCC results at various levels of truncation.
\item The angular-momentum-restored MBPT and CC theory~\cite{Duguet:2014jja} based on the three-parameter non-Abelian $SU(2)$ Lie group has recently undergone its first numerical implementation~\cite{binder15a}. Doing so, it was identified that the exact fulfillment of the second-order ODE necessary to extract the norm kernel associated with an exact restoration of $J^2$ is compromised when going beyond lowest, i.e. HF, order. The remedy to this issue requires an extension of the angular-momentum-restored MBPT and CC theory formulated in Ref.~\cite{Duguet:2014jja} in order to compute $J_z$  and $J^2$ kernels via a "symmetric", e.g. expectation value, rather than a "projective" formula. This is in fact necessary to match single-reference CC theory at zero Euler angles where properties as $J_z$ and $J^2$ are known to be best approximated from a "symmetric" formula, which is typically achieved via the $\Lambda$CC method~\cite{shavitt09a}. It is thus our goal to formulate such an extension of angular-momentum-restored MBPT and angular-momentum-restored MBPT CC theory in the near future. As for the one-parameter Abelian $U(1)$ group of present interest, while a similar extension will most probably be beneficial, it is not as critical as for $SU(2)$ given that the first-order ODE (Eq.~\ref{NormkernelODE1}) possesses a closed-form, i.e. exact, solution (Eq.~\ref{solNormkernelODE}) independently on the approximation used to compute the "projective" particle-number kernel $a(\varphi)$. Thus, one has confidence that PNR-BMBPT and PNR-BCC calculations can be safely performed on the basis of the present formulation while awaiting for their $\Lambda$-like extension in which the particle-number kernel will be evaluated via a symmetric rather than a projective formula.
\end{itemize}

\section{Implementation algorithm}

Let us eventually synthesize the steps the owner of a single-reference BCC code must follow to implement the particle-number restoration procedure\footnote{In order to actually fit with an existing SR-BCC code, one must proceed to the hermitian conjugation of the quantities and equations that are referred to.}.
\begin{enumerate}
\item Solve, e.g.,  Hartree-Fock-Bogoliubov equations in the single-particle basis of interest to obtain the reference state $| \Phi \rangle$ (Eq.~\ref{e:bogvac}), i.e. to determine the $(U,V)$ matrices.
\item Build matrix elements $\Omega^{ij}_{k_1 \ldots k_{i} k_{i+1}}$ according to App. A of Ref.~\cite{Signoracci:2014dia}. 
\item Discretize the interval of integration over the gauge angle $\varphi \in [0,2\pi]$. 
\item For each angle $\varphi$
\begin{enumerate}
\item Compute matrix $R^{--}(\varphi)$ (Eqs.~\ref{offdiaggeneralizeddensitymatrix2}-\ref{upperright}).
\item Build transformed matrix elements $\tilde{\Omega}^{ij}_{k_1 \ldots k_{i} k_{i+1}}(\varphi)$ (App.~\ref{transformME}). 
\item Initialize the cluster amplitudes through first-order perturbation theory; e.g. at the singles and doubles level, apply Eq.~\ref{1storderclusterB} for $\tau\rightarrow+\infty$ to obtain ${\cal T}^{\dagger (1)}_{k_1k_2}(\varphi)$ and ${\cal T}^{\dagger (1)}_{k_1k_2k_3k_4}(\varphi)$.
\item Run the single-reference BCC code using matrix elements $\tilde{\Omega}^{ij}_{k_1 \ldots k_{i} k_{i+1}}(\varphi)$ and the initial cluster amplitudes ${\cal T}^{\dagger (1)}_{k_1k_2}(\varphi)$ and ${\cal T}^{\dagger (1)}_{k_1k_2k_3k_4}(\varphi)$ as inputs.
\item Use the converged amplitudes, e.g. ${\cal T}^{\dagger}_{1}(\varphi)$ and ${\cal T}^{\dagger}_{2}(\varphi)$, to compute and store the linked/connected kernels $\omega(\varphi)$ and $a(\varphi)$. 
\end{enumerate}
\item Using the values of $a(\varphi)$ stored for discretized values of the gauge angle, extract the reduced norm kernel ${\cal N}(\varphi)$ via Eq.~\ref{solNormkernelODE}\footnote{As one is solely interested in the $\tau\rightarrow+\infty$ limit, the time argument can be simply ignored in Eq.~\ref{solNormkernelODE}.}. 
\item Calculate the particle-number restored energy according to Eq.~\ref{projected_energy_MBPT}.
\end{enumerate}

Two remarks can be added
\begin{itemize}
\item Step 4 can be carried out independently for each value of the gauge angle and is thus amenable to a trivial parallelization. Eventually one solely needs to retrieve and store $\omega(\varphi)$ and $a(\varphi)$.
\item In practice, the domain of integration in step 5 can be reduced thanks to symmetries of the reference state $| \Phi \rangle$. For instance, the domain can be limited to $\varphi \in [0,\pi]$ whenever using a reference state $| \Phi \rangle$ with good number parity.
\item If limiting oneself to a perturbative approach, one simply needs to replace steps 4.c-4.e by the evaluation of $\omega(\varphi)$ and $a(\varphi)$ at order $n$ in the off-diagonal BMBPT scheme.
\end{itemize}

\section{Conclusions}

Both in Paper I (Ref.~\cite{Duguet:2014jja}) and in the present work, we have addressed a long-term challenge of ab initio many-body theory, i.e. extend symmetry-unrestricted Rayleigh-Schroedinger many-body perturbation theory and coupled-cluster theory in such a way that a broken symmetry is {\it exactly} restored at any truncation order. The newly proposed symmetry-restored MBPT and CC formalisms authorize the computation of connected diagrams associated with dynamical correlations while consistently incorporating static correlations through the non-perturbative restoration of the broken symmetry.  These approaches are meant to be valid for any symmetry that can be (spontaneously) broken by the reference state and to be applicable to any system independently of its closed-shell, near degenerate or open-shell character.  

In Paper I, we focused on the breaking and the restoration of $SU(2)$ rotational symmetry associated with angular momentum conservation. The proposed scheme provides access to the yrast spectroscopy, i.e. the lowest energy for each value of the angular momentum. Standard symmetry-restricted and symmetry-unrestricted MBPT and CC theories, along with angular-momentum-projected Hartree-Fock theory, were recovered as particular cases of the newly developed many-body formalism.

In the present paper, we extend the work to $U(1)$ global gauge symmetry associated with particle-number conservation. In this case, the symmetry-unrestricted single-reference many-body methods upon which the extended formalism builds are Bogoliubov MBPT~\cite{mehta1,balian62a,henley} and Bogoliubov CC theory~\cite{StolarczykMonkhorst,Signoracci:2014dia,Henderson:2014vka} that can tackle open-shell nuclei displaying a superfluid character at the price of breaking particle-number conservation. The presently proposed formalism overcomes the latter limitation by allowing the restoration of global gauge invariance associated with particle-number conservation, which is mandatory in a finite quantum system such as the atomic nucleus. Technically speaking, the present work follows the same steps as for $SU(2)$ but requires the use of Bogoliubov algebra and thus more general diagrammatic techniques. It leads to designing particle-number-restored Bogoliubov many-body perturbation theory and particle-number-restored Bogoliubov coupled cluster theory.

As the goal is to resolve the near-degenerate nature of the ground state, the proposed many-body formalisms are necessarily of multi-reference character. However, the multi-reference nature is different from any of the multi-reference MBPT or CC methods developed in quantum chemistry~\cite{bartlett07a}, i.e. reference states are not obtained from one another via elementary (i.e. dynamical) excitations but via highly non-perturbative symmetry transformations, e.g. rotation in gauge space. Most importantly, the presently proposed method corresponds to performing a set of $N_{\text{sym}}$ single-reference-like Bogoliubov MBPT or Bogoliubov CC calculations, where $N_{\text{sym}}$ corresponds to the discretization of the integral over the gauge angle that is typically of the order of $10$. 

The present work offers a wealth of potential applications and further extensions appropriate to the ab initio description of open-shell atomic nuclei. For example, mid-mass singly open-shell nuclei are also being addressed through particle-number-breaking Green's function calculations under the form of self-consistent Gorkov Green's function theory~\cite{soma11a,Soma:2012zd,Barbieri:2012rd,Soma:2013vca,Soma:2013xha}. It is thus of interest to develop the equivalent to the particle-number-restored BCC formalism within the framework of self-consistent Gorkov-Green's function techniques.  

Last but not least, symmetry-restored MBPT and CC theories provide well-founded, formally exact, references for the so-far empirical multi-reference nuclear energy density functional (EDF) method. Multi-reference EDF calculations are known to be compromised with serious pathologies when the off-diagonal EDF kernel is not strictly computed as the matrix element of an {\it effective} Hamilton operator between a product state and its rotated partner~\cite{dobaczewski07,Lacroix:2008rj,Bender:2008rn,Duguet:2008rr,Duguet:2010cv,Duguet:2013dga}, i.e. when it does not take the strict form of an (effective) projected Hartree-Fock-Bogoliubov theory. Starting from the newly proposed many-body formalisms, one can contemplate the derivation of {\it safe} parametrizations of off-diagonal EDF kernel that go beyond projected Hartree-Fock-Bogoliubov, most probably under the form of orbital- and symmetry-angle-dependent energy functionals~\cite{Duguet:2015nna}. This remains to be investigated in the future.

\begin{acknowledgments}
T.D. wishes to thank G. Ripka very deeply for enlightening discussions that were instrumental in making this work possible. The authors thank B. Bally and V. Som\`{a} for their careful proofreading of the manuscript. This work was supported in part by the U.S. Department of Energy (Oak Ridge National Laboratory), under Grant Nos. DEFG02-96ER40963 (University of Tennessee), DE-SC0008499 (NUCLEI Sci-DAC collaboration), and the Field Work Proposal ERKBP57. 
\end{acknowledgments}

\begin{appendix}

\section{Perturbative expansion of ${\cal U}(\tau)$}
\label{perturbativeannexe}

The imaginary-time evolution operator ${\cal U}(\tau)$ can be expanded in powers of $\Omega_{1}$. Taking $\tau$ real, one writes
\begin{equation}
{\cal U}(\tau)  \equiv e^{-\tau \Omega_{0}} \, {\cal U}_1(\tau) \label{evoloperatorApp1}
\end{equation}
where
\begin{eqnarray}
{\cal U}_1(\tau)  &=& e^{\tau \Omega_{0}} \, e^{-\tau(\Omega_{0}+\Omega_{1})} \, , 
\end{eqnarray}
and thus
\begin{eqnarray}
\partial_{\tau} {\cal U}_1(\tau) &=&-e^{\tau \Omega_{0}} \, \Omega_{1} \, e^{-\tau \Omega_{0}} \, {\cal U}_1(\tau) \, \, .
\end{eqnarray}
The formal solution to the latter equation reads
\begin{eqnarray}
{\cal U}_1(\tau)  &=& \textmd{T}e^{-\int_{0}^{\tau}dt \Omega_{1}(t)} \, , \label{reducedevolutionop1} 
\end{eqnarray}
where $\textmd{T}$ is a time-ordering operator and where $\Omega_{1}(\tau)$ defines the perturbation in the interaction representation
\begin{eqnarray}
\Omega_{1}(\tau)  &\equiv& e^{\tau \Omega_{0}} \, \Omega_{1} \, e^{-\tau \Omega_{0}} \, . \label{reducedevolutionop2}
\end{eqnarray}
Eventually, the full solution reads
\begin{equation}
{\cal U}(\tau)=e^{-\tau \Omega_{0}} \, \textmd{T}e^{-\int_{0}^{\tau}d\tau \Omega_{1}(\tau)
}  \, . \label{exp1}
\end{equation}

\section{Weight of diagrams and anomalous contractions}
\label{ruleanomalouscontraction}

We exemplify the diagrammatic rule, i.e. a factor $1/2$, associated with anomalous lines contracted onto a given vertex. This rule comes in combination with the standard rule associated with equivalent lines. We presently consider the simplest case of a string of operators 
\begin{equation}
\beta^{\dagger}_1 (\tau_1)\beta^{\dagger}_2 (\tau_1) \ldots \beta^{\dagger}_i (\tau_1) \beta_{i+1} (\tau_1) \beta_{i+2} (\tau_1) \ldots \beta_{i+j} (\tau_1) \nonumber
\end{equation} 
consisting of $i$ quasiparticle creation operators and $j$ quasiparticle annihilation operators\footnote{The order in which the operators are written is important to know whether contractions arising from the string of operators are eventually non zero but is irrelevant as far as dealing with the numbering, which is what concerns us in the present proof.} belonging to a vertex at time $\tau_1$. One can then extend the proof to more general cases, i.e. cases where part of the operators  belong to one vertex at time $\tau_1$ while the others belong to another vertex at time $\tau_2$\ldots Those more general cases are key as they demonstrate how each anomalous line contracted onto a given vertex requires, once the rule for equivalent lines has been properly taken into account, an extra factor $1/2$ whereas anomalous lines linking two different vertices do not. The proof for more general cases can be straightforwardly generalized from the present one and are thus not reproduced here.

For the present scenario, we assume that $j > i$, but the derivation from $j < i$ proceeds in a similar way to the same result. One must keep track of two aspects of performing contractions: (i) the number of ways a given set of contractions can be obtained and (ii) the rule associated with equivalent lines, either anomalous or normal.

\subsection{Minimal number of anomalous contractions}

We begin by evaluating the result when all $i$ quasi-particle creation operators are members of a normal contraction with a quasi-particle annihilation 
operator, which leaves $(j-i)$ quasi-particle annihilation operators for anomalous contractions\footnote{It is easiest to assume $i$ and $j$ are both even, although this is not necessary in principle.}.  Thus, the number of ways to pair the $i+j$ operators with minimal anomalous contractions is
\begin{equation}
j(j-1)(j-2)\ldots(j-i+1) (j-i-1)!! = \frac{j!}{(j-i)!} (j-i-1)!! \, ,
\end{equation}
since the first quasi-particle creation operator can be paired with any of the $j$ quasi-particle annihilation operators, while the second can be 
paired with any of the remaining $j-1$ operators, etc., until the $i^{th}$ can be paired with any of the remaining $j-i+1$.  The remaining $j-i$ 
quasi-particle annihilation operators must contract with each other. Starting from one, it can be paired with any of the remaining $j-i-1$.  At this point, 
$j-i-2$ remain to contract with each other.  Selecting one, $j-i-3$ options are available, and once a contraction has occurred, $j-i-4$ operators remain.  
This results in the double factorial $(j-i-1)!!$.  

Applying the diagrammatic rules, one must associate a factor to equivalent lines, which here is
\begin{equation}
\left[i! \left(\frac{j-i}{2}\right)!\right]^{-1} \, ,
\end{equation}
given that the $i$ normal contractions are all equivalent while the $(j-i)/2$ anomalous contractions are also all equivalent in the present example.  In 
further considering the factor $[i!j!]^{-1}$ from the definition of the operator giving rise to the string of creation and annihilation operators, one is left with a factor
\begin{equation}
 \frac{j!}{(j-i)!} (j-i-1)!! \left[i! \left(\frac{j-i}{2}\right)!\right] \left[i! j! \right]^{-1}=\frac{[(j-i)/2]!}{(j-i)!!} \label{factor1}
\end{equation}
to account for. Given that 
\begin{eqnarray}
(2n)!! &=& 2n(2n-2)(2n-4)(2n-6)....2  \nonumber \\
       &=& 2n*2(n-1)*2(n-2)*2(n-3)....*2(1)  \nonumber \\
       &=& 2^n * n! \, ,
\end{eqnarray}
the right-hand side of Eq.~\ref{factor1} can be rewritten as $(1/2)^{(j-i)/2}$, which corresponds to a factor $1/2$ to be attributed to each of the $(j-i)/2$ anomalous lines starting and ending at the operator.

\subsection{Two additional anomalous contractions}

When one anomalous contraction occurs between two quasiparticle creation operators, the permutations possible from the pairing of operators is 
\begin{widetext}
\begin{equation}
\frac{i(i-1)}{2} j(j-1)(j-2)\ldots(j-i+3) (j-i+1)!! = \frac{j!}{(j-i)!}(j-i-1)!! \frac{i(i-1)}{2} \frac{1}{j-i+2} \; ,
\end{equation}
\end{widetext}
where on the left side of the equation, any two of the $i$ quasiparticle creation operators can be combined in the anomalous contraction (i.e., pick one 
from $i$, then one from remaining $i-1$, and divide by 2 since the order of selection is irrelevant), the remaining $i-2$ quasiparticle creation 
operators are contracted with quasi-particle annihilation operators, and the remaining $j-i+2$ quasi-particle annihilation operators are contracted 
amongst themselves.  The factors are obtained analogously to the prior subsection.  The right-hand side of the equation above is found by separating 
$(j-1+1)!!$ into $(j-i+1)(j-i-1)!!$ and multiplying by $(j-i+2)/(j-i+2)$ to put the expression in a similar form to the prior subsection.  

The factor accounted for by the rule on equivalent lines is in this example
\begin{equation}
(i-2)! \left(\frac{j-i+2}{2}\right)! = \frac{i!}{i(i-1)} \left( \frac{j-i+2}{2} \right)\left(\frac{j-i}{2}\right)! \; .
\end{equation}

Taking the ratio of the counting factor to the one obtained in the previous case gives
\begin{widetext}
\begin{equation}
\frac{ \frac{j!}{(j-i)!}(j-i-1)!! \frac{i(i-1)}{2} \frac{1}{j-i+2} \frac{i!}{i(i-1)} ( \frac{j-i+2}{2} )(\frac{j-i}{2})!}{\frac{j!}{(j-i)!} (j-i-1)!! \, i! (\frac{j-i}{2})!} =
\frac{i(i-1)}{2} \frac{1}{j-i+2} ( \frac{j-i+2}{2} )\frac{1}{i(i-1)} = \frac{1}{4} \: ,
\end{equation}
\end{widetext}
which corresponds to a factor $1/2$ for each additional anomalous contraction, since the anomalous contraction among quasi-particle creation operators has induced an additional anomalous contraction among quasi-particle annihilation operators. This again justifies the rule that can be further extended up to the point where all quasi-particle creation operators are contracted amongst themselves.

\section{$n(\tau,\varphi)$ at BMBPT(2)}
\label{diagramsN}

Applying the diagrammatic rules explicated in Sec.~\ref{diagrulenormMBPT} to the 18 first- and second-order off-diagonal BMBPT diagrams contributing to $n(\tau,\varphi)$ and displayed in Fig.~\ref{diagramsNc} gives
 \begin{widetext}
 \begin{subequations}
 \begin{align*}
  \text{PN}.1=& + \frac{1}{2} \displaystyle\sum_{k_1 k_2} \frac{\Omega^{02}_{k_1 k_2}}{E_{k_1}+E_{k_2}} \left[ 1-
  e^{-\tau(E_{k_1}+E_{k_2})} \right] R^{--}_{k_1k_2}(\varphi) \\
  \text{PN}.2=& - \frac{1}{8} \displaystyle\sum_{k_1 k_2 k_3 k_4} \frac{\Omega^{04}_{k_1k_2k_3k_4}}{E_{k_1}+E_{k_2}+E_{k_3}+E_{k_4}} \left[ 1 - e^{-\tau(E_{k_1}+E_{k_2}+E_{k_3}+E_{k_4})} \right] R^{--}_{k_1k_2}(\varphi) R^{--}_{k_3k_4}(\varphi)\\
  \text{PN}.3=&  + \frac{1}{2} \displaystyle\sum_{k_1 k_2} \frac{\Omega^{02}_{k_1k_2}\Omega^{20}_{k_1k_2}}{E_{k_1}+E_{k_2}} 
  \left[ \tau - \frac{1-e^{-\tau(E_{k_1}+E_{k_2})}}{E_{k_1}+E_{k_2}} \right]\\
  \text{PN}.4=&  +\frac{1}{2} \displaystyle\sum_{k_1 k_2 k_3} \frac{\Omega^{02}_{k_1k_3}\Omega^{11}_{k_1k_2}}{E_{k_1}-E_{k_2}} \left[ 
  \frac{1-e^{-\tau(E_{k_1}+E_{k_3})}}{E_{k_1}+E_{k_3}} - \frac{1-e^{-\tau(E_{k_2}+E_{k_3})}}{E_{k_2}+E_{k_3}} \right] R^{--}_{k_2 k_3}(\varphi)\\
 \text{PN}.5=& +\frac{1}{2}\displaystyle\sum_{k_1k_2k_3k_4}\frac{\Omega^{02}_{k_1k_2}\Omega^{02}_{k_3k_4}}{(E_{k_1}+E_{k_2})(E_{k_3}+E_{k_4})} 
  \left[ (1-e^{-\tau(E_{k_1}+E_{k_2})})(1-e^{-\tau(E_{k_3}+E_{k_4})}) \right] R^{--}_{k_1 k_4}(\varphi) R^{--}_{k_2 k_3}(\varphi) \\
  \text{PN}.6=& +\frac{1}{4!} \displaystyle\sum_{k_1 k_2 k_3 k_4} \frac{\Omega^{04}_{k_1k_2k_3k_4}\Omega^{40}_{k_1k_2k_3k_4}}{E_{k_1}+E_{k_2}+E_{k_3}+E_{k_4}} \left[ \tau - \frac{1-e^{-\tau(E_{k_1}+E_{k_2}+E_{k_3}+E_{k_4})}}{E_{k_1}+E_{k_2}+E_{k_3}+E_{k_4}} \right]\\
  \text{PN}.7=& +\frac{1}{3!} \displaystyle\sum_{\substack{k_1 k_2 k_3 k_4 \\ k_5}} \frac{\Omega^{31}_{k_1k_2k_3k_4} \Omega^{04}_{k_5k_1k_2k_3}}{E_{k_1}+E_{k_2}+E_{k_3}-E_{k_4}} \left[ \frac{1- e^{-\tau(E_{k_4}+E_{k_5})}}{E_{k_4}+E_{k_5}} - \frac{1- e^{-\tau(E_{k_1}+E_{k_2}+E_{k_3}+E_{k_5})}}{E_{k_1}+E_{k_2}+E_{k_3}+E_{k_5}}\right] R^{--}_{k_4k_5}(\varphi) \\
  \text{PN}.8=&  + \frac{1}{4} \!\!\!\displaystyle\sum_{\substack{k_1 k_2 k_3 k_4 \\ k_5 k_6}} \frac{\Omega^{22}_{k_1k_2k_3k_4} \Omega^{04}_{k_5k_6k_1k_2}}{E_{k_1}+E_{k_2}-E_{k_3}-E_{k_4}} \left[ \frac{1- e^{-\tau(E_{k_3}+E_{k_4}+E_{k_5}+E_{k_6})}}{E_{k_3}+E_{k_4}+E_{k_5}+E_{k_6}} \right.\nonumber \\
  & \hspace{7cm} \left.- \frac{1- e^{-\tau(E_{k_1}+E_{k_2}+E_{k_5}+E_{k_6})}}{E_{k_1}+E_{k_2}+E_{k_5}+E_{k_6}}\right] R^{--}_{k_3k_6}(\varphi) R^{--}_{k_4k_5}(\varphi)  \\
  \text{PN}.9=& +\frac{1}{3!} \displaystyle\sum_{\substack{k_1 k_2 k_3 k_4 \\ k_5 k_6 k_7}} \frac{\Omega^{13}_{k_1k_2k_3k_4} \Omega^{04}_{k_1k_5k_6k_7}}{E_{k_1}-E_{k_2}-E_{k_3}-E_{k_4}} \left[ \frac{1- e^{-\tau(E_{k_2}+E_{k_3}+E_{k_4}+E_{k_5}+E_{k_6}+E_{k_7})}}{E_{k_2}+E_{k_3}+E_{k_4}+E_{k_5}+E_{k_6}+E_{k_7}} \right.\nonumber \\
  & \hspace{6cm} \left.- \frac{1- e^{-\tau(E_{k_1}+E_{k_5}+E_{k_6}+E_{k_7})}}{E_{k_1}+E_{k_5}+E_{k_6}+E_{k_7}}\right] R^{--}_{k_2k_7}(\varphi)  R^{--}_{k_3k_6}(\varphi) R^{--}_{k_4k_5}(\varphi)  \\
  \text{PN}.10=& +\frac{1}{4!} \displaystyle\sum_{\substack{k_1 k_2 k_3 k_4 \\ k_5 k_6 k_7 k_8}} \frac{\Omega^{04}_{k_1k_2k_3k_4} \Omega^{04}_{k_5k_6k_7k_8}}{(E_{k_1}+E_{k_2}+E_{k_3}+E_{k_4})(E_{k_5}+E_{k_6}+E_{k_7}+E_{k_8})} \left[1- e^{-\tau(E_{k_1}+E_{k_2}+E_{k_3}+E_{k_4})} \right] \nonumber \\
  & \hspace{4cm}  \times \left[1- e^{-\tau(E_{k_5}+E_{k_6}+E_{k_7}+E_{k_8})} \right] R^{--}_{k_1k_8}(\varphi)  R^{--}_{k_2k_7}(\varphi) R^{--}_{k_3k_6}(\varphi) R^{--}_{k_4k_5}(\varphi) \\
  \text{PN}.11=& +\frac{1}{4} \displaystyle\sum_{k_1 k_2 k_3 k_4} \frac{\Omega^{22}_{k_1k_2k_3k_4} \Omega^{02}_{k_1k_2}}{E_{k_1}+E_{k_2}-E_{k_3}-E_{k_4}} \left[ \frac{1- e^{-\tau(E_{k_3}+E_{k_4})}}{E_{k_3}+E_{k_4}} - \frac{1- e^{-\tau(E_{k_1}+E_{k_2})}}{E_{k_1}+E_{k_2}}\right] R^{--}_{k_4k_3}(\varphi) \\
  \text{PN}.12=& -\frac{1}{2} \displaystyle\sum_{\substack{k_1 k_2 k_3 k_4 \\ k_5}} \frac{\Omega^{13}_{k_1 k_2 k_3 k_4} \Omega^{02}_{k_1 k_5}}{E_{k_1}-E_{k_2}-E_{k_3}-E_{k_4}} \left[ \frac{1- e^{-\tau(E_{k_1}+E_{k_5})}}{E_{k_1}+E_{k_5}} - \frac{1- e^{-\tau(E_{k_2}+E_{k_3}+E_{k_4}+E_{k_5})}}{E_{k_2}+E_{k_3}+E_{k_4}+E_{k_5}}\right] R^{--}_{k_4k_5}(\varphi) R^{--}_{k_3k_2}(\varphi) \\
  \text{PN}.13=& +\frac{1}{4} \displaystyle\sum_{\substack{k_1 k_2 k_3 k_4 \\ k_5 k_6}} \frac{\Omega^{02}_{k_1 k_2} \Omega^{04}_{k_3 k_4 k_5 k_6}}{E_{k_1}+E_{k_2}} \left[ \frac{1- e^{-\tau(E_{k_3}+E_{k_4}+E_{k_5}+E_{k_6})}}{E_{k_3}+E_{k_4}+E_{k_5}+E_{k_6}}\right.\nonumber \\
  & \hspace{5cm} \left. - \frac{1- e^{-\tau(E_{k_1}+E_{k_2}+E_{k_3}+E_{k_4}+E_{k_5}+E_{k_6})}}{E_{k_1}+E_{k_2}+E_{k_3}+E_{k_4}+E_{k_5}+E_{k_6}}\right] 
  R^{--}_{k_2 k_3}(\varphi) R^{--}_{k_1 k_4}(\varphi) R^{--}_{k_6 k_5}(\varphi) \\
  \text{PN}.14=& +\frac{1}{4} \displaystyle\sum_{k_1 k_2 k_3 k_4} \frac{\Omega^{20}_{k_1k_2} \Omega^{04}_{k_3 k_4 k_1 k_2}}{E_{k_1}+E_{k_2}} \left[ \frac{1- e^{-\tau(E_{k_3}+E_{k_4})}}{E_{k_3}+E_{k_4}} - \frac{1- e^{-\tau(E_{k_1}+E_{k_2}+E_{k_3}+E_{k_4})}}{E_{k_1}+E_{k_2}+E_{k_3}+E_{k_4}}\right] R^{--}_{k_4k_3}(\varphi) \\
  \text{PN}.15=& +\frac{1}{2} \displaystyle\sum_{\substack{k_1 k_2 k_3 k_4 \\ k_5}} \frac{\Omega^{11}_{k_1 k_2} \Omega^{04}_{k_3 k_4 k_1 k_5}}{E_{k_1}-E_{k_2}} \left[ \frac{1- e^{-\tau(E_{k_2}+E_{k_3}+E_{k_4}+E_{k_5})}}{E_{k_2}+E_{k_3}+E_{k_4}+E_{k_5}} \right.\nonumber \\
  & \hspace{6cm} \left.- \frac{1- e^{-\tau(E_{k_1}+E_{k_3}+E_{k_4}+E_{k_5})}}{E_{k_1}+E_{k_3}+E_{k_4}+E_{k_5}}\right] R^{--}_{k_2 k_5}(\varphi) R^{--}_{k_4 k_3}(\varphi) \\
  \text{PN}.16=& + \frac{1}{8} \!\!\!\displaystyle\sum_{\substack{k_1 k_2 k_3 k_4 \\ k_5 k_6}} \frac{\Omega^{22}_{k_1k_2k_3k_4} \Omega^{04}_{k_5k_6k_1k_2}}{E_{k_1}+E_{k_2}-E_{k_3}-E_{k_4}} \left[ \frac{1- e^{-\tau(E_{k_3}+E_{k_4}+E_{k_5}+E_{k_6})}}{E_{k_3}+E_{k_4}+E_{k_5}+E_{k_6}} \right.\nonumber \\
  & \hspace{7cm} \left.- \frac{1- e^{-\tau(E_{k_1}+E_{k_2}+E_{k_5}+E_{k_6})}}{E_{k_1}+E_{k_2}+E_{k_5}+E_{k_6}}\right] R^{--}_{k_3k_4}(\varphi) R^{--}_{k_5k_6}(\varphi) \\
  \text{PN}.17=& +\frac{1}{4} \displaystyle\sum_{\substack{k_1 k_2 k_3 k_4 \\ k_5 k_6 k_7}} \frac{\Omega^{13}_{k_1k_2k_3k_4} \Omega^{04}_{k_1k_5k_6k_7}}{E_{k_1}-E_{k_2}-E_{k_3}-E_{k_4}} \left[ \frac{1- e^{-\tau(E_{k_2}+E_{k_3}+E_{k_4}+E_{k_5}+E_{k_6}+E_{k_7})}}{E_{k_2}+E_{k_3}+E_{k_4}+E_{k_5}+E_{k_6}+E_{k_7}} \right.\nonumber \\
  & \hspace{6cm} \left.- \frac{1- e^{-\tau(E_{k_1}+E_{k_5}+E_{k_6}+E_{k_7})}}{E_{k_1}+E_{k_5}+E_{k_6}+E_{k_7}}\right] R^{--}_{k_2k_7}(\varphi)  R^{--}_{k_3k_4}(\varphi) R^{--}_{k_5k_6}(\varphi)  \\
  \text{PN}.18=& +\frac{1}{8} \displaystyle\sum_{\substack{k_1 k_2 k_3 k_4 \\ k_5 k_6 k_7 k_8}} \frac{\Omega^{04}_{k_1k_2k_3k_4} \Omega^{04}_{k_5k_6k_7k_8}}{(E_{k_1}+E_{k_2}+E_{k_3}+E_{k_4})(E_{k_5}+E_{k_6}+E_{k_7}+E_{k_8})} \left[1- e^{-\tau(E_{k_1}+E_{k_2}+E_{k_3}+E_{k_4})} \right] \nonumber \\
  & \hspace{4cm}  \times \left[1- e^{-\tau(E_{k_5}+E_{k_6}+E_{k_7}+E_{k_8})} \right] R^{--}_{k_1k_8}(\varphi)  R^{--}_{k_2k_7}(\varphi) R^{--}_{k_3k_4}(\varphi) R^{--}_{k_5k_6}(\varphi) \, , 
\end{align*}
\end{subequations}
which reduce in the infinite $\tau$ limit to
 \begin{subequations}
 \begin{align*}
  \text{PN}.1=&  +\frac{1}{2} \displaystyle\sum_{k_1 k_2} \frac{\Omega^{02}_{k_1 k_2}}{E_{k_1}+E_{k_2}} R^{--}_{k_1k_2}(\varphi) \\
  \text{PN}.2=& -\frac{1}{8} \displaystyle\sum_{k_1 k_2 k_3 k_4} \frac{\Omega^{04}_{k_1k_2k_3k_4}}{E_{k_1}+E_{k_2}+E_{k_3}+E_{k_4}} R^{--}_{k_1k_2}(\varphi) R^{--}_{k_3k_4}(\varphi) \\
  \text{PN}.3=& +\frac{1}{2} \displaystyle\sum_{k_1 k_2} \frac{\Omega^{02}_{k_1k_2}\Omega^{20}_{k_1k_2}}{E_{k_1}+E_{k_2}} \left[ \tau - \frac{1}{E_{k_1}+E_{k_2}} \right] \\
  \text{PN}.4=&  -\frac{1}{2} \displaystyle\sum_{k_1 k_2 k_3} \frac{\Omega^{02}_{k_1k_3}\Omega^{11}_{k_1k_2}}{(E_{k_1}+E_{k_2})(E_{k_2}+E_{k_3})} R^{--}_{k_2 k_3}(\varphi)\\
 \text{PN}.5=& +\frac{1}{2}\displaystyle\sum_{k_1k_2k_3k_4}\frac{\Omega^{02}_{k_1k_2}\Omega^{02}_{k_3k_4}}{(E_{k_1}+E_{k_2})(E_{k_3}+E_{k_4})} 
R^{--}_{k_1 k_4}(\varphi) R^{--}_{k_2 k_3}(\varphi) \\
  \text{PN}.6=& +\frac{1}{4!} \displaystyle\sum_{k_1 k_2 k_3 k_4} \frac{\Omega^{04}_{k_1k_2k_3k_4}\Omega^{40}_{k_1k_2k_3k_4}}{E_{k_1}+E_{k_2}+E_{k_3}+E_{k_4}} \left[ \tau - \frac{1}{E_{k_1}+E_{k_2}+E_{k_3}+E_{k_4}} \right] \\
  \text{PN}.7=& +\frac{1}{3!} \displaystyle\sum_{\substack{k_1 k_2 k_3 k_4 \\ k_5}} \frac{\Omega^{31}_{k_1k_2k_3k_4} \Omega^{04}_{k_5k_1k_2k_3}}{(E_{k_4}+E_{k_5})(E_{k_1}+E_{k_2}+E_{k_3}+E_{k_5})}R^{--}_{k_4k_5}(\varphi) \\
  \text{PN}.8=& + \frac{1}{4} \displaystyle\sum_{\substack{k_1 k_2 k_3 k_4 \\ k_5 k_6}} \frac{\Omega^{22}_{k_1k_2k_3k_4} \Omega^{04}_{k_5k_6k_1k_2}}{(E_{k_3}+E_{k_4}+E_{k_5}+E_{k_6})(E_{k_1}+E_{k_2}+E_{k_5}+E_{k_6})}  R^{--}_{k_3k_6}(\varphi) R^{--}_{k_4k_5}(\varphi) \\
  \text{PN}.9=& +\frac{1}{3!} \displaystyle\sum_{\substack{k_1 k_2 k_3 k_4 \\ k_5 k_6 k_7}} \frac{\Omega^{13}_{k_1k_2k_3k_4} \Omega^{04}_{k_1k_5k_6k_7}}{(E_{k_2}+E_{k_3}+E_{k_4}+E_{k_5}+E_{k_6}+E_{k_7})(E_{k_1}+E_{k_5}+E_{k_6}+E_{k_7})}  R^{--}_{k_2k_7}(\varphi)  R^{--}_{k_3k_6}(\varphi) R^{--}_{k_4k_5}(\varphi)  \\
  \text{PN}.10=& +\frac{1}{4!} \displaystyle\sum_{\substack{k_1 k_2 k_3 k_4 \\ k_5 k_6 k_7 k_8}} \frac{\Omega^{04}_{k_1k_2k_3k_4} \Omega^{04}_{k_5k_6k_7k_8}}{(E_{k_1}+E_{k_2}+E_{k_3}+E_{k_4})(E_{k_5}+E_{k_6}+E_{k_7}+E_{k_8})}  R^{--}_{k_1k_8}(\varphi)  R^{--}_{k_2k_7}(\varphi) R^{--}_{k_3k_6}(\varphi) R^{--}_{k_4k_5}(\varphi) \\
  \text{PN}.11=& +\frac{1}{4} \displaystyle\sum_{k_1 k_2 k_3 k_4} \frac{\Omega^{22}_{k_1k_2k_3k_4} \Omega^{02}_{k_1k_2}}
  {(E_{k_3}+E_{k_4})(E_{k_1}+E_{k_2})} R^{--}_{k_4k_3}(\varphi) \\
  \text{PN}.12=& +\frac{1}{2} \displaystyle\sum_{\substack{k_1 k_2 k_3 k_4 \\ k_5}} \frac{\Omega^{13}_{k_1 k_2 k_3 k_4} \Omega^{02}_{k_1 k_5}}{(E_{k_1}+E_{k_5})(E_{k_2}+E_{k_3}+E_{k_4}+E_{k_5})} R^{--}_{k_4k_5}(\varphi) R^{--}_{k_3k_2}(\varphi) \\
  \text{PN}.13=& +\frac{1}{4} \displaystyle\sum_{\substack{k_1 k_2 k_3 k_4 \\ k_5 k_6}} \frac{\Omega^{02}_{k_1 k_2} \Omega^{04}_{k_3 k_4 k_5 k_6}}{(E_{k_3}+E_{k_4}+E_{k_5}+E_{k_6})(E_{k_1}+E_{k_2}+E_{k_3}+E_{k_4}+E_{k_5}+E_{k_6})} R^{--}_{k_2 k_3}(\varphi) R^{--}_{k_1 k_4}(\varphi) R^{--}_{k_6 k_5}(\varphi) \\
  \text{PN}.14=& +\frac{1}{4} \displaystyle\sum_{k_1 k_2 k_3 k_4} \frac{\Omega^{20}_{k_1k_2} \Omega^{04}_{k_3 k_4 k_1 k_2}}{(E_{k_3}+E_{k_4})(E_{k_1}+E_{k_2}+E_{k_3}+E_{k_4})} R^{--}_{k_4k_3}(\varphi) \\
  \text{PN}.15=& +\frac{1}{2} \displaystyle\sum_{\substack{k_1 k_2 k_3 k_4 \\ k_5}} \frac{\Omega^{11}_{k_1 k_2} \Omega^{04}_{k_3 k_4 k_1 k_5}}{(E_{k_2}+E_{k_3}+E_{k_4}+E_{k_5})(E_{k_1}+E_{k_3}+E_{k_4}+E_{k_5})} R^{--}_{k_2 k_5}(\varphi) R^{--}_{k_4 k_3}(\varphi) \\
  \text{PN}.16=& + \frac{1}{8} \displaystyle\sum_{\substack{k_1 k_2 k_3 k_4 \\ k_5 k_6}} \frac{\Omega^{22}_{k_1k_2k_3k_4} \Omega^{04}_{k_5k_6k_1k_2}}{(E_{k_3}+E_{k_4}+E_{k_5}+E_{k_6})(E_{k_1}+E_{k_2}+E_{k_5}+E_{k_6})}  R^{--}_{k_3k_4}(\varphi) R^{--}_{k_5k_6}(\varphi)  \\
  \text{PN}.17=& +\frac{1}{4} \displaystyle\sum_{\substack{k_1 k_2 k_3 k_4 \\ k_5 k_6 k_7}} \frac{\Omega^{13}_{k_1k_2k_3k_4} \Omega^{04}_{k_1k_5k_6k_7}}{(E_{k_2}+E_{k_3}+E_{k_4}+E_{k_5}+E_{k_6}+E_{k_7})(E_{k_1}+E_{k_5}+E_{k_6}+E_{k_7})}  R^{--}_{k_2k_7}(\varphi)  R^{--}_{k_3k_4}(\varphi) R^{--}_{k_5k_6}(\varphi)  \\
  \text{PN}.18=& +\frac{1}{8} \displaystyle\sum_{\substack{k_1 k_2 k_3 k_4 \\ k_5 k_6 k_7 k_8}} \frac{\Omega^{04}_{k_1k_2k_3k_4} \Omega^{04}_{k_5k_6k_7k_8}}{(E_{k_1}+E_{k_2}+E_{k_3}+E_{k_4})(E_{k_5}+E_{k_6}+E_{k_7}+E_{k_8})}  R^{--}_{k_1k_8}(\varphi)  R^{--}_{k_2k_7}(\varphi) R^{--}_{k_3k_4}(\varphi) R^{--}_{k_5k_6}(\varphi) \, .
\end{align*}
\end{subequations}

\end{widetext}

\section{$\omega(\tau,\varphi)$ at BMBPT(1)}
\label{diagramsE}

Applying diagrammatic rules to the 20 zero- and first-order off-diagonal BMBPT connected/linked diagrams contributing to $\omega(\tau,\varphi)$ and displayed in Fig.~\ref{diagramsTL} gives
 \begin{widetext}
 \begin{subequations}
 \begin{align*}
 \text{PE}.1=& + \Omega^{00}  \, , \\
 \text{PE}.2=& + \frac{1}{2} \displaystyle\sum_{k_1 k_2} \Omega^{02}_{k_1 k_2} R^{--}_{k_2k_1}(\varphi) \, , \\
 \text{PE}.3=& + \frac{1}{8} \displaystyle\sum_{k_1 k_2 k_3 k_4} \Omega^{04}_{k_1 k_2 k_3 k_4} R^{--}_{k_2k_1}(\varphi) R^{--}_{k_4k_3}(\varphi)  \, , \\
 \text{PE}.4=& - \frac{1}{2} \displaystyle\sum_{k_1 k_2} \frac{\Omega^{02}_{k_1 k_2} \Omega^{20}_{k_1 k_2}}{E_{k_1}+E_{k_2}} \left[ 1- e^{-\tau(E_{k_1}+E_{k_2})}\right]  \, , \\
 \text{PE}.5=& -  \displaystyle\sum_{k_1 k_2 k_3} \frac{\Omega^{02}_{k_1 k_2} \Omega^{11}_{k_1 k_3}}{E_{k_1}+E_{k_2}} \left[ 1- e^{-\tau(E_{k_1}+E_{k_2})}\right] R^{--}_{k_2k_3}(\varphi)  \, , \\
 \text{PE}.6=& - \frac{1}{2} \displaystyle\sum_{k_1 k_2 k_3 k_4} \frac{\Omega^{02}_{k_1 k_2} \Omega^{02}_{k_3 k_4}}{E_{k_1}+E_{k_2}} \left[ 1- e^{-\tau(E_{k_1}+E_{k_2})}\right] R^{--}_{k_1k_4}(\varphi) R^{--}_{k_2k_3}(\varphi)  \, , \\
 \text{PE}.7=&  -\frac{1}{4!}  \displaystyle\sum_{k_1 k_2 k_3 k_4} \frac{\Omega^{04}_{k_1 k_2 k_3 k_4}\Omega^{40}_{k_1 k_2 k_3 k_4}}{E_{k_1}+E_{k_2}+E_{k_3}+E_{k_4}} \left[ 1- e^{-\tau(E_{k_1}+E_{k_2}+E_{k_3}+E_{k_4})}\right]  \, , \\
 \text{PE}.8=&  -\frac{1}{3!} \displaystyle\sum_{\substack{k_1 k_2 k_3 k_4 \\ k_5}} \frac{\Omega^{04}_{k_1k_2k_3k_4} \Omega^{31}_{k_1k_2k_3k_5} }{E_{k_1}+E_{k_2}+E_{k_3}+E_{k_4}} \left[ 1- e^{-\tau(E_{k_1}+E_{k_2}+E_{k_3}+E_{k_4})}\right] R^{--}_{k_4k_5}(\varphi) \, , \\
 \text{PE}.9=& -\frac{1}{4}  \displaystyle\sum_{\substack{k_1 k_2 k_3 k_4 \\ k_5 k_6}} \frac{\Omega^{04}_{k_1 k_2 k_3 k_4}\Omega^{22}_{k_1 k_2 k_5 k_6}}{E_{k_1}+E_{k_2}+E_{k_3}+E_{k_4}} \left[ 1- e^{-\tau(E_{k_1}+E_{k_2}+E_{k_3}+E_{k_4})}\right]  R^{--}_{k_3k_6}(\varphi) R^{--}_{k_4k_5}(\varphi) \, , \\
 \text{PE}.10=& -\frac{1}{3!}  \displaystyle\sum_{\substack{k_1 k_2 k_3 k_4 \\ k_5 k_6 k_7}} \frac{\Omega^{04}_{k_1 k_2 k_3 k_4}\Omega^{13}_{k_1 k_5 k_6 k_7}}{E_{k_1}+E_{k_2}+E_{k_3}+E_{k_4}}   \left[ 1- e^{-\tau(E_{k_1}+E_{k_2}+E_{k_3}+E_{k_4})}\right]  R^{--}_{k_2k_8}(\varphi) R^{--}_{k_3k_6}(\varphi)  R^{--}_{k_4k_5}(\varphi)  \, , \\
 \text{PE}.11=& -\frac{1}{4!}  \displaystyle\sum_{\substack{k_1 k_2 k_3 k_4 \\ k_5 k_6 k_7 k_8}} \frac{\Omega^{04}_{k_1 k_2 k_3 k_4}\Omega^{04}_{k_5 k_6 k_7 k_8}}{E_{k_1}+E_{k_2}+E_{k_3}+E_{k_4}}   \left[ 1- e^{-\tau(E_{k_1}+E_{k_2}+E_{k_3}+E_{k_4})}\right]  R^{--}_{k_1k_8}(\varphi) R^{--}_{k_2k_7}(\varphi)  R^{--}_{k_3k_6}(\varphi) R^{--}_{k_4k_5}(\varphi)   \, , \\
 \text{PE}.12=&  - \frac{1}{4} \displaystyle\sum_{k_1 k_2 k_3 k_4} \frac{\Omega^{02}_{k_1 k_2} \Omega^{22}_{k_1 k_2 k_3 k_4}}{E_{k_1}+E_{k_2}} \left[ 1- e^{-\tau(E_{k_1}+E_{k_2})}\right] R^{--}_{k_4k_3}(\varphi) \, , \\
 \text{PE}.13=&  - \frac{1}{2} \displaystyle\sum_{\substack{k_1 k_2 k_3 k_4 \\ k_5}} \frac{\Omega^{02}_{k_1 k_2} \Omega^{13}_{k_1 k_3 k_4 k_5}}{E_{k_1}+E_{k_2}} \left[ 1- e^{-\tau(E_{k_1}+E_{k_2})}\right] R^{--}_{k_2k_5}(\varphi)  R^{--}_{k_4k_3}(\varphi)  \, , \\
 \text{PE}.14=&  - \frac{1}{4} \displaystyle\sum_{\substack{k_1 k_2 k_3 k_4 \\ k_5 k_6}} \frac{\Omega^{02}_{k_1 k_2} \Omega^{04}_{k_3 k_4 k_5 k_6}}{E_{k_1}+E_{k_2}} \left[ 1- e^{-\tau(E_{k_1}+E_{k_2})}\right] R^{--}_{k_1k_6}(\varphi)  R^{--}_{k_2k_5}(\varphi)  R^{--}_{k_4k_3}(\varphi)  \, , \\
 \text{PE}.15=&  - \frac{1}{4} \displaystyle\sum_{k_1 k_2 k_3 k_4} \frac{\Omega^{04}_{k_1 k_2 k_3 k_4} \Omega^{20}_{k_1 k_2}}{E_{k_1}+E_{k_2}+E_{k_3}+E_{k_4}} \left[ 1- e^{-\tau(E_{k_1}+E_{k_2}+E_{k_3}+E_{k_4})}\right] R^{--}_{k_4k_3}(\varphi)  \, , \\
 \text{PE}.16=&  -\frac{1}{2} \displaystyle\sum_{\substack{k_1 k_2 k_3 k_4 \\ k_5}} \frac{\Omega^{04}_{k_1k_2k_3k_4} \Omega^{11}_{k_1k_5} }{E_{k_1}+E_{k_2}+E_{k_3}+E_{k_4}} \left[ 1- e^{-\tau(E_{k_1}+E_{k_2}+E_{k_3}+E_{k_4})}\right] R^{--}_{k_2k_5}(\varphi) R^{--}_{k_4k_3}(\varphi) \, , \\
 \text{PE}.17=&  -\frac{1}{4}  \displaystyle\sum_{\substack{k_1 k_2 k_3 k_4 \\ k_5 k_6}} \frac{\Omega^{04}_{k_1 k_2 k_3 k_4}\Omega^{02}_{k_5 k_6}}{E_{k_1}+E_{k_2}+E_{k_3}+E_{k_4}}   \left[ 1- e^{-\tau(E_{k_1}+E_{k_2}+E_{k_3}+E_{k_4})}\right]  R^{--}_{k_1k_6}(\varphi) R^{--}_{k_2k_5}(\varphi)  R^{--}_{k_4k_3}(\varphi) \, , \\
 \text{PE}.18=&  -\frac{1}{8}  \displaystyle\sum_{\substack{k_1 k_2 k_3 k_4 \\ k_5 k_6}} \frac{\Omega^{04}_{k_1 k_2 k_3 k_4}\Omega^{22}_{k_1 k_2 k_5 k_6}}{E_{k_1}+E_{k_2}+E_{k_3}+E_{k_4}} \left[ 1- e^{-\tau(E_{k_1}+E_{k_2}+E_{k_3}+E_{k_4})}\right]  R^{--}_{k_3k_4}(\varphi) R^{--}_{k_5k_6}(\varphi) \, , \\
 \text{PE}.19=&  -\frac{1}{4}  \displaystyle\sum_{\substack{k_1 k_2 k_3 k_4 \\ k_5 k_6 k_7}} \frac{\Omega^{04}_{k_1 k_2 k_3 k_4}\Omega^{13}_{k_1 k_5 k_6 k_7}}{E_{k_1}+E_{k_2}+E_{k_3}+E_{k_4}}   \left[ 1- e^{-\tau(E_{k_1}+E_{k_2}+E_{k_3}+E_{k_4})}\right]  R^{--}_{k_2k_8}(\varphi) R^{--}_{k_3k_4}(\varphi)  R^{--}_{k_5k_6}(\varphi)  \, , \\
 \text{PE}.20=& -\frac{1}{8}  \displaystyle\sum_{\substack{k_1 k_2 k_3 k_4 \\ k_5 k_6 k_7 k_8}} \frac{\Omega^{04}_{k_1 k_2 k_3 k_4}\Omega^{04}_{k_5 k_6 k_7 k_8}}{E_{k_1}+E_{k_2}+E_{k_3}+E_{k_4}}   \left[ 1- e^{-\tau(E_{k_1}+E_{k_2}+E_{k_3}+E_{k_4})}\right]  R^{--}_{k_1k_8}(\varphi) R^{--}_{k_2k_7}(\varphi)  R^{--}_{k_3k_4}(\varphi) R^{--}_{k_5k_6}(\varphi)    \, , 
\label{PE1}
\end{align*}
\end{subequations}
which reduce in the infinite $\tau$ limit to
\begin{subequations}
  \begin{align*}
 \text{PE}.1=& + \Omega^{00}  \, , \\
 \text{PE}.2=& + \frac{1}{2} \displaystyle\sum_{k_1 k_2} \Omega^{02}_{k_1 k_2} R^{--}_{k_2k_1}(\varphi)  \, , \\
 \text{PE}.3=& + \frac{1}{8} \displaystyle\sum_{k_1 k_2 k_3 k_4} \Omega^{04}_{k_1 k_2 k_3 k_4} R^{--}_{k_2k_1}(\varphi) R^{--}_{k_4k_3}(\varphi) \, , \\
 \text{PE}.4=& - \frac{1}{2} \displaystyle\sum_{k_1 k_2} \frac{\Omega^{02}_{k_1 k_2} \Omega^{20}_{k_1 k_2}}{E_{k_1}+E_{k_2}}   \, , \\
 \text{PE}.5=& -  \displaystyle\sum_{k_1 k_2 k_3} \frac{\Omega^{02}_{k_1 k_2} \Omega^{11}_{k_1 k_3}}{E_{k_1}+E_{k_2}} R^{--}_{k_2k_3}(\varphi) \, , \\
 \text{PE}.6=& - \frac{1}{2} \displaystyle\sum_{k_1 k_2 k_3 k_4} \frac{\Omega^{02}_{k_1 k_2} \Omega^{02}_{k_3 k_4}}{E_{k_1}+E_{k_2}}  R^{--}_{k_1k_4}(\varphi) R^{--}_{k_2k_3}(\varphi)  \, , \\
 \text{PE}.7=& -\frac{1}{4!}  \displaystyle\sum_{k_1 k_2 k_3 k_4} \frac{\Omega^{04}_{k_1 k_2 k_3 k_4}\Omega^{40}_{k_1 k_2 k_3 k_4}}{E_{k_1}+E_{k_2}+E_{k_3}+E_{k_4}}   \, , \\
 \text{PE}.8=&  -\frac{1}{3!} \displaystyle\sum_{\substack{k_1 k_2 k_3 k_4 \\ k_5}} \frac{\Omega^{04}_{k_1k_2k_3k_4} \Omega^{31}_{k_1k_2k_3k_5} }{E_{k_1}+E_{k_2}+E_{k_3}+E_{k_4}} R^{--}_{k_4k_5}(\varphi) \, , \\
 \text{PE}.9=&  -\frac{1}{4}  \displaystyle\sum_{\substack{k_1 k_2 k_3 k_4 \\ k_5 k_6}} \frac{\Omega^{04}_{k_1 k_2 k_3 k_4}\Omega^{22}_{k_1 k_2 k_5 k_6}}{E_{k_1}+E_{k_2}+E_{k_3}+E_{k_4}} R^{--}_{k_3k_6}(\varphi) R^{--}_{k_4k_5}(\varphi) \, , \\
 \text{PE}.10=&  -\frac{1}{3!}  \displaystyle\sum_{\substack{k_1 k_2 k_3 k_4 \\ k_5 k_6 k_7}} \frac{\Omega^{04}_{k_1 k_2 k_3 k_4}\Omega^{13}_{k_1 k_5 k_6 k_7}}{E_{k_1}+E_{k_2}+E_{k_3}+E_{k_4}}  R^{--}_{k_2k_8}(\varphi) R^{--}_{k_3k_6}(\varphi)  R^{--}_{k_4k_5}(\varphi) \, , \\
 \text{PE}.11=& -\frac{1}{4!}  \displaystyle\sum_{\substack{k_1 k_2 k_3 k_4 \\ k_5 k_6 k_7 k_8}} \frac{\Omega^{04}_{k_1 k_2 k_3 k_4}\Omega^{04}_{k_5 k_6 k_7 k_8}}{E_{k_1}+E_{k_2}+E_{k_3}+E_{k_4}}   R^{--}_{k_1k_8}(\varphi) R^{--}_{k_2k_7}(\varphi)  R^{--}_{k_3k_6}(\varphi) R^{--}_{k_4k_5}(\varphi)   \, , \\
 \text{PE}.12=& - \frac{1}{4} \displaystyle\sum_{k_1 k_2 k_3 k_4} \frac{\Omega^{02}_{k_1 k_2} \Omega^{22}_{k_1 k_2 k_3 k_4}}{E_{k_1}+E_{k_2}} R^{--}_{k_4k_3}(\varphi)  \, , \\
 \text{PE}.13=&  - \frac{1}{2} \displaystyle\sum_{\substack{k_1 k_2 k_3 k_4 \\ k_5}} \frac{\Omega^{02}_{k_1 k_2} \Omega^{13}_{k_1 k_3 k_4 k_5}}{E_{k_1}+E_{k_2}} R^{--}_{k_2k_5}(\varphi)  R^{--}_{k_4k_3}(\varphi)  \, , \\
 \text{PE}.14=&   - \frac{1}{4} \displaystyle\sum_{\substack{k_1 k_2 k_3 k_4 \\ k_5 k_6}} \frac{\Omega^{02}_{k_1 k_2} \Omega^{04}_{k_3 k_4 k_5 k_6}}{E_{k_1}+E_{k_2}} R^{--}_{k_1k_6}(\varphi)  R^{--}_{k_2k_5}(\varphi)  R^{--}_{k_4k_3}(\varphi)  \, , \\
 \text{PE}.15=&  - \frac{1}{4} \displaystyle\sum_{k_1 k_2 k_3 k_4} \frac{\Omega^{04}_{k_1 k_2 k_3 k_4} \Omega^{20}_{k_1 k_2}}{E_{k_1}+E_{k_2}+E_{k_3}+E_{k_4}}  R^{--}_{k_4k_3}(\varphi)   \, , \\
 \text{PE}.16=&  -\frac{1}{2} \displaystyle\sum_{\substack{k_1 k_2 k_3 k_4 \\ k_5}} \frac{\Omega^{04}_{k_1k_2k_3k_4} \Omega^{11}_{k_1k_5} }{E_{k_1}+E_{k_2}+E_{k_3}+E_{k_4}} R^{--}_{k_2k_5}(\varphi) R^{--}_{k_4k_3}(\varphi) \, , \\
 \text{PE}.17=& -\frac{1}{4}  \displaystyle\sum_{\substack{k_1 k_2 k_3 k_4 \\ k_5 k_6}} \frac{\Omega^{04}_{k_1 k_2 k_3 k_4}\Omega^{02}_{k_5 k_6}}{E_{k_1}+E_{k_2}+E_{k_3}+E_{k_4}}   R^{--}_{k_1k_6}(\varphi) R^{--}_{k_2k_5}(\varphi)  R^{--}_{k_4k_3}(\varphi)   \, , \\
 \text{PE}.18=&   -\frac{1}{8}  \displaystyle\sum_{\substack{k_1 k_2 k_3 k_4 \\ k_5 k_6}} \frac{\Omega^{04}_{k_1 k_2 k_3 k_4}\Omega^{22}_{k_1 k_2 k_5 k_6}}{E_{k_1}+E_{k_2}+E_{k_3}+E_{k_4}}  R^{--}_{k_3k_4}(\varphi) R^{--}_{k_5k_6}(\varphi) \, , \\
 \text{PE}.19=& -\frac{1}{4}  \displaystyle\sum_{\substack{k_1 k_2 k_3 k_4 \\ k_5 k_6 k_7}} \frac{\Omega^{04}_{k_1 k_2 k_3 k_4}\Omega^{13}_{k_1 k_5 k_6 k_7}}{E_{k_1}+E_{k_2}+E_{k_3}+E_{k_4}}   R^{--}_{k_2k_8}(\varphi) R^{--}_{k_3k_4}(\varphi)  R^{--}_{k_5k_6}(\varphi)  \, , \\
 \text{PE}.20=&  -\frac{1}{8}  \displaystyle\sum_{\substack{k_1 k_2 k_3 k_4 \\ k_5 k_6 k_7 k_8}} \frac{\Omega^{04}_{k_1 k_2 k_3 k_4}\Omega^{04}_{k_5 k_6 k_7 k_8}}{E_{k_1}+E_{k_2}+E_{k_3}+E_{k_4}}   R^{--}_{k_1k_8}(\varphi) R^{--}_{k_2k_7}(\varphi)  R^{--}_{k_3k_4}(\varphi) R^{--}_{k_5k_6}(\varphi)  \, .
\end{align*}
\end{subequations}
\end{widetext}

\section{$\omega(\tau,\varphi)$ from BCC}
\label{CCenergycontrib}

The algebraic expressions of the twenty off-diagonal BCC diagrams contributing to $\omega(\tau, \varphi)$ and displayed in Fig.~\ref{T1contribtokinetic} are
\begin{widetext}
\begin{subequations}
\label{variousEcontrib}
\begin{align*}
\text{E}.1=& +\Omega^{00} \, , \\
\text{E}.2=& + \frac{1}{2} \displaystyle\sum_{k_1 k_2} \Omega^{02}_{k_1 k_2} R^{--}_{k_2 k_1}(\varphi) \, , \\
\text{E}.3=& +\frac{1}{8} \displaystyle\sum_{k_1 k_2 k_3 k_4} \Omega^{04}_{k_1 k_2 k_3 k_4} R^{--}_{k_2 k_1}(\varphi)R^{--}_{k_4 k_3}(\varphi) \, , \\
\text{E}.4 =& + \frac{1}{2} \displaystyle\sum_{k_1 k_2} \mathcal{T}^{\dagger}_{k_1 k_2}(\tau,\varphi) \, \Omega^{20}_{k_1 k_2} \, , \\
\text{E}.5 =& +\displaystyle\sum_{k_1 k_2 k_3}  \mathcal{T}^{\dagger}_{k_1 k_2}(\tau,\varphi) \, \Omega^{11}_{k_1 k_3} R^{--}_{k_2 k_3} (\varphi) \, , \\
\text{E}.6 =& +\frac{1}{2} \displaystyle\sum_{k_1 k_2 k_3 k_4}  \mathcal{T}^{\dagger}_{k_1 k_2}(\tau,\varphi) \, \Omega^{02}_{k_3 k_4} R^{--}_{k_1 k_4} (\varphi) R^{--}_{k_2 k_3} (\varphi) \, , \\
\text{E}.7 =& + \frac{1}{4!} \displaystyle\sum_{k_1 k_2 k_3 k_4} \mathcal{T}^{\dagger}_{k_1 k_2 k_3 k_4}(\tau,\varphi) \, \Omega^{40}_{k_1 k_2 k_3 k_4} \, , \\
\text{E}.8 =& +\frac{1}{3!} \displaystyle\sum_{\substack{k_1 k_2 k_3 k_4 \\ k_5}} \mathcal{T}^{\dagger}_{k_1 k_2 k_3 k_4}(\tau,\varphi) \, \Omega^{31}_{k_1 k_2 k_3 k_5}
R^{--}_{k_4 k_5}(\varphi) \, , \\
\text{E}.9 =& + \frac{1}{4} \displaystyle\sum_{\substack{k_1 k_2 k_3 k_4 \\ k_5 k_6}} \mathcal{T}^{\dagger}_{k_1 k_2 k_3 k_4}(\tau,\varphi) \, \Omega^{22}_{k_1 k_2 k_5 k_6}
R^{--}_{k_4 k_5}(\varphi) R^{--}_{k_3 k_6}(\varphi) \, , \\
\text{E}.10 =& +\frac{1}{3!} \displaystyle\sum_{\substack{k_1 k_2 k_3 k_4\\ k_5 k_6 k_7}} \mathcal{T}^{\dagger}_{k_1 k_2 k_3 k_4}(\tau,\varphi) \, \Omega^{13}_{k_1 k_5 k_6 k_7} R^{--}_{k_4 k_5}(\varphi) R^{--}_{k_3 k_6}(\varphi) R^{--}_{k_2 k_7}(\varphi) \, , \\
\text{E}.11 =& +\frac{1}{4!} \displaystyle\sum_{\substack{k_1 k_2 k_3 k_4\\k_5 k_6 k_7 k_8}} \mathcal{T}^{\dagger}_{k_1 k_2 k_3 k_4}(\tau,\varphi) \, \Omega^{04}_{k_5 k_6 k_7 k_8} R^{--}_{k_4 k_5}(\varphi) R^{--}_{k_3 k_6}(\varphi)R^{--}_{k_2 k_7}(\varphi) R^{--}_{k_1 k_8}(\varphi) \, , \\
\text{E}.12 =& +\frac{1}{8} \displaystyle\sum_{k_1 k_2 k_3 k_4} \mathcal{T}^{\dagger}_{k_1 k_2}(\tau,\varphi) \mathcal{T}^{\dagger}_{k_3 k_4}(\tau,\varphi) \, \Omega^{40}_{k_1 k_2 k_3 k_4} \, , \\
\text{E}.13 =& +\frac{1}{2} \displaystyle\sum_{\substack{k_1 k_2 k_3 k_4 \\ k_5}} \mathcal{T}^{\dagger}_{k_1 k_2}(\tau,\varphi) \mathcal{T}^{\dagger}_{k_3 k_4}(\tau,\varphi) \, \Omega^{31}_{k_1 k_2 k_3 k_5} R^{--}_{k_4 k_5}(\varphi) \, , \\
\text{E}.14 =& +\frac{1}{4} \displaystyle\sum_{\substack{k_1 k_2 k_3 k_4 \\ k_5 k_6}} \mathcal{T}^{\dagger}_{k_1 k_2}(\tau,\varphi) \mathcal{T}^{\dagger}_{k_3 k_4}(\tau,\varphi) \, \Omega^{22}_{k_1 k_2 k_5 k_6} R^{--}_{k_4 k_5}(\varphi) R^{--}_{k_3 k_6}(\varphi) \, , \\
\text{E}.15 =& +\frac{1}{2} \displaystyle\sum_{\substack{k_1 k_2 k_3 k_4 \\ k_5 k_6}} \mathcal{T}^{\dagger}_{k_1 k_2}(\tau,\varphi) \mathcal{T}^{\dagger}_{k_3 k_4}(\tau,\varphi) \, \Omega^{22}_{k_1 k_3 k_5 k_6} R^{--}_{k_2 k_5}(\varphi) R^{--}_{k_4 k_6}(\varphi) \, , \\
\text{E}.16 =& +\frac{1}{2} \displaystyle\sum_{\substack{k_1 k_2 k_3 k_4\\ k_5 k_6 k_7}} \mathcal{T}^{\dagger}_{k_1 k_2}(\tau,\varphi) \mathcal{T}^{\dagger}_{k_3 k_4}(\tau,\varphi) \, \Omega^{13}_{k_1 k_5 k_6 k_7} R^{--}_{k_4 k_5}(\varphi) R^{--}_{k_3 k_6}(\varphi) R^{--}_{k_2 k_7}(\varphi) \, , \\
\text{E}.17 =& +\frac{1}{8} \displaystyle\sum_{\substack{k_1 k_2 k_3 k_4\\k_5 k_6 k_7 k_8}} \mathcal{T}^{\dagger}_{k_1 k_2}(\tau,\varphi) \mathcal{T}^{\dagger}_{k_3 k_4}(\tau,\varphi) \, \Omega^{04}_{k_5 k_6 k_7 k_8} R^{--}_{k_4 k_5}(\varphi) R^{--}_{k_3 k_6}(\varphi)R^{--}_{k_2 k_7}(\varphi) R^{--}_{k_1 k_8}(\varphi) \\
\text{E}.18 =& + \frac{1}{4} \displaystyle\sum_{k_1 k_2 k_3 k_4}  \mathcal{T}^{\dagger}_{k_1 k_2}(\tau,\varphi) \, \Omega^{22}_{k_1 k_2 k_3 k_4}
R^{--}_{k_4 k_3} (\varphi) \, , \\
\text{E}.19 =& +\frac{1}{2} \displaystyle\sum_{\substack{k_1 k_2 k_3 k_4 \\ k_5}}  \mathcal{T}^{\dagger}_{k_1 k_2}(\tau,\varphi) \, \Omega^{13}_{k_1 k_3 k_4 k_5}
 R^{--}_{k_2 k_5} (\varphi) R^{--}_{k_4 k_3} (\varphi) \, , \\
\text{E}.20 =&  +\frac{1}{4} \displaystyle\sum_{\substack{k_1 k_2 k_3 k_4 \\ k_5 k_6}}  \mathcal{T}^{\dagger}_{k_1 k_2}(\tau,\varphi) \, \Omega^{04}_{k_3 k_4 k_5 k_6}
 R^{--}_{k_1 k_6} (\varphi) R^{--}_{k_2 k_5} (\varphi) R^{--}_{k_4 k_3} (\varphi) \, .
\end{align*}
\end{subequations}
\end{widetext}
The infinite time limit of these expressions is simply obtained by replacing $\mathcal{T}^{\dagger}_{n}(\tau,\varphi)$ everywhere by $\mathcal{T}^{\dagger}_{n}(\varphi)$.

\section{Off-diagonal BCC amplitude equations}
\label{amplitudeequations}

Starting from Eq.~\ref{dynamicalkernels}, we derive the equations of motion (Eqs.~\ref{reduceddynamicalkernels} and~\ref{CCamplitudekernels}) satisfied by matrix elements of the off-diagonal cluster operators ${\cal T}^{\dagger}_{n}(\tau,\varphi)$. The derivations below are obtained by adapting to $n$-tuply excited grand-potential and norm kernels the steps taken in Secs.~\ref{energykernel} and~\ref{energykernelMBPTtoCC} for the non-excited grand potential kernel. Being formally similar, those steps are not detailed here. 

\subsection{Grand potential equation}

With ${\cal B}_{k_1 k_2 \ldots}=\bbone$, Eq.~\ref{dynamicalkernels} provides 
\begin{equation}
\Omega(\tau,\varphi) = -\partial_{\tau} N(\tau,\varphi) \, , \label{nonconnectedenergy}
\end{equation}
which expresses the off-diagonal grand-potential kernel as the (imaginary-)time derivative of the off-diagonal norm kernel.

\subsection{Single amplitude equation}
\label{singles}

Considering the operator ${\cal B}_{k_1 k_2}=\beta^{\dagger}_{k_1} \beta^{\dagger}_{k_2}$ that creates a two-quasiparticle (i.e. single) excitation, Eq.~\ref{dynamicalkernels} 
\begin{equation}
\Omega_{k_1 k_2}(\tau,\varphi) = -\partial_{\tau} N_{k_1 k_2}(\tau,\varphi) \, . \label{nonconnectedsingleamplitude}
\end{equation}

Let us start with $N_{k_1 k_2}(\tau,\varphi)$. Rewriting $| \Psi (\tau) \rangle$ in terms of ${\cal U}(\tau)$, expanding the latter through perturbation theory and applying off-diagonal Wick's theorem~\cite{balian69a}, one obtains the factorization of the singly-excited norm kernel as
\begin{equation}
N_{k_1 k_2}(\tau,\varphi) = n_{k_1 k_2}(\tau,\varphi) \, N(\tau,\varphi) \, , \label{factorizesinglenormkernel}
\end{equation}
where 
\begin{equation}
n_{k_1 k_2}(\tau,\varphi) \equiv \frac{\langle \Phi | {\cal U}(\tau)   {\cal B}_{k_1 k_2} | \Phi(\varphi) \rangle_{c}}{\langle \Phi |  \Phi(\varphi) \rangle} \,  \label{factorizesinglenormkernela}
\end{equation}
contains the complete set of BMBPT connected vacuum-to-vacuum diagrams {\it linked} to ${\cal B}_{k_1 k_2}$. In the next step, this complete set of diagrams can be rewritten as
\begin{subequations}
\label{singleconnectednormkernel}
\begin{eqnarray}
n_{k_1 k_2}(\tau,\varphi) &=& \frac{\langle \Phi | {\cal T}^{\dagger}_{1}(\tau,\varphi)   {\cal B}_{k_1 k_2} | \Phi(\varphi) \rangle_{c}}{\langle \Phi |  \Phi(\varphi) \rangle} \label{singleconnectednormkernel2} \\
&=& {\cal T}^{\dagger}_{k_1 k_2}(\tau,\varphi) \label{singleconnectednormkernel1} \\
&=& \frac{\langle \Phi | e^{{\cal T}^{\dagger}(\tau,\varphi)}   {\cal B}_{k_1 k_2} | \Phi(\varphi) \rangle_{c}}{\langle \Phi |  \Phi(\varphi) \rangle} \label{singleconnectednormkernel3} \, ,
\end{eqnarray}
\end{subequations}
where the rule is that no contraction is to be considered among cluster operators or within a cluster operator when expanding the exponential. Off-diagonal contractions within the operator ${\cal B}_{k_1 k_2 \ldots}$ are zero (Eq.~\ref{propagatorsB3}).  Expression~\ref{singleconnectednormkernel3} can be equated at no cost to Eq.~\ref{singleconnectednormkernel2} by virtue of the linked/connected character of the kernel.

Let us now come to $\Omega_{k_1 k_2}(\tau,\varphi)$. Because of the presence of two fixed-time operators ${\cal B}_{k_1 k_2}$ and $\Omega$ in the matrix elements, perturbation theory leads to the typical structure
\begin{equation}
\Omega_{k_1 k_2}(\tau,\varphi) = \omega_{k_1 k_2}(\tau,\varphi) \, N(\tau,\varphi)  + n_{k_1 k_2}(\tau,\varphi) \, \Omega(\tau,\varphi) \, . \label{factorizesingleenergykernel}
\end{equation}
In Eq.~\ref{factorizesingleenergykernel} was introduced the kernel
\begin{subequations}
\label{singleconnectedenergykernel}
\begin{eqnarray}
\omega_{k_1 k_2}(\tau,\varphi) &\equiv& \frac{\langle \Phi | {\cal U}(\tau) \Omega  {\cal B}_{k_1 k_2} | \Phi(\varphi) \rangle_{c}}{\langle \Phi |  \Phi(\varphi) \rangle} \label{singleconnectedenergykernel1} \\
&=& \frac{\langle \Phi | e^{{\cal T}^{\dagger}(\tau,\varphi)} \Omega  {\cal B}_{k_1 k_2} | \Phi(\varphi) \rangle_{c}}{\langle \Phi |  \Phi(\varphi) \rangle} \label{singleconnectedenergykernel2} \,  ,
\end{eqnarray}
\end{subequations}
where operators in the matrix element are all connected together by strings of contractions and where no contraction is to be considered among cluster operators or within a cluster operator. A crucial remark is here in order. In Eq.~\ref{singleconnectedenergykernel2}, the only cluster operator that could contract exclusively with the operators entering ${\cal B}_{k_1 k_2}$ is ${\cal T}^{\dagger}_{1}(\tau,\varphi)$. However, if it were to happen, the product ${\cal T}^{\dagger}_{1}(\tau,\varphi){\cal B}_{k_1 k_2}$ would be disconnected from $\Omega$ and from the other allowed ${\cal T}^{\dagger}_{n}(\tau,\varphi)$, which would contradict the fact that the matrix elements are connected, i.e. such contractions actually contribute to the second term on the right-hand side of Eq.~\ref{factorizesingleenergykernel}. Consequently, all allowed cluster operators are only partially contracted with ${\cal B}_{k_1 k_2}$ and are thus necessarily contracted with $\Omega$. Eventually, this is the actual meaning carried by the label $c$ in Eq.~\ref{singleconnectedenergykernel2}. This result allows us to recover the natural termination of the expanded exponential at play in standard BCC theory. 

Inserting Eqs.~\ref{factorizesinglenormkernel} and~\ref{factorizesingleenergykernel} into Eq.~\ref{nonconnectedsingleamplitude}, utilizing Eq.~\ref{singleconnectednormkernel1} and combining the result with Eq.~\ref{nonconnectedenergy} eventually leads to the single amplitude equation under the practical form of Eq.~\ref{reduceddynamicalkernels}, i.e. 
\begin{equation}
\omega_{k_1 k_2}(\tau,\varphi) = -\partial_{\tau} {\cal T}^{\dagger}_{k_1 k_2}(\tau,\varphi)  \, . \label{reducedsingleamplitudeequation}
\end{equation}

\subsection{Double amplitude equation}
\label{doubles}

Considering the operator ${\cal B}_{k_1 k_2 k_3 k_4}=\beta^{\dagger}_{k_1} \beta^{\dagger}_{k_2} \beta^{\dagger}_{k_3} \beta^{\dagger}_{k_4}$ that creates a four-quasiparticle (i.e. double) excitation, Eq.~\ref{dynamicalkernels} provides 
\begin{equation}
\Omega_{k_1 k_2 k_3 k_4}(\tau,\varphi) = -\partial_{\tau} N_{k_1 k_2 k_3 k_4}(\tau,\varphi) \, . \label{nonconnecteddoubleamplitude}
\end{equation}
Following the same steps as before, one first obtains
\begin{equation}
N_{k_1 k_2 k_3 k_4}(\tau,\varphi) = n_{k_1 k_2 k_3 k_4}(\tau,\varphi) \, N(\tau,\varphi) \, , \label{factorizedoublenormkernel}
\end{equation}
along with
\begin{widetext}
\begin{subequations}
\label{doubleconnectednormkernel}
\begin{eqnarray}
n_{k_1 k_2 k_3 k_4}(\tau,\varphi) &=& \frac{\langle \Phi | \Big[{\cal T}^{\dagger}_{2}(\tau,\varphi) +  \frac{1}{2} {\cal T}^{\dagger\, 2}_{1}(\tau,\varphi)\Big] {\cal B}_{k_1 k_2 k_3 k_4} | \Phi(\varphi) \rangle_{c}}{\langle \Phi |  \Phi(\varphi) \rangle} \label{doubleconnectednormkernel1} \\
&=& {\cal T}^{\dagger}_{k_1 k_2 k_3 k_4}(\tau,\varphi) + {\cal T}^{\dagger}_{k_1 k_2}(\tau,\varphi) \, {\cal T}^{\dagger}_{k_3 k_4}(\tau,\varphi) - {\cal T}^{\dagger}_{k_1 k_3}(\tau,\varphi) \, {\cal T}^{\dagger}_{k_2 k_4}(\tau,\varphi) +  {\cal T}^{\dagger}_{k_1 k_4}(\tau,\varphi) \, {\cal T}^{\dagger}_{k_2 k_3}(\tau,\varphi) \label{doubleconnectednormkernel2} \\
&=& \frac{\langle \Phi | e^{{\cal T}^{\dagger}(\tau,\varphi)}   {\cal B}_{k_1 k_2 k_3 k_4} | \Phi(\varphi) \rangle_{c}}{\langle \Phi |  \Phi(\varphi) \rangle} \label{doubleconnectednormkernel3} \, ,
\end{eqnarray}
\end{subequations}
where the same rules and explanations as before apply.
\end{widetext}

Coming to $\Omega_{k_1 k_2 k_3 k_4}(\tau,\varphi)$, perturbation theory leads once again to the typical structure
\begin{eqnarray}
\Omega_{k_1 k_2 k_3 k_4}(\tau,\varphi) &=& \frac{\langle \Phi | {\cal U}(\tau) \Omega  {\cal B}_{k_1 k_2 k_3 k_4} | \Phi(\varphi) \rangle_{c}}{\langle \Phi |  \Phi(\varphi) \rangle} \, N(\tau,\varphi)  \nonumber \\
&& + n_{k_1 k_2 k_3 k_4}(\tau,\varphi) \, \Omega(\tau,\varphi) \, , \label{factorizedoubleenergykernel}
\end{eqnarray}
where operators in the first kernel on the right-hand side are all connected together by strings of contractions. Following the same steps as before, one proceeds to the identification of the clusters, which leads to
\begin{eqnarray}
\frac{\langle \Phi | {\cal U}(\tau) \Omega  {\cal B}_{k_1 k_2 k_3 k_4} | \Phi(\varphi) \rangle_{c}}{\langle \Phi |  \Phi(\varphi) \rangle} &=& \frac{\langle \Phi | e^{{\cal T}^{\dagger}(\tau,\varphi)} \Omega  {\cal B}_{k_1 k_2 k_3 k_4} | \Phi(\varphi) \rangle_{c}}{\langle \Phi |  \Phi(\varphi) \rangle} \, , \nonumber \\
&& \label{doubleconnectedenergykernelintermediate}
\end{eqnarray}
where no contraction is to be considered among cluster operators or within a cluster operator.

Again, it is essential to detail the connected structure of this kernel. At this point, it can only be stated that the operators at play on the righthand side of Eq.~\ref{doubleconnectedenergykernelintermediate} are all connected together through strings of contractions by virtue of the connected character of the associated diagrams. A priori, this leaves the possibility that a cluster operator is solely, and thus entirely, connected to ${\cal B}_{k_1 k_2 k_3 k_4}$, i.e. that it is not connected to $\Omega$. In the present case, it can at most happen for ${\cal T}^{\dagger}_{1}(\tau,\varphi)$ or ${\cal T}^{\dagger}_{2}(\tau,\varphi)$. Contracting fully ${\cal T}^{\dagger}_{2}(\tau,\varphi)$ with ${\cal B}_{k_1 k_2 k_3 k_4}$ leaves no possibility for the latter to further connect to $\Omega$ and contradicts the fact that all the operators are connected together through strings of contractions, i.e. such a contribution is already included in the second term on the right-hand side of Eq.~\ref{factorizedoubleenergykernel}. As for ${\cal T}^{\dagger}_{1}(\tau,\varphi)$, the situation is more subtle. Let us thus consider contributions to Eq.~\ref{doubleconnectedenergykernelintermediate} where ${\cal T}^{\dagger}_{1}(\tau,\varphi)$ is fully contracted with ${\cal B}_{k_1 k_2 k_3 k_4}$. This leaves two quasi-particle creation operators originating from ${\cal B}_{k_1 k_2 k_3 k_4}$, i.e. a single-excitation operator ${\cal B}_{j_1 j_2}$ with $j_1, j_2$ to be chosen among the four $k_i$ indices to operate further contractions with $\Omega$. For each term with $p\geq 1$ powers of ${\cal T}^{\dagger}_{1}(\tau,\varphi)$ in the exponential, i.e. terms proportional  to ${\cal T}^{\dagger \, p}_{1}(\tau,\varphi)/p!$, there are $p$ possibilities to fully contract a ${\cal T}^{\dagger}_{1}(\tau,\varphi)$ operator with ${\cal B}_{k_1 k_2 k_3 k_4}$, which leaves ${\cal T}^{\dagger \, p\!-\!1}_{1}(\tau,\varphi)/(p\!-\!1)!$ for further contractions. Summing over all terms stemming from  the exponential, one can eventually re-factorize each time the full contribution of ${\cal T}^{\dagger}_{1}(\tau,\varphi)$ to the exponential. Performing the algebraic manipulations in details, one eventually arrives at
\begin{eqnarray}
\frac{\langle \Phi | e^{{\cal T}^{\dagger}(\tau,\varphi)} \Omega  {\cal B}_{k_1 k_2 k_3 k_4} | \Phi(\varphi) \rangle_{c}}{\langle \Phi |  \Phi(\varphi) \rangle} &=&  \omega_{k_1 k_2 k_3 k_4}(\tau,\varphi) \label{doubleconnectedenergykernel}  \\
&& + \omega_{k_1 k_2}(\tau,\varphi) \, {\cal T}^{\dagger}_{k_3 k_4}(\tau,\varphi) \nonumber \\
&& - \omega_{k_1 k_3}(\tau,\varphi) \, {\cal T}^{\dagger}_{k_2 k_4}(\tau,\varphi) \nonumber \\
&& + \omega_{k_1 k_4}(\tau,\varphi) \, {\cal T}^{\dagger}_{k_2 k_3}(\tau,\varphi) \nonumber \\
&& + \omega_{k_2 k_3}(\tau,\varphi) \, {\cal T}^{\dagger}_{k_1 k_4}(\tau,\varphi) \nonumber \\
&& - \omega_{k_2 k_4}(\tau,\varphi) \, {\cal T}^{\dagger}_{k_1 k_3}(\tau,\varphi) \nonumber \\
&& + \omega_{k_3 k_4}(\tau,\varphi) \, {\cal T}^{\dagger}_{k_1 k_2}(\tau,\varphi) \nonumber \, , 
\end{eqnarray}
where $\omega_{k_1 k_2 k_3 k_4}(\tau,\varphi)$ denotes the contributions to the matrix elements where all cluster operators are {\it necessarily} contracted with $\Omega$, which ultimately leads to the usual termination of the exponential. The last six terms in Eq.~\ref{doubleconnectedenergykernel} gather all the contributions where a ${\cal T}^{\dagger}_{1}(\tau,\varphi)$ was fully contracted to ${\cal B}_{k_1 k_2 k_3 k_4}$.

To eventually obtain the practical form of the double amplitude equation (Eq.~\ref{reduceddynamicalkernels}), one needs not only to insert  Eqs.~\ref{doubleconnectednormkernel2} and~\ref{doubleconnectedenergykernel} into Eq.~\ref{nonconnecteddoubleamplitude}, but one must also invoke the single amplitude equation (Eq.~\ref{reducedsingleamplitudeequation}) along with the norm equation (Eq.~\ref{nonconnectedenergy}). In doing so, one finally arrives at the equation of motion for double amplitudes under the desired form
\begin{equation}
\omega_{k_1 k_2 k_3 k_4}(\tau,\varphi) = -\partial_{\tau} {\cal T}^{\dagger}_{k_1 k_2 k_3 k_4}(\tau,\varphi)  \, . \label{reduceddoubleamplitudeequation}
\end{equation}

\subsection{$n$-tuple amplitude equation}
\label{ntuple}

As for single and double amplitude equations, the derivation of the $n$-tuple amplitude equation invokes all the amplitude equations of lower rank. Reasoning by recurrence, one can prove Eq.~\ref{reduceddynamicalkernels} for any $n$-tuply excited amplitude.

\section{BCCSD contributions to $\omega_{k_1k_2}(\tau, \varphi)$}
\label{CCsinglecontrib}

The fifty-six off-diagonal BCCSD diagrams contributing to $\omega_{k_1k_2}(\tau, \varphi)$ are given by
 \begin{widetext}
 \begin{eqnarray*}
 \text{S}.1&=& + \Omega^{02}_{\alpha \beta}  \, , \\
 \text{S}.2&=& + \frac{1}{2} \sum_{k_1 k_2} \Omega^{04}_{\alpha \beta k_1 k_2} R^{--}_{k_2 k_1}(\varphi)   \, , \\
 \text{S}.3&=& + \frac{1}{2} \displaystyle\sum_{k_1 k_2} \mathcal{T}^{\dagger}_{\alpha \beta k_1 k_2}(\tau,\varphi) \Omega^{20}_{k_1 k_2} \, , \\
\text{S}.4 &=&  + \frac{1}{2} P(\alpha / \beta) \displaystyle\sum_{k_1 k_2} \mathcal{T}^{\dagger}_{\alpha k_1}(\tau,\varphi)
 \mathcal{T}^{\dagger}_{k_2 \beta}(\tau,\varphi) \Omega^{20}_{k_1 k_2} \, , \\
  \text{S}.5&=& + P(\alpha / \beta) \displaystyle\sum_{k_1} \mathcal{T}^{\dagger}_{\alpha k_1}(\tau,\varphi) \Omega^{11}_{k_1 \beta} \, , \\
 \text{S}.6&=& +\displaystyle\sum_{k_1 k_2 k_3} \mathcal{T}^{\dagger}_{\alpha \beta k_1 k_2}(\tau,\varphi) \Omega^{11}_{k_1 k_3} R^{--}_{k_2 k_3}(\varphi) \, , \\
 \text{S}.7&=& - P(\alpha / \beta) \displaystyle\sum_{k_1 k_2 k_3} \mathcal{T}^{\dagger}_{\alpha k_1}(\tau,\varphi) 
\mathcal{T}^{\dagger}_{k_3 \beta}(\tau,\varphi) \Omega^{11}_{k_3 k_2} R^{--}_{k_1 k_2}(\varphi)  \, , \\
 \text{S}.8&=& +P(\alpha / \beta) \displaystyle\sum_{k_1 k_2} \mathcal{T}^{\dagger}_{\alpha k_1}(\tau,\varphi) 
\Omega^{02}_{\beta k_2} R^{--}_{k_1 k_2}(\varphi) \, , \\
 \text{S}.9&=& +\frac{1}{2} \displaystyle\sum_{k_1 k_2 k_3 k_4} \mathcal{T}^{\dagger}_{\alpha \beta k_1 k_2}(\tau,\varphi) \Omega^{02}_{k_4 k_3} R^{--}_{k_1 k_3}(\varphi) R^{--}_{k_2 k_4}(\varphi) \, , \\
 \text{S}.10&=& +\frac{1}{2} P(\alpha / \beta) \displaystyle\sum_{k_1 k_2 k_3 k_4} \mathcal{T}^{\dagger}_{\alpha k_1}(\tau,\varphi) 
\mathcal{T}^{\dagger}_{k_2 \beta}(\tau,\varphi) \Omega^{02}_{k_4 k_3} R^{--}_{k_1 k_3}(\varphi) R^{--}_{k_2 k_4}(\varphi) \, , \\
 \text{S}.11&=& +\frac{1}{4} \displaystyle\sum_{k_1 k_2 k_3 k_4} \mathcal{T}^{\dagger}_{\alpha \beta k_1 k_2}(\tau,\varphi)
 \mathcal{T}^{\dagger}_{k_3 k_4}(\tau,\varphi) \Omega^{40}_{k_1 k_2 k_3 k_4} \, , \\
 \text{S}.12&=& +\frac{1}{3!} P(\alpha / \beta) \displaystyle\sum_{k_1 k_2 k_3 k_4} \mathcal{T}^{\dagger}_{\alpha k_1 k_2 k_3}(\tau,\varphi)
 \mathcal{T}^{\dagger}_{k_4 \beta}(\tau,\varphi) \Omega^{40}_{k_1 k_2 k_3 k_4} \, , \\
 \text{S}.13&=& +\frac{1}{4} P(\alpha / \beta) \displaystyle\sum_{k_1 k_2 k_3 k_4} \mathcal{T}^{\dagger}_{\alpha k_1}(\tau,\varphi) 
 \mathcal{T}^{\dagger}_{k_2 k_3}(\tau,\varphi) \mathcal{T}^{\dagger}_{k_4 \beta}(\tau,\varphi) \Omega^{40}_{k_1 k_2 k_3 k_4} \, , \\
 \text{S}.14&=& +\frac{1}{6} P(\alpha / \beta) \displaystyle\sum_{k_1 k_2 k_3} \mathcal{T}^{\dagger}_{\alpha k_1 k_2 k_3}(\tau,\varphi)
 \Omega^{31}_{k_1 k_2 k_3 \beta} \, , \\
 \text{S}.15&=& +\frac{1}{2} P(\alpha / \beta) \displaystyle\sum_{k_1 k_2 k_3} \mathcal{T}^{\dagger}_{\alpha k_1}(\tau,\varphi)
 \mathcal{T}^{\dagger}_{k_2 k_3}(\tau,\varphi) \Omega^{31}_{k_1 k_2 k_3 \beta} \, , \\
 \text{S}.16&=& +\frac{1}{2} \displaystyle\sum_{k_1 k_2 k_3 k_4 k_5} \mathcal{T}^{\dagger}_{\alpha \beta k_1 k_2}(\tau,\varphi)
 \mathcal{T}^{\dagger}_{k_3 k_4}(\tau,\varphi) \Omega^{31}_{k_1 k_2 k_3 k_5} R^{--}_{k_4 k_5}(\varphi) \, , \\
 \text{S}.17&=& +\frac{1}{2} \displaystyle\sum_{k_1 k_2 k_3 k_4 k_5} \mathcal{T}^{\dagger}_{\alpha \beta k_1 k_2}(\tau,\varphi)
 \mathcal{T}^{\dagger}_{k_3 k_4}(\tau,\varphi) \Omega^{31}_{k_1 k_3 k_4 k_5} R^{--}_{k_2 k_5}(\varphi) \, , \\
 \text{S}.18&=& +\frac{1}{3!} P(\alpha / \beta) \displaystyle\sum_{k_1 k_2 k_3 k_4 k_5} \mathcal{T}^{\dagger}_{\alpha k_1 k_2 k_3}(\tau,\varphi)
 \mathcal{T}^{\dagger}_{k_4 \beta}(\tau,\varphi) \Omega^{31}_{k_1 k_2 k_3 k_5} R^{--}_{k_4 k_5}(\varphi) \, , \\
 \text{S}.19&=& +\frac{1}{2} P(\alpha / \beta) \displaystyle\sum_{k_1 k_2 k_3 k_4 k_5} \mathcal{T}^{\dagger}_{\alpha k_1 k_2 k_3}(\tau,\varphi)
 \mathcal{T}^{\dagger}_{k_4 \beta}(\tau,\varphi) \Omega^{31}_{k_1 k_3 k_4 k_5} R^{--}_{k_2 k_5}(\varphi) \, , \\
 \text{S}.20&=& +\frac{1}{2} P(\alpha / \beta) \displaystyle\sum_{k_1 k_2 k_3 k_4 k_5} \mathcal{T}^{\dagger}_{\alpha k_1}(\tau, \varphi)
 \mathcal{T}^{\dagger}_{k_2 k_3}(\tau,\varphi) \mathcal{T}^{\dagger}_{k_4 \beta}(\tau,\varphi) \Omega^{31}_{k_1 k_2 k_3 k_5} R^{--}_{k_4 k_5}(\varphi) \, , \\
 \text{S}.21&=& +\frac{1}{2} P(\alpha / \beta) \displaystyle\sum_{k_1 k_2 k_3 k_4 k_5} \mathcal{T}^{\dagger}_{\alpha k_1}(\tau, \varphi)
 \mathcal{T}^{\dagger}_{k_2 k_3}(\tau,\varphi) \mathcal{T}^{\dagger}_{k_4 \beta}(\tau,\varphi) \Omega^{31}_{k_1 k_3 k_4 k_5} R^{--}_{k_2 k_5}(\varphi) \, , \\
 \text{S}.22&=& +\frac{1}{2} \displaystyle\sum_{k_1 k_2} \mathcal{T}^{\dagger}_{k_1 k_2}(\tau,\varphi) \Omega^{22}_{k_1 k_2 \alpha \beta} \, , \\
 \text{S}.23&=& +\frac{1}{4} \displaystyle\sum_{k_1 k_2 k_3 k_4} \mathcal{T}^{\dagger}_{\alpha \beta k_1 k_2}(\tau,\varphi)
\Omega^{22}_{k_1 k_2 k_3 k_4} R^{--}_{k_4 k_3} (\varphi) \, , \\
 \text{S}.24&=& +\frac{1}{2} P(\alpha / \beta) \displaystyle\sum_{k_1 k_2 k_3 k_4} \mathcal{T}^{\dagger}_{\alpha k_1 k_2 k_3}(\tau,\varphi) 
\Omega^{22}_{k_1 k_2 \beta k_4}R^{--}_{k_3 k_4}(\varphi) \, , \\
 \text{S}.25&=& +\frac{1}{4} P(\alpha / \beta) \displaystyle\sum_{k_1 k_2 k_3 k_4} \mathcal{T}^{\dagger}_{\alpha k_1}(\tau,\varphi) 
 \mathcal{T}^{\dagger}_{k_2 \beta}(\tau,\varphi) \Omega^{22}_{k_1 k_2 k_3 k_4} R^{--}_{k_4 k_3}(\varphi) \, , \\
 \text{S}.26&=& +P(\alpha / \beta) \displaystyle\sum_{k_1 k_2 k_3 k_4} \mathcal{T}^{\dagger}_{\alpha k_1}(\tau,\varphi) 
 \mathcal{T}^{\dagger}_{k_2 k_3}(\tau,\varphi) \Omega^{22}_{k_1 k_2 \beta k_4} R^{--}_{k_3 k_4}(\varphi) \, , \\
 \text{S}.27&=& +\frac{1}{2} P(\alpha / \beta) \displaystyle\sum_{k_1 k_2 k_3 k_4} \mathcal{T}^{\dagger}_{\alpha k_1}(\tau,\varphi) 
 \mathcal{T}^{\dagger}_{k_2 k_3}(\tau,\varphi) \Omega^{22}_{k_2 k_3 \beta k_4} R^{--}_{k_1 k_4}(\varphi) \, , \\
 \text{S}.28&=& +\frac{1}{4}  \displaystyle\sum_{\substack{k_1 k_2 k_3 k_4 \\ k_5 k_6}} \mathcal{T}^{\dagger}_{\alpha \beta k_1 k_2}(\tau,\varphi)
 \mathcal{T}^{\dagger}_{k_3 k_4}(\tau,\varphi) \Omega^{22}_{k_1 k_2 k_5 k_6} R^{--}_{k_3 k_6}(\varphi) R^{--}_{k_4 k_5}(\varphi) \, , \\
 \text{S}.29&=& +\frac{1}{4}  \displaystyle\sum_{\substack{k_1 k_2 k_3 k_4 \\ k_5 k_6}} \mathcal{T}^{\dagger}_{\alpha \beta k_1 k_2}(\tau,\varphi)
 \mathcal{T}^{\dagger}_{k_3 k_4}(\tau,\varphi) \Omega^{22}_{k_3 k_4 k_5 k_6} R^{--}_{k_1 k_6}(\varphi) R^{--}_{k_2 k_5}(\varphi) \, , \\
 \text{S}.30&=&  +\displaystyle\sum_{\substack{k_1 k_2 k_3 k_4 \\ k_5 k_6}} \mathcal{T}^{\dagger}_{\alpha \beta k_1 k_2}(\tau,\varphi)
 \mathcal{T}^{\dagger}_{k_3 k_4}(\tau,\varphi) \Omega^{22}_{k_1 k_4 k_5 k_6} R^{--}_{k_2 k_6}(\varphi) R^{--}_{k_3 k_5}(\varphi) \, , \\
 \text{S}.31&=& +\frac{1}{2} P(\alpha / \beta)  \displaystyle\sum_{\substack{k_1 k_2 k_3 k_4 \\ k_5 k_6}} \mathcal{T}^{\dagger}_{\alpha k_1 k_2 k_3}(\tau,\varphi)
 \mathcal{T}^{\dagger}_{k_4 \beta}(\tau,\varphi) \Omega^{22}_{k_1 k_2 k_5 k_6} R^{--}_{k_3 k_6}(\varphi) R^{--}_{k_4 k_5}(\varphi) \, , \\
 \text{S}.32&=& +\frac{1}{2} P(\alpha / \beta)  \displaystyle\sum_{\substack{k_1 k_2 k_3 k_4 \\ k_5 k_6}} \mathcal{T}^{\dagger}_{\alpha k_1 k_2 k_3}(\tau,\varphi)
 \mathcal{T}^{\dagger}_{k_4 \beta}(\tau,\varphi) \Omega^{22}_{k_1 k_4 k_5 k_6} R^{--}_{k_2 k_6}(\varphi) R^{--}_{k_3 k_5}(\varphi) \, , \\
 \text{S}.33&=& +P(\alpha / \beta)  \displaystyle\sum_{\substack{k_1 k_2 k_3 k_4 \\ k_5 k_6}} \mathcal{T}^{\dagger}_{\alpha k_1}(\tau, \varphi)
 \mathcal{T}^{\dagger}_{k_2 k_3}(\tau,\varphi) \mathcal{T}^{\dagger}_{k_4 \beta}(\tau,\varphi) \Omega^{22}_{k_1 k_2 k_5 k_6} 
 R^{--}_{k_3 k_6}(\varphi) R^{--}_{k_4 k_5}(\varphi) \, , \\
 \text{S}.34&=& +\frac{1}{4} P(\alpha / \beta)  \displaystyle\sum_{\substack{k_1 k_2 k_3 k_4 \\ k_5 k_6}} \mathcal{T}^{\dagger}_{\alpha k_1}(\tau, \varphi)
 \mathcal{T}^{\dagger}_{k_2 k_3}(\tau,\varphi) \mathcal{T}^{\dagger}_{k_4 \beta}(\tau,\varphi) \Omega^{22}_{k_1 k_4 k_5 k_6} 
 R^{--}_{k_2 k_6}(\varphi) R^{--}_{k_3 k_5}(\varphi) \, , \\
 \text{S}.35&=& +\frac{1}{4} P(\alpha / \beta)  \displaystyle\sum_{\substack{k_1 k_2 k_3 k_4 \\ k_5 k_6}} \mathcal{T}^{\dagger}_{\alpha k_1}(\tau, \varphi)
 \mathcal{T}^{\dagger}_{k_2 k_3}(\tau,\varphi) \mathcal{T}^{\dagger}_{k_4 \beta}(\tau,\varphi) \Omega^{22}_{k_2 k_3 k_5 k_6} 
 R^{--}_{k_1 k_6}(\varphi) R^{--}_{k_4 k_5}(\varphi) \, , \\
 \text{S}.36&=& +\frac{1}{2} P(\alpha / \beta) \displaystyle\sum_{k_1 k_2 k_3} \mathcal{T}^{\dagger}_{\alpha k_1}(\tau,\varphi) 
 \Omega^{13}_{k_1 \beta k_2 k_3} R^{--}_{k_3 k_2}(\varphi) \, , \\
 \text{S}.37&=& +\displaystyle\sum_{k_1 k_2 k_3} \mathcal{T}^{\dagger}_{k_1 k_2}(\tau,\varphi) \Omega^{13}_{k_1 \alpha \beta k_3} R^{--}_{k_2 k_3}(\varphi) \, , \\
 \text{S}.38&=& +\frac{1}{2} \displaystyle\sum_{k_1 k_2 k_3 k_4 k_5} \mathcal{T}^{\dagger}_{\alpha \beta k_1 k_2}(\tau,\varphi)
\Omega^{13}_{k_1 k_3 k_4 k_5} R^{--}_{k_2 k_5}(\varphi) R^{--}_{k_4 k_3}(\varphi) \, , \\
 \text{S}.39&=& +\frac{1}{2} P(\alpha / \beta) \displaystyle\sum_{k_1 k_2 k_3 k_4 k_5} \mathcal{T}^{\dagger}_{\alpha k_1 k_2 k_3}(\tau,\varphi) 
\Omega^{13}_{k_1 \beta k_4 k_5} R^{--}_{k_2 k_5}(\varphi) R^{--}_{k_3 k_4}(\varphi) \, , \\
 \text{S}.40&=& +\frac{1}{2} P(\alpha / \beta) \displaystyle\sum_{k_1 k_2 k_3 k_4 k_5} \mathcal{T}^{\dagger}_{\alpha k_1}(\tau,\varphi) 
 \mathcal{T}^{\dagger}_{k_2 \beta}(\tau,\varphi) \Omega^{13}_{k_1 k_3 k_4 k_5} R^{--}_{k_2 k_5}(\varphi) R^{--}_{k_4 k_3}(\varphi) \, , \\
 \text{S}.41&=& +\frac{1}{2} P(\alpha / \beta) \displaystyle\sum_{k_1 k_2 k_3 k_4 k_5} \mathcal{T}^{\dagger}_{\alpha k_1}(\tau,\varphi) 
 \mathcal{T}^{\dagger}_{k_2 k_3}(\tau,\varphi) \Omega^{13}_{k_1 \beta k_4 k_5} R^{--}_{k_2 k_5}(\varphi) R^{--}_{k_3 k_4}(\varphi) \, , \\
 \text{S}.42&=& - P(\alpha / \beta) \displaystyle\sum_{k_1 k_2 k_3 k_4 k_5} \mathcal{T}^{\dagger}_{\alpha k_1}(\tau,\varphi) 
 \mathcal{T}^{\dagger}_{k_2 k_3}(\tau,\varphi) \Omega^{13}_{k_2 \beta k_4 k_5} R^{--}_{k_1 k_5}(\varphi) R^{--}_{k_3 k_4}(\varphi) \, , \\
 \text{S}.43&=& +\frac{1}{2} \displaystyle\sum_{\substack{k_1 k_2 k_3 k_4 \\ k_5 k_6 k_7}} \mathcal{T}^{\dagger}_{\alpha \beta k_1 k_2}(\tau,\varphi)
 \mathcal{T}^{\dagger}_{k_3 k_4}(\tau,\varphi) \Omega^{13}_{k_1 k_5 k_6 k_7} R^{--}_{k_2 k_7}(\varphi) R^{--}_{k_3 k_6}(\varphi) R^{--}_{k_4 k_5}(\varphi) \, , \\
 \text{S}.44&=& +\frac{1}{2} \displaystyle\sum_{\substack{k_1 k_2 k_3 k_4 \\ k_5 k_6 k_7}} \mathcal{T}^{\dagger}_{\alpha \beta k_1 k_2}(\tau,\varphi)
 \mathcal{T}^{\dagger}_{k_3 k_4}(\tau,\varphi) \Omega^{13}_{k_3 k_5 k_6 k_7} R^{--}_{k_1 k_7}(\varphi) R^{--}_{k_2 k_6}(\varphi)R^{--}_{k_4 k_5}(\varphi) \, , \\
 \text{S}.45&=& +\frac{1}{2} P(\alpha / \beta)  \displaystyle\sum_{\substack{k_1 k_2 k_3 k_4 \\ k_5 k_6 k_7}} 
 \mathcal{T}^{\dagger}_{\alpha k_1 k_2 k_3}(\tau,\varphi) \mathcal{T}^{\dagger}_{k_4 \beta}(\tau,\varphi) 
 \Omega^{13}_{k_1 k_5 k_6 k_7} R^{--}_{k_2 k_7}(\varphi) R^{--}_{k_3 k_6}(\varphi) R^{--}_{k_4 k_5}(\varphi) \, , \\
 \text{S}.46&=& - \frac{1}{3!} P(\alpha / \beta)  \displaystyle\sum_{\substack{k_1 k_2 k_3 k_4 \\ k_5 k_6 k_7}} 
 \mathcal{T}^{\dagger}_{\alpha k_1 k_2 k_3}(\tau,\varphi) \mathcal{T}^{\dagger}_{k_4 \beta}(\tau,\varphi) 
 \Omega^{13}_{k_4 k_5 k_6 k_7} R^{--}_{k_2 k_7}(\varphi) R^{--}_{k_3 k_6}(\varphi) R^{--}_{k_4 k_5}(\varphi) \, , \\
 \text{S}.47&=& +\frac{1}{2} P(\alpha / \beta) \displaystyle\sum_{\substack{k_1 k_2 k_3 k_4 \\ k_5 k_6 k_7}}
  \mathcal{T}^{\dagger}_{\alpha k_1}(\tau, \varphi) \mathcal{T}^{\dagger}_{k_2 k_3}(\tau,\varphi) \mathcal{T}^{\dagger}_{k_4 \beta}(\tau,\varphi)
 \Omega^{13}_{k_1 k_5 k_6 k_7} R^{--}_{k_2 k_7}(\varphi) R^{--}_{k_3 k_6}(\varphi) R^{--}_{k_4 k_5}(\varphi) \, , \\
 \text{S}.48&=& +\frac{1}{2} P(\alpha / \beta) \displaystyle\sum_{\substack{k_1 k_2 k_3 k_4 \\ k_5 k_6 k_7}}
 \mathcal{T}^{\dagger}_{\alpha k_1}(\tau, \varphi) \mathcal{T}^{\dagger}_{k_2 k_3}(\tau,\varphi) \mathcal{T}^{\dagger}_{k_4 \beta}(\tau,\varphi)
\Omega^{13}_{k_3 k_5 k_6 k_7} R^{--}_{k_1 k_7}(\varphi) R^{--}_{k_2 k_6}(\varphi) R^{--}_{k_4 k_5}(\varphi) \, , \\
 \text{S}.49&=& +\frac{1}{2} P(\alpha / \beta) \displaystyle\sum_{k_1 k_2 k_3 k_4} \mathcal{T}^{\dagger}_{\alpha k_1}(\tau,\varphi) 
 \Omega^{04}_{\beta k_2 k_3 k_4} R^{--}_{k_1 k_4}(\varphi) R^{--}_{k_3 k_2}(\varphi) \, , \\
 \text{S}.50&=& +\frac{1}{2} \displaystyle\sum_{k_1 k_2 k_3 k_4} \mathcal{T}^{\dagger}_{k_1 k_2}(\tau,\varphi) 
 \Omega^{04}_{\alpha \beta k_3 k_4} R^{--}_{k_1 k_4}(\varphi) R^{--}_{k_2 k_3}(\varphi) \, , \\
 \text{S}.51&=& +\frac{1}{4} \displaystyle\sum_{\substack{k_1 k_2 k_3 k_4 \\ k_5 k_6}} \mathcal{T}^{\dagger}_{\alpha \beta k_1 k_2}(\tau,\varphi)
\Omega^{04}_{k_3 k_4 k_5 k_6} R^{--}_{k_1 k_6}(\varphi) R^{--}_{k_2 k_5}(\varphi) R^{--}_{k_4 k_3}(\varphi) \, , \\
 \text{S}.52&=& +\frac{1}{3!} P(\alpha / \beta) \displaystyle\sum_{\substack{k_1 k_2 k_3 k_4 \\ k_5 k_6}} 
 \mathcal{T}^{\dagger}_{\alpha k_1 k_2 k_3}(\tau,\varphi)  \Omega^{04}_{\beta k_4 k_5 k_6} 
 R^{--}_{k_1 k_6} R^{--}_{k_2 k_5}(\varphi) R^{--}_{k_3 k_4}(\varphi) \, , \\
 \text{S}.53&=& +\frac{1}{4} P(\alpha / \beta) \displaystyle\sum_{\substack{k_1 k_2 k_3 k_4 \\ k_5 k_6}} \mathcal{T}^{\dagger}_{\alpha k_1}(\tau,\varphi) 
 \mathcal{T}^{\dagger}_{k_2 \beta}(\tau,\varphi) \Omega^{04}_{k_3 k_4 k_5 k_6} R^{--}_{k_1 k_6}(\varphi) R^{--}_{k_2 k_5}(\varphi) R^{--}_{k_4 k_3}(\varphi) \, , \\
 \text{S}.54&=& +\frac{1}{2} P(\alpha / \beta) \displaystyle\sum_{\substack{k_1 k_2 k_3 k_4 \\ k_5 k_6}} \mathcal{T}^{\dagger}_{\alpha k_1}(\tau,\varphi) 
 \mathcal{T}^{\dagger}_{k_2 k_3}(\tau,\varphi) \Omega^{04}_{\beta k_4 k_5 k_6} R^{--}_{k_1 k_6}(\varphi)  R^{--}_{k_2 k_5}(\varphi) R^{--}_{k_3 k_4}(\varphi) \, , \\
 \text{S}.55&=& +\frac{1}{4} \displaystyle\sum_{\substack{k_1 k_2 k_3 k_4 \\ k_5 k_6 k_7 k_8}} \mathcal{T}^{\dagger}_{\alpha \beta k_1 k_2}(\tau,\varphi)
 \mathcal{T}^{\dagger}_{k_3 k_4}(\tau,\varphi) \Omega^{04}_{k_5 k_6 k_7 k_8} R^{--}_{k_1 k_8}(\varphi) R^{--}_{k_2 k_7}(\varphi) R^{--}_{k_3 k_6}(\varphi) R^{--}_{k_4 k_5}(\varphi) \, , \\
 \text{S}.56&=& +\frac{1}{3!} P(\alpha / \beta)  \displaystyle\sum_{\substack{k_1 k_2 k_3 k_4 \\ k_5 k_6 k_7 k_8}} 
 \mathcal{T}^{\dagger}_{\alpha k_1 k_2 k_3}(\tau,\varphi) \mathcal{T}^{\dagger}_{k_4 \beta}(\tau,\varphi) 
 \Omega^{04}_{k_5 k_6 k_7 k_8} R^{--}_{k_1 k_8}(\varphi) R^{--}_{k_2 k_7}(\varphi) R^{--}_{k_3 k_6}(\varphi) R^{--}_{k_4 k_5}(\varphi) \, , \\
 \text{S}.57&=& +\frac{1}{4} P(\alpha / \beta) \displaystyle\sum_{\substack{k_1 k_2 k_3 k_4 \\ k_5 k_6 k_7 k_8}}
  \mathcal{T}^{\dagger}_{\alpha k_1}(\tau, \varphi) \mathcal{T}^{\dagger}_{k_2 k_3}(\tau,\varphi) \mathcal{T}^{\dagger}_{k_4 \beta}(\tau,\varphi)
 \Omega^{04}_{k_5 k_6 k_7 k_8} R^{--}_{k_1 k_8}(\varphi) R^{--}_{k_2 k_7}(\varphi) R^{--}_{k_3 k_6}(\varphi) R^{--}_{k_4 k_5}(\varphi) \, .
\end{eqnarray*}
\end{widetext}
The infinite time limit of these expressions is simply obtained by replacing $\mathcal{T}^{\dagger}_{n}(\tau,\varphi)$ everywhere by $\mathcal{T}^{\dagger}_{n}(\varphi)$.

\section{Matrix elements of $\tilde{\Omega}(\varphi)$ and $\tilde{A}(\varphi)$}
\label{transformME}

The matrix elements of the various normal-ordered contributions to the transformed grand potential $\tilde{\Omega}(\varphi)$ are expressed in terms of those of $\Omega$ through 
\begin{widetext}
\begin{subequations}
\label{e:me3defas}
\begin{align*}
\tilde{\Omega}^{00}(\varphi) & \equiv \tilde{\Omega}^{00(00)}(\varphi) + \tilde{\Omega}^{00(02)}(\varphi) + \tilde{\Omega}^{00(04)} (\varphi) \\ 
& = \Omega^{00} + \frac{1}{2} \displaystyle\sum_{k_1 k_2} \Omega^{02}_{k_1 k_2} R^{--}_{k_2 k_1}(\varphi)
+ \frac{1}{8} \displaystyle\sum_{k_1 k_2 k_3 k_4} \Omega^{04}_{k_1 k_2 k_3 k_4} R^{--}_{k_4 k_3}(\varphi) R^{--}_{k_2 k_1}(\varphi) \, , \nonumber \\
 \tilde{\Omega}^{11}_{k_1 k_2}(\varphi)  & \equiv  \tilde{\Omega}^{11(11)}_{k_1 k_2}(\varphi) + \tilde{\Omega}^{11(02)}_{k_1 k_2}(\varphi) + 
 \tilde{\Omega}^{11(13)}_{k_1 k_2}(\varphi) + \tilde{\Omega}^{11(04)}_{k_1 k_2}(\varphi) \\
& = \Omega^{11}_{k_1 k_2} + \displaystyle\sum_{k_3} \Omega^{02}_{k_2 k_3} R^{--}_{k_1 k_3}(\varphi) + \frac{1}{2} \displaystyle\sum_{k_3 k_4}
\Omega^{13}_{k_1 k_2 k_3 k_4} R^{--}_{k_4 k_3}(\varphi) + \frac{1}{2} \displaystyle\sum_{k_3 k_4 k_5} \Omega^{04}_{k_2 k_3 k_4 k_5} 
R^{--}_{k_1 k_5}(\varphi) R^{--}_{k_4 k_3}(\varphi)  \, ,  \nonumber \\
 \tilde{\Omega}^{20}_{k_1 k_2}(\varphi) & \equiv  \tilde{\Omega}^{20(20)}_{k_1 k_2}(\varphi) + \tilde{\Omega}^{20(11)}_{k_1 k_2}(\varphi) + 
 \tilde{\Omega}^{20(02)}_{k_1 k_2}(\varphi) + \tilde{\Omega}^{20(22)}_{k_1 k_2}(\varphi) +
 \tilde{\Omega}^{20(13)}_{k_1 k_2}(\varphi) + \tilde{\Omega}^{20(04)}_{k_1 k_2}(\varphi) \\
 \nonumber
& = \Omega^{20}_{k_1 k_2} + \displaystyle\sum_{k_3} \big[ \Omega^{11}_{k_1 k_3}R^{--}_{k_2 k_3}(\varphi) - 
\Omega^{11}_{k_2 k_3}R^{--}_{k_1 k_3}(\varphi) \big] + \displaystyle\sum_{k_3 k_4} \Omega^{02}_{k_4 k_3} R^{--}_{k_1 k_3}(\varphi)R^{--}_{k_2 k_4} (\varphi) \\
\nonumber 
& + \frac{1}{2} \displaystyle\sum_{k_3 k_4} \Omega^{22}_{k_1 k_2 k_3 k_4} R^{--}_{k_4 k_3}(\varphi) 
+ \frac{1}{2} \displaystyle\sum_{k_3 k_4 k_5} \big[\Omega^{13}_{k_1 k_3 k_4 k_5} R^{--}_{k_2 k_5}(\varphi) R^{--}_{k_4 k_3}(\varphi) - 
\Omega^{13}_{k_2 k_3 k_4 k_5}R^{--}_{k_1 k_5}(\varphi) R^{--}_{k_4 k_3}(\varphi) \big] \\
& + \frac{1}{2} \displaystyle\sum_{k_3 k_4 k_5 k_6}
\Omega^{04}_{k_3 k_4 k_5 k_6} R^{--}_{k_1 k_6}(\varphi)R^{--}_{k_2 k_5}(\varphi) R^{--}_{k_4 k_3}(\varphi)  \, ,  \nonumber \\
 \tilde{\Omega}^{02}_{k_1 k_2}(\varphi) & \equiv \tilde{\Omega}^{02(02)}_{k_1 k_2}(\varphi) + \tilde{\Omega}^{02(04)}_{k_1 k_2}(\varphi) \\
& = \Omega^{02}_{k_1 k_2} + \frac{1}{2} \displaystyle\sum_{k_3 k_4} \Omega^{04}_{k_1 k_2 k_3 k_4} R^{--}_{k_4 k_3} (\varphi) \, ,  \nonumber \\
\tilde{\Omega}^{22}_{k_1 k_2 k_3 k_4}(\varphi) &\equiv \tilde{\Omega}^{22(22)}_{k_1 k_2 k_3 k_4}(\varphi) + \tilde{\Omega}^{22(13)}_{k_1 k_2 k_3 k_4}(\varphi) + \tilde{\Omega}^{22(04)}_{k_1 k_2 k_3 k_4}(\varphi)  \\
	 & = \Omega^{22}_{k_1 k_2 k_3 k_4} + \displaystyle\sum_{k_5} \big[ \Omega^{13}_{k_1 k_3 k_4 k_5} R^{--}_{k_2 k_5}(\varphi) - \Omega^{13}_{k_2 k_3 k_4 k_5} R^{--}_{k_1 k_5}(\varphi) \big] + \displaystyle\sum_{k_5 k_6} \Omega^{04}_{k_3 k_4 k_5 k_6} R^{--}_{k_1 k_6}(\varphi)R^{--}_{k_2 k_5}(\varphi) \, , \nonumber  \\
 \tilde{\Omega}^{31}_{k_1 k_2 k_3 k_4}(\varphi) &\equiv  \tilde{\Omega}^{31(31)}_{k_1 k_2 k_3 k_4}(\varphi) + \tilde{\Omega}^{31(22)}_{k_1 k_2 k_3 k_4}(\varphi) + 
 \tilde{\Omega}^{31(13)}_{k_1 k_2 k_3 k_4}(\varphi) + \tilde{\Omega}^{31(04)}_{k_1 k_2 k_3 k_4}(\varphi)\\
\nonumber
& = \Omega^{31}_{k_1 k_2 k_3 k_4} + \displaystyle\sum_{k_5} \big[ \Omega^{22}_{k_2 k_3 k_4 k_5} R^{--}_{k_1 k_5} (\varphi) + \Omega^{22}_{k_1 k_2 k_4 k_5} R^{--}_{k_3 k_5}(\varphi) - \Omega^{22}_{k_1 k_3 k_4 k_5} R^{--}_{k_2 k_5}(\varphi) \big] \\
\nonumber
& + \displaystyle\sum_{k_5 k_6} \big[ \Omega^{13}_{k_1 k_4 k_5 k_6} R^{--}_{k_2 k_6} (\varphi) R^{--}_{k_3 k_5} (\varphi) + \Omega^{13}_{k_3 k_4 k_5 k_6} R^{--}_{k_1 k_6} (\varphi) R^{--}_{k_2 k_5}(\varphi) - \Omega^{13}_{k_2 k_4 k_5 k_6} R^{--}_{k_1 k_6} (\varphi) R^{--}_{k_3 k_5}(\varphi) \big] \\
& + \displaystyle\sum_{k_5 k_6 k_7} \Omega^{04}_{k_4 k_5 k_6 k_7} R^{--}_{k_1 k_7}(\varphi) R^{--}_{k_2 k_6} (\varphi)  R^{--}_{k_3 k_5}(\varphi) \, , \nonumber  \\
\tilde{\Omega}^{13}_{k_1 k_2 k_3 k_4}(\varphi) &\equiv  \tilde{\Omega}^{13(13)}_{k_1 k_2 k_3 k_4}(\varphi) + \tilde{\Omega}^{13(04)}_{k_1 k_2 k_3 k_4}(\varphi) \\
& = \Omega^{13}_{k_1 k_2 k_3 k_4} + \displaystyle\sum_{k_5} \Omega^{04}_{k_2 k_3 k_4 k_5} R^{--}_{k_1 k_5} (\varphi) \, , \nonumber  \\
\tilde{\Omega}^{40}_{k_1 k_2 k_3 k_4}(\varphi) &\equiv  \tilde{\Omega}^{40(40)}_{k_1 k_2 k_3 k_4}(\varphi) + \tilde{\Omega}^{40(31)}_{k_1 k_2 k_3 k_4}(\varphi) + \tilde{\Omega}^{40(22)}_{k_1 k_2 k_3 k_4}(\varphi) +  \tilde{\Omega}^{40(13)}_{k_1 k_2 k_3 k_4}(\varphi) + \tilde{\Omega}^{40(04)}_{k_1 k_2 k_3 k_4}(\varphi) \\
\nonumber
& = \Omega^{40}_{k_1 k_2 k_3 k_4} + 
\displaystyle\sum_{k_5} \big[ \Omega^{31}_{k_1 k_2 k_3 k_5} R^{--}_{k_4 k_5} (\varphi) - \Omega^{31}_{k_2 k_3 k_4 k_5} R^{--}_{k_1 k_5}(\varphi) - \Omega^{31}_{k_1 k_2 k_4 k_5} R^{--}_{k_3 k_5}(\varphi) + \Omega^{31}_{k_1 k_3 k_4 k_5} R^{--}_{k_2 k_5}(\varphi) \big]  \\
\nonumber 
& + \displaystyle\sum_{k_5 k_6} \big[ \Omega^{22}_{k_1 k_4 k_5 k_6} R^{--}_{k_2 k_6} (\varphi) R^{--}_{k_3 k_5} (\varphi) + \Omega^{22}_{k_4 k_3 k_5 k_6} R^{--}_{k_2 k_6} (\varphi) R^{--}_{k_1 k_5}(\varphi) - \Omega^{22}_{k_4 k_2 k_5 k_6} R^{--}_{k_3 k_6} (\varphi) R^{--}_{k_1 k_5}(\varphi) \\
\nonumber
&+ \Omega^{22}_{k_1 k_2 k_5 k_6} R^{--}_{k_3 k_6} (\varphi) R^{--}_{k_4 k_5} (\varphi) + \Omega^{22}_{k_2 k_3 k_5 k_6} R^{--}_{k_1 k_6} (\varphi) R^{--}_{k_4 k_5}(\varphi) - \Omega^{22}_{k_1 k_3 k_5 k_6} R^{--}_{k_2 k_6} (\varphi) R^{--}_{k_4 k_5}(\varphi) \big] \\
\nonumber
&+ \displaystyle\sum_{k_5 k_6 k_7} \big[ \Omega^{13}_{k_3 k_5 k_6 k_7} R^{--}_{k_1 k_7} (\varphi) R^{--}_{k_2 k_6} (\varphi) R^{--}_{k_4 k_5} (\varphi) - \Omega^{13}_{k_2 k_5 k_6 k_7} R^{--}_{k_1 k_7} (\varphi) R^{--}_{k_3 k_6}(\varphi) R^{--}_{k_4 k_5} (\varphi) \\
\nonumber
& + \Omega^{13}_{k_1 k_5 k_6 k_7} R^{--}_{k_2 k_7} (\varphi) R^{--}_{k_3 k_6} (\varphi) R^{--}_{k_4 k_5} (\varphi) - \Omega^{13}_{k_4 k_5 k_6 k_7} R^{--}_{k_2 k_7} (\varphi) R^{--}_{k_3 k_6}(\varphi) R^{--}_{k_1 k_5} (\varphi) \big] \\
& + \displaystyle\sum_{k_5 k_6 k_7 k_8} \Omega^{04}_{k_5 k_6 k_7 k_8} R^{--}_{k_1 k_8}(\varphi) R^{--}_{k_2 k_7} (\varphi)  R^{--}_{k_3 k_6}(\varphi) R^{--}_{k_4 k_5} (\varphi) \, , \nonumber  \\
   \tilde{\Omega}^{04}_{k_1 k_2 k_3 k_4}(\varphi) &\equiv  \tilde{\Omega}^{04(04)}_{k_1 k_2 k_3 k_4}(\varphi) \\
   &= \Omega^{04}_{k_1 k_2 k_3 k_4} \, .  \nonumber 
\end{align*}
\end{subequations}

Similarly for $A$
\begin{subequations}
\label{AA}
\begin{align*}
\tilde{A}^{00}(\varphi) & \equiv \tilde{A}^{00(00)}(\varphi) + \tilde{A}^{00(02)}(\varphi) \\ 
& = A^{00} + \frac{1}{2} \displaystyle\sum_{k_1 k_2} A^{02}_{k_1 k_2} R^{--}_{k_2 k_1}(\varphi) \, , \nonumber  \\
 \tilde{A}^{11}_{k_1 k_2}(\varphi)  & \equiv  \tilde{A}^{11(11)}_{k_1 k_2}(\varphi) + \tilde{A}^{11(02)}_{k_1 k_2}(\varphi)  \\
& = A^{11}_{k_1 k_2} + \displaystyle\sum_{k_3} A^{02}_{k_2 k_3} R^{--}_{k_1 k_3}(\varphi)  \, , \nonumber  \\
 \tilde{A}^{20}_{k_1 k_2}(\varphi) & \equiv  \tilde{A}^{20(20)}_{k_1 k_2}(\varphi) + \tilde{A}^{20(11)}_{k_1 k_2}(\varphi) + 
 \tilde{A}^{20(02)}_{k_1 k_2}(\varphi)  \\
 \nonumber
& = A^{20}_{k_1 k_2} + \displaystyle\sum_{k_3} \big[ A^{11}_{k_1 k_3}R^{--}_{k_2 k_3}(\varphi) - 
A^{11}_{k_2 k_3}R^{--}_{k_1 k_3}(\varphi) \big] + \displaystyle\sum_{k_3 k_4} A^{02}_{k_4 k_3} R^{--}_{k_1 k_3}(\varphi)R^{--}_{k_2 k_4} (\varphi) \, ,  \nonumber \\
 \tilde{A}^{02}_{k_1 k_2}(\varphi) & \equiv \tilde{A}^{02(02)}_{k_1 k_2}(\varphi)  \\
& = A^{02}_{k_1 k_2}   \, .  \nonumber 
\end{align*}
\end{subequations}
\end{widetext}

\section{Useful identities}
\label{usefulID}

\begin{widetext}
\begin{subequations}
\label{integral} 
\begin{eqnarray}
\int_{0}^{\tau}d\tau_1 \, e^{a\tau_1} &=& \frac{1}{a}\Big(e^{\tau a}-1\Big) \, , \label{integral1} \\
\int_{0}^{\tau}d\tau_{1}d\tau_{2}\,\theta\left(  \tau_{1}-\tau_{2}\right)
e^{a\left(  \tau_{1}-\tau_{2}\right)  } &=&\int_{0}^{\tau}d\tau_{1} \, e^{a\tau_{1}}\int_{0}^{\tau_{1}}d\tau
_{2} \, e^{-a\tau_{2}} = -\frac{\tau}{a}+\frac{1}{a^{2}%
}\Big(  e^{\tau a}-1\Big) \, , \label{integral2} \\
\int_{0}^{\tau}d\tau_{1}d\tau_{2}\,\theta\left(  \tau_{1}-\tau_{2}\right)
e^{a\tau_{1}+b\tau_{2}} &=& \int_{0}^{\tau}d\tau_{1} \, e^{a\tau_{1}}\int_{0}^{\tau_{1}}d\tau
_{2} \, e^{b\tau_{2}} = \frac{1}{b\left(  a+b\right)  } \Big(
e^{\tau\left(  a+b\right)  }-1\Big)   -\frac{1}{ab}\Big(  e^{\tau a}-1\Big) \,  . \label{integral3} 
\end{eqnarray}
\end{subequations}
\end{widetext}
Given that such integrals only appear in the theory with $a<0$ and $a+b<0$, one obtains 
\begin{subequations}
\label{limitintegral} 
\begin{eqnarray}
\lim\limits_{\tau \to \infty} \int_{0}^{\tau}d\tau \, e^{a\tau} &=& -\frac{1}{a} \, , \label{limitintegral1} \\
\lim\limits_{\tau \to \infty} \int_{0}^{\tau}d\tau_{1}d\tau_{2}\,\theta\left(  \tau_{1}-\tau_{2}\right)
e^{a\left(  \tau_{1}-\tau_{2}\right)  } &=& -\frac{\tau}{a}-\frac{1}{a^{2}}  \, , \label{limitintegral2} \\
\lim\limits_{\tau \to \infty}  \int_{0}^{\tau}d\tau_{1}d\tau_{2}\,\theta\left(  \tau_{1}-\tau_{2}\right)
e^{a\tau_{1}+b\tau_{2}} &=& \frac{1}{a(a+b)} \,  , \label{limitintegral3} 
\end{eqnarray}
\end{subequations}
where the first and third result are necessarily positive.

\end{appendix}

\bibliography{PNR-BCC}

\end{document}